\documentstyle[12pt]{article}
\input amssym.def
\input amssym.tex

\newcommand{\qed}{\rule{3mm}{3mm}}
\newcommand{\eto}[1]{\mbox{$e^{\displaystyle #1}$}}

\newcommand{\g}{\mbox{\boldmath$g$}}
\newcommand{\G}{\mbox{\boldmath$G$}}

\newcommand{\rH}{{\rm H}}

\newcommand{\ga}{{\goth a}}
\newcommand{\gb}{{\goth b}}
\newcommand{\gc}{{\goth c}}
\newcommand{\gd}{{\goth d}}

\newcommand{\bM}{{\bf M}}
\newcommand{\bN}{{\bf N}}

\newcommand{\ba}{{\bf a}}
\newcommand{\bb}{{\bf b}}
\newcommand{\bc}{{\bf c}}
\newcommand{\bd}{{\bf d}}
\newcommand{\bg}{{\bf g}}

\newcommand{\bq}{{\bf q}}

\newcommand{\bu}{{\bf u}}
\newcommand{\bv}{{\bf v}}
\newcommand{\bw}{{\bf w}}
\newcommand{\bx}{{\bf x}}

\newcommand{\mbA}{\mbox{\boldmath$A$}}
\newcommand{\mbB}{\mbox{\boldmath$B$}}
\newcommand{\mbC}{\mbox{\boldmath$C$}}
\newcommand{\mbD}{\mbox{\boldmath$D$}}

\newcommand{\mbL}{\mbox{\boldmath$L$}}

\newcommand{\mba}{\mbox{\boldmath$a$}}

\newcommand{\cB}{{\cal B}}

\newcommand{\cK}{{\cal K}}
\newcommand{\cL}{{\cal L}}
\newcommand{\cM}{{\cal M}}

\newcommand{\cP}{{\cal P}}
\newcommand{\cR}{{\cal R}}

\newcommand{\cT}{{\cal T}}
\newcommand{\cV}{{\cal V}}
\newcommand{\cW}{{\cal W}}
\newcommand{\cX}{{\cal X}}

\newcommand{\wx}{\widetilde{x}}
\newcommand{\wa}{\widetilde{a}}
\newcommand{\wb}{\widetilde{b}}
\newcommand{\wc}{\widetilde{c}}
\newcommand{\wid}{\widetilde{d}}
\newcommand{\wq}{\widetilde{q}}
\newcommand{\wv}{\widetilde{v}}
\newcommand{\wu}{\widetilde{u}}
\newcommand{\ww}{\widetilde{w}}

\newcommand{\wL}{\widetilde{L}}
\newcommand{\wT}{\widetilde{T}}
\newcommand{\wU}{\widetilde{U}}
\newcommand{\wV}{\widetilde{V}}
\newcommand{\wW}{\widetilde{W}}

\newtheorem{theorem}{Theorem}[section]

\begin{document}
\begin{titlepage}
\title{Integrable discretizations for lattice systems:\\
local equations of motion \\and their Hamiltonian properties}
\author{Yuri B. SURIS\\Center for Complex Systems and Visualization\\
Universit\"at Bremen, Universit\"atsallee 29\\
28359 Bremen, GERMANY}
\end{titlepage}
\maketitle

{\small {\bf Abstract.} We develop the approach to the problem of integrable
discretization based on the notion of $r$--matrix hierarchies. One of
its basic features is the coincidence of Lax matrices of discretized systems
with the Lax matrices of the underlying continuous time systems. A common 
feature of the discretizations obtained in this approach is non--locality. 
We demonstrate how to overcome this drawback. Namely, we introduce the notion 
of localizing changes of variables and construct such changes of variables 
for a large number of examples, including
the Toda and the relativistic Toda lattices, the Volterra lattice and its 
integrable perturbation, the second flows of the Toda and of the Volterra 
hierarchies, the modified Volterra lattice, the Belov--Chaltikian lattice, 
the Bogoyavlensky lattices, the Bruschi--Ragnisco lattice. We also introduce 
a novel class of constrained lattice KP systems, discretize all of them, 
and find the corresponding localizing change of variables. Pulling back the 
differential equations of motion under the localizing changes of variables, 
we find also (sometimes novel) integrable one--parameter deformations of 
integrable lattice systems. Poisson properties of the localizing changes of 
variables are also studied: they produce interesting one--parameter 
deformations of the known Poisson algebras.}

\newpage
\tableofcontents
\setcounter{equation}{0}
\section{Introduction}\label{Introduction}
This paper deals with some aspects of the following general problem: 
how to discretize one or several of independent variables in a given
integrable system, maintaining the integrability property? We call
this {\it the problem of integrable discretization}. 

To assure the coincidence of the qualitative properties of the discretized
models with that of the continuous ones becomes one of the central ideas
of the modern numerical analysis, which therefore comes to a close
interplay with different aspects of the theory of dynamical systems. One of
the most advanced examples of this approach is the symplectic integration, 
on which recently the first monograph appeared \cite{SSC}. 

The problem of integrable discretization constitutes another aspect of
this general line of thinking. It arose in the course
of development of the theory of solitons. This theory, born exactly 30
years ago, has grown so tremendously, that it is difficult to keep an
overview of the whole variety of different notions and contexts of 
integrability, not to say about the concrete results. (Recall that 
recently there appeared a thick book under the title ''What is 
integrability?'' \cite{Zakh}).

Correspondingly, various approaches to the problem of integrable discretization 
are currently available. They began to be discussed sporadically in the soliton 
literature starting from the mid--70s. Following ones should be mentioned:
\begin{enumerate}
\item An approach based on the representation of an integrable
system as a compatibility condition of two auxiliary
linear problems. A natural proposition is to discretize one or the both
of them \cite{AL1}. This, however, can be made in a great 
variety of ways, cf., for instance, different spatial discretizations
of the nonlinear Schr\"odinger equation and of the sine--Gordon equation, 
found in \cite{AL1} and in \cite{IK}. One attempt of fixing discretization 
crystallized with the development of the Hamiltonian approach. Namely,  
Faddeev and Takhtajan, based on the experience of the Leningrad soliton school,
formulated in \cite{FT2} the following rule for a transition from models 
with one continuous space variable to lattice models: the $r$--matrix 
should be preserved, the linear Poisson bracket being replaced by the quadratic
one. See \cite{FT2} for a collection of examples showing productivity of this 
approach.
\item One of the most intriguing and universal approaches is the Hirota's 
one \cite{H}, based on the notion of the $\tau$--function and on a bilinear 
representation of integrable systems. It seems to be able to produce discrete 
versions of the majority of soliton equations, but still remains somewhat 
mysterious, and the mechanism behind it is yet to be fully understood. One 
successful way to do it was proposed in \cite{DJM}, where also a large number 
of integrable discretizations was derived. Among the most interesting products
of this approach is the so called Hirota--Miwa equation \cite{H1}, \cite{Mi},
which is sometimes claimed to contain ''everything'', i.e. the majority if not 
all soliton equations (continuous and discrete) are particular or limiting cases 
of this single equation, cf. \cite{Zab}.
\item A fruitful method is based on the ''direct linearization'' \cite{NCW},
\cite{QNCV}, \cite{CWN}, \cite{WC}, \cite{NPCQ}, \cite{NC}. Its basic idea 
is to derive integrable nonlinear differential equations 
which are satisfied by the solutions of certain linear integral equations. 
A large variety of continuous and discrete soliton equations has been obtained 
on this way. 
\item Approach based on the variational principle (discrete Lagrangian 
equations), combined with matrix factorizations \cite{V}, \cite{MV},
\cite{DLT2}. Historically, it was the work of Veselov and Moser that
consolidated the more or less isolated results to a separate branch
of the theory of integrable systems.
\item Considering stationary and restricted flows of soliton hierarchies,
closely related to the ''nonlinearization'' of spectral problems, often
leads to interesting discrete equations \cite{QRT}, \cite{R}, \cite{RR}.
\item Differential equations describing various geometric problem (surfaces of
the constant mean curvature, motion of the curve in the space, etc.) turn
out to be integrable \cite{Sym}, \cite{B1}, \cite{B2}. Correspondingly, a 
discretization of geometric notions naturally leads to discrete integrable 
equations \cite{BP1}, \cite{BP2}, \cite{DS}, \cite{Do}.
\item There exist integrable discretizations which
belong to the most beautiful examples, but were derived by guess, without any
systematical approach \cite{S1}, \cite{S5}.
\item Last but not least we menton an approach to the temporal discretization
in which the auxiliary spectral problem is not discretized at all. In other
words, the basic feature of this approach is maintaining the Lax matrix of
the continuous time system. The first example of this approach is the work
by Ablowitz--Ladik \cite{AL2}, futher developed towards the practical algorithms
in \cite{TA}. This feature was also put in the basis of the work by Gibbons and
Kupershmidt \cite{GK}, \cite{K2}. The discretizations found in all these papers
were somewhat unsatisfactory from the esthetical point of view, namely they 
suffered from being nonlocal, as opposed to the underlying continuous time 
systems. Morever, these authors did not recognize the connection with the 
factorization problem, which did not allow them to identify their 
discretizations as certain members of the corresponding hierarchies and to 
establish the Poisson properties of these discretizations. This led Gibbons 
and Kupershmidt to call this method ''the method of the bizarre ansatz''.
\end{enumerate}

Recently the author pushed forward the last mentioned approach to the 
problem of integrable discretization, putting it in a connection with the
$r$--matrix theory of integrable hierarchies (see \cite{RSTS} for a review 
of this theory). In this context the method could be understood properly, 
and became rather natural and simple. It was applied to a number of integrable
lattice systems \cite{S6}--\cite{S11}, \cite{S13}, \cite{S14}. Its clear
advantage is universality.  The method is in principle applicable 
to any system admitting an $r$--matrix interpretation, which is the 
common feature of the great majority of the known integrable systems. 
As for the drawback of nonlocality, there exist several ways to repair it.
The first one, connected with the notion of discrete time Newtonian 
equations of motion, was followed in \cite{S6}, \cite{S7}, \cite{S10}, 
\cite{S11}. A splitting of complicated flows into superpositions of more
simple ones was used in \cite{S13}, \cite{S14}. The present work is devoted 
to another way connected with the so called localizing changes of variables.

The paper has the following structure. In Sect. \ref{Sect problem} we give
an accurate formulation of the problem of integrable discretization.
Sect. \ref{Sect Lax},\ref{Sect factorization} are devoted to a general framework 
of integrable $r$--matrix hierarchies of Lax equations on associative algebras.
In Sect. \ref{Sect recipe} we formulate a general recipe of integrable 
discretization, and in Sect. \ref{Sect local} we introduce the notion of 
localizing changes of variables and discuss their general properties. 
Sect. \ref{Sect notations} contains the description of algebras used for
the analysis of all integrable lattice systems in this paper.
The rest of the paper is devoted to a detailed elaboration of a number of
examples, including the most prominent ones, such as the Toda lattice and the
Volterra lattice, and less well known ones, such as the Belov--Chaltikian 
lattice.


\setcounter{equation}{0}
\section{The problem of integrable discretization}\label{Sect problem}

Let us formulate the problem of integrable discretization more precisely.
Let $\cX$ be a Poisson manifold with a Poisson bracket $\{\cdot,\cdot\}$.
Let $H$ be a completely integrable Hamilton function on $\cX$, i.e. the
system 
\begin{equation}\label{Ham syst}
\dot{x}=\{H,x\}
\end{equation}
possesses many enough functionally independent integrals $I_k(x)$ in 
involution. 

The problem consists in finding a map $\cX\mapsto\cX$ described by a formula
\begin{equation}\label{Ham map}
\wx=\Phi(x;h)
\end{equation}
depending on a small parameter $h>0$, and satisfying the following requirements: 
\begin{enumerate}
\item The map (\ref{Ham map}) is a discrete time approximation for the flow 
(\ref{Ham syst}) in the following sense:
\begin{equation}\label{approx}
\Phi(x;h)=x+h\{H,x\}+O(h^2)
\end{equation}
(of course, one might require also a higher order of approximation). In all
our considerations and formulas we pay a special attention to a simple
and transparent control of the continuous limit $h\to 0$.
\item The map (\ref{Ham map}) is Poisson with respect to the bracket 
$\{\cdot,\cdot\}$ on $\cX$ or with respect to some its deformation
$\{\cdot,\cdot\}_h$ such that $\{\cdot,\cdot\}_h=\{\cdot,\cdot\}+O(h)$.
\item The map (\ref{Ham map}) is integrable, i.e. possesses the necessary
number of independent integrals in involution $I_k(x;h)$ approximating the
integrals of the original system: $I_k(x;h)=I_k(x)+O(h)$.
\end{enumerate}


\setcounter{equation}{0}
\section{Lax representations}\label{Sect Lax}
Our approach to the problem of integrable discretization is applicable to any 
system allowing an $r$--matrix interpretation, but we formulate the basic 
recipe in a simplified form, applicable to systems with
a Lax representation of one of the following types:
\begin{equation}\label{Lax in recipe}
\dot{L}=\Big[\,L,\pi_+(f(L))\,\Big]=-\Big[\,L,\pi_-(f(L))\,\Big]
\end{equation}
or
\begin{equation}\label{Lax triads in recipe}
\dot{L}_j\; =\; L_j\cdot\pi_+\Big(f(T_{j-1})\Big)-\pi_+\Big(f(T_j)\Big)\cdot L_j
\; = \; -\,L_j\cdot\pi_-\Big(f(T_{j-1})\Big)+\pi_-\Big(f(T_j)\Big)\cdot L_j
\end{equation}
Let us discuss the notations. 

Let $\g$ be an associative algebra, supplied with a nondegenerate scalar 
product, which allows to identify $\g^*$ with $\g$. One can introduce in $\g$ 
the structure of Lie algebra in a standard way. Let $\g_+$, $\g_-$ be
two subalgebras such that as a vector space $\g$ is a direct sum
$\g=\g_+\oplus \g_-$. Denote by $\pi_{\pm}:\g\mapsto\g_{\pm}$ the corresponding
projections. Finally, let $f:\g\mapsto\g$ be an ${\rm Ad}^*$--covariant
function on $\g$, and let $L$ stand for a generic element of $\g$. 
Then (\ref{Lax in recipe}) is a certain differential equation on $\g$. 

Further, let $\bg=\bigotimes_{j=1}^m\g$ be a direct product of $m$ copies
of the algebra $\g$. A generic element of $\bg$ is denoted by
$\mbL=(L_1,\ldots,L_m)$. We use also the notation
\begin{equation}
T_j=T_j(\mbL)=L_j\cdot\ldots\cdot L_1\cdot L_m\cdot\ldots\cdot L_{j+1}
\end{equation}
Then (\ref{Lax triads in recipe}) is a certain differential equation on $\bg$.
Such equations are sometimes called {\it Lax triads}.

One says that (\ref{Lax in recipe}), resp. (\ref{Lax triads
in recipe}), is a Lax representation of the flow (\ref{Ham syst}), if
there exists a map $L:\cX\mapsto\g$ (resp. $\mbL:\cX\mapsto\bg$)
such that the former equations of motion are equivalent to the latter ones.
Let us stress that when considering equations (\ref{Lax in recipe}), resp. 
(\ref{Lax triads in recipe}) in the role of Lax representation, the letter 
$L$ (resp. $\mbL$) does not stand for a generic element of the corresponding 
algebra any more; rather, it represents the elements of the images of the maps 
$L:\cX\mapsto\g$ and $\mbL:\cX\mapsto\bg$, correspondingly. The elements $L(x)$, resp. $\mbL(x)$ 
(and the map $L$, resp. $\mbL$, itself) are called {\it Lax matrices}. 

Equations (\ref{Lax in recipe}) and (\ref{Lax triads in recipe}) have several
remarkable features. First of all, they are Hamiltonian with respect to a 
certain {\it linear $r$--matrix Poisson bracket} on $\g$ (resp. on $\bg$) 
\cite{STS}. Moreover, under some natural conditions (for example, if $\g_+$ 
and $\g_-$ serve as orthogonal complements to each other) one can define 
{\it quadratic} and {\it cubic} $r$--matrix brackets on $\g$, compatible 
with the linear one, such that the equations above are Hamiltonian with 
respect to all three brackets \cite{STS}, \cite{LP}, \cite{OR}, \cite{S6}, 
\cite{S12}.

The Lax representations are especially useful, if the maps $L$, resp. $\mbL$,
corresponding to the Lax matrices, are Poisson with respect to one of the
above mentioned $r$--matrix brackets; then the manifold consisting of the Lax
matrices is a Poisson submanifold. In such cases one says that the Lax 
representation admits an $r$--matrix interpretation.


\setcounter{equation}{0}
\section{Factorization theorems}\label{Sect factorization}

As a further remarkable feature of the  equations (\ref{Lax in recipe}) and 
(\ref{Lax triads in recipe}) we consider the possibility to solve them
explicitly in terms of a certain factorization problem 
in the Lie group $\G$ corresponding to $\g$ \cite{Sy}, \cite{STS}, 
\cite{RSTS}. (Actually, this can be done even in a more general situation 
of hierarchies governed by $R$--operators satisfying the so--called {\it 
modified Yang--Baxter equation}, see \cite{RSTS}). The factorization problem 
is described by the equation
\begin{equation}\label{fact problem}
U=\Pi_+(U)\Pi_-(U),\quad U\in\G,\quad\Pi_{\pm}(U)\in \G_{\pm}
\end{equation}
where $\G_{\pm}$ are two subgroups of $\G$ with the Lie algebras $\g_{\pm}$, 
respectively. This problem has a unique solution in a certain neighbourhood
of the group unit. In what follows we suppose that $\G$ is a matrix group, and 
write the coadjoint action of the group elements on $\g^*\simeq\g$ as a 
conjugation by the corresponding matrices. [In this context we write 
$\Pi_{\pm}^{-1}(U)$ for $\Big(\Pi_{\pm}(U)\Big)^{-1}$]. Correspondingly, 
we call ${\rm Ad}^*$--covariant functions $\g\mapsto\g$ also ''conjugation
covariant''. This notation has an additional advantage of being applicable
also to functions $\g\mapsto\G$. 

For the history of the following fundamental theorem and its different proofs
the reader is referred to \cite{RSTS}.

\begin{theorem} \label{split kinematics}
Let $f:\g\mapsto\g$ be a conjugation covariant function. Then 
the solution of the differential equation {\rm (\ref{Lax in recipe})} 
with the initial condition $L(0)=L_0$ is given, at least for $t$ small
enough, by
\begin{equation}\label{solution cont}
L(t) = \Pi_+^{-1}\!\left(\eto{tf(L_0)}\right)L_0\,
\Pi_+\!\left(\eto{tf(L_0)}\right)
= \Pi_-\!\left(\eto{tf(L_0)}\right)L_0\,\Pi_-^{-1}\!\left(\eto{tf(L_0)}\right)
\end{equation}
\end{theorem}
{\bf Proof. } We give a proof based on direct and simple verification.
Denote
\[
{\cal L}(t)=\Pi_+\!\left(\eto{tf(L_0)}\right),\qquad
{\cal R}(t)=\Pi_-\!\left(\eto{tf(L_0)}\right)
\]
so that
\begin{equation}\label{e to tf}
\eto{tf(L_0)}={\cal L}(t)\,{\cal R}(t),\qquad {\cal L}(t)\in\G_+,\quad
{\cal R}(t)\in\G_- 
\end{equation}
Now we set
\begin{equation}\label{def T cont}
L(t)={\cal L}^{-1}(t)\,L_0\,{\cal L}(t)={\cal R}(t)\,L_0\,{\cal R}^{-1}(t)
\end{equation}
(these two expressions for $L(t)$ are equal due to ${\rm Ad^*}$--covariance of 
$f(L)$), and check by direct calculation that this $L(t)$ satisfies the 
differential equation (\ref{Lax in recipe}). The theorem will follow by the 
uniqueness of solution. We see immediately that $L(t)$ satisfies the following 
Lax type equation:
\[
\dot{L}=[L,{\cal L}^{-1}\dot{{\cal L}}]=-[L,\dot{{\cal R}}\,{\cal R}^{-1}]
\]
and it remains to show that
\[
{\cal L}^{-1}\dot{{\cal L}}=\pi_+(f(L)),\qquad
\dot{{\cal R}}\,{\cal R}^{-1}=\pi_-(f(L))
\]
Since, obviously, $\cL^{-1}\dot{\cL}\in\g_+$,$\;\dot{\cR}\,\cR^{-1}\in\g_-$,
we need to demonstrate only that
\begin{equation}\label{L,U}
{\cal L}^{-1}\dot{{\cal L}}+\dot{{\cal R}}\,{\cal R}^{-1}=f(L)
\end{equation}
To do this, we differentiate (\ref{e to tf}) and derive, using 
${\rm Ad^*}$--covariance of $f$ and the definition (\ref{def T cont}):
\[
\dot{{\cal L}}\,{\cal R}+{\cal L}\,\dot{{\cal R}}=\eto{tf(L_0)}f(L_0)=
{\cal L}\,{\cal R}f(L_0)={\cal L}\,f(L)\,{\cal R}
\]
This is equivalent to (\ref{L,U}). \qed
\vspace{2mm}

For an arbitrary conjugation covariant function $F:\g\mapsto\G$ one can 
define the map ${\rm B}_F:\g\mapsto\g$ according to the formula 
\begin{equation}\label{BT}
\wL={\rm B}_F(L)=\Pi_+^{-1}\left(F(L)\right)\cdot L\cdot\Pi_+\left(F(L)\right)
= \Pi_-\left(F(L)\right)\cdot L\cdot \Pi_-^{-1}\left(F(L)\right)
\end{equation}
(''B'' is for ''B\"acklund'').
Theorem \ref{split kinematics} shows that the flows defined by the 
differential equations (\ref{Lax in recipe}) consist of maps having such a
form with $F(L)=\eto{tf(L)}$. The very remarkable feature of such maps is 
their commutativity for different $F$'s. 
\begin{theorem}\label{superposition of BTs}
For two arbitrary conjugation covariant functions $F_1,F_2:\g\mapsto\G$
\begin{equation}\label{BT superp}
{\rm B}_{F_2}\circ{\rm B}_{F_1}={\rm B}_{F_2F_1}
\end{equation}
and therefore the maps ${\rm B}_{F_1}$, ${\rm B}_{F_2}$ commute.
\end{theorem}
{\bf Proof.} Denote:
\[
L_1={\rm B}_{F_1}(L),\qquad L_2={\rm B}_{F_2}\circ{\rm B}_{F_1}(L)=
{\rm B}_{F_2}(L_1)
\]
So, by definition we have:
\begin{equation}\label{BT aux1}
L_1=\cL_1^{-1}L\cL_1=\cR_1 L\cR_1^{-1},\qquad
L_2=\cL_2^{-1}L_1\cL_2=\cR_2 L_1\cR_2^{-1}
\end{equation}
where the matrices $\cL_i\in\G_+$, $\cR_i\in\G_-$ $(i=1,2)$ come from the 
following factorizations:
\[
F_1(L)=\cL_1\cR_1, \qquad F_2(L_1)=\cL_2\cR_2
\]
From (\ref{BT aux1}) we have:
\begin{equation}\label{BT aux2}
L_2=\cL^{-1}L\cL=\cR L\cR^{-1},\quad{\rm where}\quad\cL=\cL_1\cL_2\in\G_+,
\quad \cR=\cR_2\cR_1\in\G_-
\end{equation}
Now the following chain of equalities holds:
\[
\cL\cR=\cL_1\cL_2\cR_2\cR_1=\cL_1 F_2(L_1)\cR_1=
\cL_1 F_2(\cL_1^{-1}L\cL_1)\cR_1=F_2(L)\cL_1\cR_1=F_2(L)F_1(L)
\]
In view of (\ref{BT aux2}) we get:
\[
\cL=\Pi_+\Big(F_2(L)F_1(L)\Big),\qquad \cR=\Pi_-\Big(F_2(L)F_1(L)\Big)
\]
and the theorem is proved. \qed
\vspace{2mm}

Theorem \ref{superposition of BTs} implies that the flows of two
arbitrary differential equations of the form (\ref{Lax in recipe}) commute. 
Another important consequence of Theorem \ref{superposition of BTs} is the 
following discrete--time counterpart of Theorem \ref{split kinematics}, going
back to \cite{Sy}.
\begin{theorem} \label{discr split kinematics} 
Let $F:\g\mapsto\G$ be a conjugation covariant function.
Then the solution of the difference equation
\begin{equation}\label{dLax in recipe}
\wL=\Pi_+^{-1}(F(L))\cdot L\cdot\Pi_+(F(L))=\Pi_-(F(L))\cdot L\cdot
\Pi_-^{-1}(F(L))
\end{equation}
where $L=L(n)$, $\wL=L(n+1)$, with the initial condition $L(0)=L_0$, is given by
\begin{equation}\label{solution discr}
L(n) = \Pi_+^{-1}\left(F^n(L_0)\right)\cdot L_0\cdot\Pi_+\left(F^n(L_0)\right)
= \Pi_-\left(F^n(L_0)\right)\cdot L_0\cdot\Pi_-^{-1}\left(F^n(L_0)\right)
\end{equation}
\end{theorem}
{\bf Proof.} From Theorem \ref{superposition of BTs} there follows by induction
that $({\rm B}_F)^n={\rm B}_{F^n}$. \qed
\vspace{2mm}

Comparing the formulas (\ref{solution discr}), (\ref{solution cont}), we see 
that the map (\ref{dLax in recipe}) is the time $h$ shift along the trajectories
of the flow (\ref{Lax in recipe}) with
\[
f(L)=h^{-1}\log(F(L))
\]

The above results are purely kinematic, in the sense that no additional
Hamiltonian structure is necessary neither to formulate nor to prove them.
However, as mentioned above, the equations (\ref{Lax in recipe}) often 
admit a Hamiltonian or even a multi--Hamiltonian interpretation. If this is
the case, then we get some useful additional information. In particular, 
all maps (\ref{dLax in recipe}) are Poisson with respect to the invariant 
Poisson bracket of the hierarchy (\ref{Lax in recipe}), being shifts along the
trajectories of Hamiltonian flows. Further, if 
the set of Lax matrices $L({\cal X})$ for the system at hand forms a Poisson 
submanifold for one of the $r$--matrix brackets on $\g$, then this manifold 
is left invariant by the flows (\ref{Lax in recipe}) and by the maps
(\ref{dLax in recipe}). The functions on $\cX$ of the form $I\circ L$, where 
$I$ are conjugation invariants of $\g$, are integrals of motion of the 
corresponding systems and in involution with respect to $\{\cdot,\cdot\}$.

We close this section by giving analogous results for the Lax equations 
on the direct products $\bg=\bigotimes_{j=1}^m\g$. 
\begin{theorem}\label{split Lax on big algebras}
For a conjugation covariant function $f:\g\mapsto\g$ the solution of the
equation {\rm(\ref{Lax triads in recipe})} with the initial value $\mbL(0)$ 
is given, at least for $t$ small enough, by the formula
\begin{eqnarray}
L_j(t) & = & \Pi_+^{-1}\left(\eto{tf(T_j(0))}\right)\cdot
L_j(0)\cdot\Pi_+\left(\eto{tf(T_{j-1}(0))}\right)
\nonumber\\
& = & \Pi_-\left(\eto{tf(T_j(0))}\right)\cdot
L_j(0)\cdot\Pi_-^{-1}\left(\eto{tf(T_{j-1}(0))}\right)
\label{split big Lax solution}
\end{eqnarray}
\end{theorem}
\begin{theorem}\label{split dLax on big algebras}
For a conjugation covariant function $F:g\mapsto G$ consider the following
system of difference equations on $\bg$:
\begin{equation}\label{dLax triads in recipe}
\wL_j \;= \; \Pi_+^{-1}\!\left(F(T_j)\right)\cdot
L_j\cdot\Pi_+\!\left(F(T_{j-1})\right)
\; = \; \Pi_-\!\left(F(T_j)\right)\cdot
L_j\cdot\Pi_-^{-1}\!\left(F(T_{j-1})\right)
\end{equation}
Its solution with the initial value $\mbL(0)$ is given by the formula
\begin{eqnarray}
L_j(n) & = & \Pi_+^{-1}\Big(F^n(T_j(0))\Big)\cdot
L_j(0)\cdot\Pi_+\Big(F^n(T_{j-1}(0))\Big)
\nonumber\\
& = & \Pi_-\Big(F^n(T_j(0))\Big)\cdot
L_j(0)\cdot\Pi_-^{-1}\Big(F^n(T_{j-1}(0))\Big)
\label{split big dLax solution}
\end{eqnarray}
\end{theorem}
The {\bf proofs} are kinematic 
and absolutely parallel to the case of the ''small'' algebras $\g$.


\setcounter{equation}{0}
\section{Recipe for integrable discretization}\label{Sect recipe}
The results of the previous section inspire the following recipe for 
integrable discretization, clearly formulated for the first time
in \cite{S6}, \cite{S7}.
\vspace{2mm}

{\bf Recipe.} Suppose you are looking for an integrable discretization
of an integrable system (\ref{Ham syst}) allowing a Lax representation 
of the form (\ref{Lax in recipe}). Then as a solution of your task you may 
take the difference equation {\rm(\ref{dLax in recipe})}
with the same Lax matrix $L$ and some conjugation covariant function
$F:\g\mapsto\G$ such that
\[
F(L)=I+hf(L)+O(h^2)
\]
Analogously, if your system has a Lax representation of the form (\ref{Lax 
triads in recipe}) on the algebra $\bg$, then you may 
take as its integrable discretization the difference Lax equation
(\ref{dLax triads in recipe}) with $F$ as above.
\vspace{2mm}

Of course, this prescription makes sense only if the corresponding factors 
$\Pi_{\pm}(F(L))$ [resp. $\Pi_{\pm}(F(T_j))$] admit more or less explicit 
expressions, allowing to write down the corresponding difference equations in 
a more or less closed form. 
The choice of $F$ is a transcendent problem, which however turns out to be 
solvable for many of (hopefully, for the majority of or even for all) the known 
integrable systems. The simplest possible choice $F(L)=I+hf(L)$
works perfectly well for a vast set of examples considered below.

Let us stress the advantages of this approach to the problem of
integrable discretization.
\begin{itemize}
\item Although we formulate our recipe only for systems with Lax 
representations of the particular form, it is in fact more universal.
Almost without changes it may be applied to any system whose Lax representation
is governed by an $R$--operator satisfying the modified Yang--Baxter equation. 
\item The discretizations obtained in this way share the Lax matrix and 
therefore the integrals of motion with their underlying continuous time systems.
\item If the Lax representation (\ref{Lax in recipe}) [resp. (\ref{Lax triads 
in recipe})] allows an $r$--matrix interpretation, then our discretizations 
share also the invariant Poisson bracket with the underlying continuous time
systems. In particular, if the Lax matrices $L$ [resp. $\mbL$] form a Poisson 
submanifold for some $r$--matrix bracket, then this submanifold is left 
invariant by the corresponding Poisson map (\ref{dLax in recipe}) [resp. 
(\ref{dLax triads in recipe})].
\item The initial value problem for our discrete time equations can be solved
in terms of the same factorization in a Lie group as the initial value problem
for the continuous time system.
\item Interpolating Hamiltonian flows also belong to the set of granted 
by--products of this approach.
\end{itemize}


\setcounter{equation}{0}
\section{Localizing changes of variables}\label{Sect local}
Along with these advantageous properties our recipe has also an important
drawback: it produces, as a rule, nonlocal difference equations, when applied
to lattice systems with local interactions. Under {\it locality} we understand
the following property: in some coordinates $(x_1,\ldots, x_N)$ on $\cX$
the equations of motion (\ref{Ham syst}) have the form
\begin{equation}\label{loc syst}
\dot{x}_k=\phi_k(x)=\phi_k(x_k,x_{k\pm 1},\ldots,x_{k\pm s})
\end{equation}
with a {\it fixed} $s\in{\Bbb N}$. Nonlocal difference equations produced 
by our scheme have the form
\begin{equation}\label{nonloc discr}
\wx_k=x_k+h\Phi_k(x;h), \qquad \Phi_k(x,0)=\phi_k(x)
\end{equation}
where $\Phi_k$ depends explicitly on all $x_j$, not only on $2s$ nearest 
neighbours of $x_k$. The aim of the present paper is to demonstrate on a
large number of examples how this drawback can be overcome, i.e. how to bring
the latter difference equations into a local form; the price we have to pay
is that they become implicit. 

The general strategy will be to find {\it localizing changes of 
variables} $\cX(\bx)\mapsto\cX(x)$ such that in the variables $\bx$ the map
(\ref{nonloc discr}) may be written as
\begin{equation}\label{loc discr}
\widetilde{\bx}_k=\bx_k+h\Psi_k(\bx,\widetilde{\bx};h), \qquad 
\Psi_k(x,x;0)=\phi_k(x)
\end{equation}
where $\Psi_k$ depends only on the $\bx_j$'s and $\widetilde{\bx}_j$'s 
with correct indices $|j-k|\le s$. Such implicit local equations of motion 
are much better suited for the purposes of numerical simulation and are much 
more satisfactory from the esthetical point of view. Moreover, in all our 
examples the functions $\Psi_k$ actually depend only on $\bx_j$'s 
with $k\le j\le k+s$ and on $\widetilde{\bx}_j$'s with $k-s\le j\le k$, which
makes the practical implementation of the corresponding difference equations
even more effective (if, for instance, one uses the Newton's iterative 
method to solve (\ref{loc discr}) for $\widetilde{\bx}$, then one has to solve 
only linear systems whose matrices are {\it triangular} and have a band 
structure, i.e. only $s$ nonzero diagonals). Further, it has to be remarked 
that, when considered as equations on the lattice $(t,k)$, the equations 
(\ref{loc discr}) often allow transformations of independent variables 
(mixing $t$ and $k$) bringing these equations into local {\it explicit} 
form (cf. \cite{PNC}). This last remark will be the subject of a separate 
publication.
 
It is by no means evident that such localizing changes of variables exist, but 
we give in this paper a large number of examples which, hopefully, will convince 
the reader that this is indeed the case. The probably first examples appeared
in the context of the Bogoyavlensky lattices in \cite{S8}, and will be reproduced
here in a clarified form for the sake of completeness.

The localizing changes of variables turn out to have many additional 
remarkable properties. They are always given by the formulas
\begin{equation}\label{loc map}
x_k=\bx_k+h\Xi_k(\bx;h)
\end{equation}
with {\it local} functions $\Xi_k$. However, the inverse change of variables
is always nonlocal. Therefore nothing guarantees {\it a priori} that the 
pull--back of the differential equations of motion (\ref{loc syst}) under the
change of variables (\ref{loc map}) will be given by local formulas. 
Nevertheless, this turns out to be the case. This gives a way of producing 
(sometimes novel) one--parameter families of integrable local deformations
of lattice systems (see \cite{K1} for a general concept and some examples 
of integrable deformations).

The system (\ref{loc syst}) often admits one or several invariant {\it local}
Poisson brackets. Nothing guarantees {\it a priori} that the pull--backs of
these brackets under the change of variables (\ref{loc map}) are also given by 
local formulas. Indeed, as a rule these pull--backs are non--local. However, 
in the multi--Hamiltonian case it often turns out that pull--backs of 
certain linear combinations of invariant Poisson brackets are local again!

These facts still wait to be completely understood. It seems that
the remarkable properties of the localizing maps have the same nature as
that of the Miura maps (''miraculous cancellations''). Moreover, actually
our localizing changes of variables {\it are} Miura maps, and in this image
some of them already appeared in \cite{K1}. However, the observation that
they bring integrable discretizations into a local form, seems to be completely
new. The scope of the present work is restricted to elaborating a large number 
of examples of localizing changes of variables, along with their Poisson 
properties, in a hope to attract the attention of the soliton community to 
these fascinating and beautiful objects.


\setcounter{equation}{0}
\section{Basic algebras and operators}\label{Sect notations}

Two concrete algebras play the basic role in our presentation. 
They are well suited to describe various lattice systems with the so called 
open--end and periodic boundary conditions, respectively. Here are
the relevant definitions.

For the {\it open--end case} we always set $\g=gl(N)$, the algebra of
$N\times N$ matrices with the usual matrix product, the Lie bracket 
$[u,v]=uv-vu$, and the nondegenerate bi--invariant scalar product
$\langle u,v\rangle={\rm tr}(u\cdot v)$.  As a linear space, $\g$ may be 
presented as a direct sum
\[
\g=\g_+\oplus\g_-
\]
where the subalgebras $\g_+$ and $\g_-$ consist of lower triangular and of 
strictly upper triangular matrices, respectively. 

The Lie group $\G$ corresponding to the Lie algebra $\g$ is $GL(N)$, the group
of $N\times N$ nondegenerate matrices. The subgroups $\G_+$, $\G_-$ 
corresponding to the Lie algebras $\g_+$, $\g_-$ consist of nondegenerate
lower triangular matrices and of upper triangular matrices with unit diagonal,
respectively. The $\Pi_+\Pi_-$ factorization is well known in the linear 
algebra under the name of the $LU$ {\it factorization}.

In the {\it periodic case} we always choose as $\g$ a certain {\it twisted 
loop algebra} over $gl(N)$. A loop algebra over $gl(N)$ is an algebra of 
Laurent polynomials with coefficients from $gl(N)$ and a natural commutator
$[u\lambda^j,v\lambda^k]=[u,v]\lambda^{j+k}$. Our twisted algebra $\g$ is a 
subalgebra singled out by the additional condition
\[
\g=\left\{u(\lambda)\in gl(N)[\lambda,\lambda^{-1}]:
\Omega u(\lambda)\Omega^{-1}=u(\omega\lambda)\right\},
\]
where $\Omega={\rm diag}(1,\omega,\ldots,\omega^{N-1})$, 
$\omega=\exp(2\pi i/N)$. In other words, elements of $\g$ satisfy
\begin{equation}
u(\lambda)=
\sum_{p}
\sum_{j-k\equiv p \atop({\rm mod}\,N)}
\lambda^p u_{jk}^{(p)}E_{jk}
\end{equation}
(Here and below $E_{jk}$ stands for the matrix whose only nonzero entry is on 
the intersection of the $j$th row and the $k$th column and is equal to 1).
The nondegenerate bi--invariant scalar product is chosen as 
\begin{equation}
\langle u(\lambda),\,v(\lambda)\rangle={\rm tr}(u(\lambda)\cdot v(\lambda))_0
\end{equation}
the subscript 0 denoting the free term of the formal Laurent series. This 
scalar product allows to identify $\g^*$ with $\g$.

As a linear space, $\g$ is again a direct sum
\[
\g=\g_+\oplus\g_-
\]
with the subalgebras
\begin{equation}
\g_+=\bigoplus_{k\ge 0}\lambda^k\g_k,\qquad 
\g_-=\bigoplus_{k<0}\lambda^k\g_k
\end{equation}

The group $\G$ corresponding to the Lie algebra $\g$ is a {\it twisted loop 
group}, consisting of $GL(N)$--valued functions $U(\lambda)$ of the complex 
parameter $\lambda$, regular in ${\Bbb C}P^1\backslash\{0,\infty\}$ and 
satisfying $\Omega U(\lambda)\Omega^{-1}=U(\omega\lambda)$. Its subgroups
$\G_+$ and $\G_-$ corresponding to the Lie algebras $\g_+$ and $\g_-$,  
are singled out by the following conditions: 
\begin{itemize}
\item $U(\lambda)\in\G_+$ are regular in the neighbourhood of $\lambda=0$;
\item $U(\lambda)\in\G_-$ are regular in the neighbourhood of 
$\lambda=\infty$ and  $U(\infty)=I$.
\end{itemize}
We call the corresponding $\Pi_+\Pi_-$ factorization the {\it generalized
$LU$ factorization}. It is uniquely defined in a certain neighbourhood of
the unit element of $\G$. As opposed to the open--end case, finding the
generalized $LU$ factorization is a problem of the Riemann--Hilbert type
which is solved in terms of algebraic geometry rather than in terms of linear 
algebra.


\setcounter{equation}{0}
\section{Toda lattice}\label{Chapter Toda}

\subsection{Equations of motion and tri--Hamiltonian structure}
\label{Sect Toda motivation}

The equations of motion of the Toda lattice (hereafter TL) read:
\begin{equation}\label{TL}
\dot{a}_k=a_k(b_{k+1}-b_k),\qquad \dot{b}_k=a_k-a_{k-1},
\qquad 1\le k\le N,
\end{equation}
with one of the two types of boundary conditions: open--end
($a_0=a_N=0$), or periodic (all subscripts are taken (mod $N$), so that
$a_0\equiv a_N$, $b_{N+1}\equiv b_1$). 

The phase space of the TL in the case of the periodic boundary conditions is
\begin{equation}\label{TL phase sp}
\cT={\Bbb R}^{2N}(b_1,a_1,\ldots,b_N,a_N)
\end{equation}
There exist three compatible local Poisson brackets on $\cT$ 
such that the system TL is Hamiltonian with respect to each one of them
\cite{A}, \cite{K1}, see also \cite{D}. 
We adopt once and forever the following conventions: the Poisson 
brackets will be defined by writing down {\it all nonvanishing} brackets between 
the coordinate functions; the indices in the corresponding formulas are
taken (mod $N$).

The ''linear'' Poisson structure on $\cT$ is defined by the brackets
\begin{equation}\label{TL l br}
\{b_k,a_k\}_1=-a_k, \qquad \{a_k,b_{k+1}\}_1=-a_k 
\end{equation}
the corresponding Hamilton function for the flow TL is given by:
\begin{equation}\label{TL H2}
\rH_2(a,b)=\frac{1}{2}\sum_{k=1}^N b_k^2+\sum_{k=1}^{N}a_k
\end{equation}
The ''quadratic'' Poisson structure has the following definition:
\begin{equation}\label{TL q br}
\begin{array}{cclcccl}
\{b_k,a_k\}_2 & = & -a_kb_k,&          \quad & 
\{a_k,b_{k+1}\}_2 & = & -a_kb_{k+1}  \\ \\
\{a_k,a_{k+1}\}_2 & = & -a_{k+1}a_k,& \quad & 
\{b_k,b_{k+1}\}_2 & = & -a_k 
\end{array}
\end{equation}
The Hamilton function generating TL in this bracket is:
\begin{equation}\label{TL H1}
\rH_1(a,b)=\sum_{k=1}^N b_k
\end{equation}
Finally, the ''cubic'' bracket on $\cT$ is given by the relations
\begin{equation}\label{TL c br}
\begin{array}{cclcccl}
\{b_k,a_k\}_3     & = & -a_k(b_k^2+a_k),     & \quad &
\{a_k,b_{k+1}\}_3 & = & -a_k(b_{k+1}^2+a_k), \\ \\
\{a_k,a_{k+1}\}_3 & = & -2a_ka_{k+1}b_{k+1}, & \quad &
\{b_k,b_{k+1}\}_3 & = & -a_k(b_k+b_{k+1}),   \\ \\
\{a_k,b_{k+2}\}_3 & = & -a_ka_{k+1},         & \quad & 
\{b_k,a_{k+1}\}_3 & = & -a_ka_{k+1}
\end{array}
\end{equation}
The expression for the corresponding Hamilton function, suitable in both the 
periodic and the open--end case, is nonlocal in the coordinates $(a,b)$. 
However, in the periodic case one has an alternative Hamilton function:
\begin{equation}\label{TL H0}
\rH_0(a,b)=\frac{1}{2}\sum_{k=1}^N \log(a_k)
\end{equation}

\subsection{Lax representation}\label{Sect Toda Lax}

The Lax representation of the Toda lattice \cite{F}, \cite{M} lives in the 
algebra $\g$ introduced in Sect. \ref{Sect notations}. 
Actually, there exist different versions of the Lax representation connected
with different ways to represent the algebra $\g$ as a direct sum of its two
subalgebras \cite{DLT1}. We discuss here only the $LU$ version which 
corresponds in the open--end case to the decomposition of an arbitrary matrix 
into the sum of a lower triangular and a strictly upper triangular matrices.

The Lax matrix $T:\cT\mapsto\g$ of the TL corresponding to the 
generalized $LU$ decomposition is:
\begin{equation}\label{TL T}
T(a,b,\lambda)=\lambda^{-1}\sum_{k=1}^{N} a_kE_{k,k+1}+\sum_{k=1}^N b_kE_{k,k}
+\lambda\sum_{k=1}^{N} E_{k+1,k}
\end{equation}
We use here and below a convention according to which in the periodic 
case $E_{N+1,N}=E_{1,N}$, $E_{N,N+1}=E_{N,1}$; in the open--end case
$E_{N+1,N}=E_{N,N+1}=0$ and we may set $\lambda=1$.

The equations of motion {\rm (\ref{TL})} are equivalent to the Lax equations
\begin{equation}\label{TL Lax}
\dot{T}=[T,B_+]=-[T,B_-]
\end{equation}
with
\begin{eqnarray}
B_+(a,b,\lambda) & = & \pi_+(T(a,b,\lambda))\;=\;
\sum_{k=1}^Nb_kE_{k,k}+\lambda\sum_{k=1}^{N} E_{k+1,k}
\label{TL B+}\\
B_-(a,b,\lambda) & = & 
\pi_-(T(a,b,\lambda))\;=\;\lambda^{-1}\sum_{k=1}^{N} a_kE_{k,k+1}
\label{TL B-}
\end{eqnarray}
where $\pi_{\pm}\,:\,\g\mapsto\g_{\pm}$ are the projections to the subalgebras
$\g_{\pm}$ defined as in Sect. \ref{Sect notations} 
(the generalized $LU$ decomposition).

Spectral invariants of the Lax matrix $T(a,b,\lambda)$ serve as integrals of 
motion of this system. Note that all Hamilton functions in different 
Hamiltonian formulations belong to these spectral invariants. For instance,
\[
\rH_2(a,b)=\frac{1}{2}\Big({\rm tr}\, T^2(a,b,\lambda)\Big)_0,\qquad
\rH_1(a,b)=\Big({\rm tr}\, T(a,b,\lambda)\Big)_0
\]
where the subscript ''0'' is used to denote the free term of the corresponding
Laurent series.
 
{\it All} spectral invariants turn out to be in involution with respect to
each of the Poisson brackets (\ref{TL l br}), (\ref{TL q br}), (\ref{TL c br}). 
Most directly it follows from the $r$--matrix interpretation of the Lax 
equation (\ref{TL Lax}), which can be given for all three brackets \cite{AM},
\cite{DLT1}, \cite{OR}, \cite{S5}, \cite{MP}. 

\subsection{Discretization}
\label{Sect discretization TL}

In order to find an integrable time discretization for the flow TL, we apply 
the recipe of Sect. \ref{Sect recipe} with $F(T)=I+hT$, i.e. we take as 
a solution of this problem the map described by the discrete time Lax equation
\begin{equation}\label{dTL Lax}
\wT=\mbB_+^{-1}T\mbB_+=\mbB_-T\mbB_-^{-1} \quad{\rm with}\quad 
\mbB_{\pm}=\Pi_{\pm}(I+hT)
\end{equation}

\begin{theorem}\label{discrete TL} {\rm\cite{S6} (see also 
\cite{GK})}. 
The discrete time Lax equation {\rm(\ref{dTL Lax})}
is equivalent to the map $(a,b)\mapsto(\wa,\wb)$ described by the following
equations:
\begin{equation}\label{dTL}
\wa_k=a_k\;\frac{\beta_{k+1}}{\beta_k}, \qquad 
\wb_k=b_k+h\left(\frac{a_k}{\beta_k}-\frac{a_{k-1}}{\beta_{k-1}}\right)
\end{equation}
where the functions $\beta_k=\beta_k(a,b)=1+O(h)$ are uniquely defined by
the recurrent relation
\begin{equation}\label{dTL beta}
\beta_k=1+hb_k-\frac{h^2a_{k-1}}{\beta_{k-1}}
\end{equation}
and have the asymptotics
\begin{equation}\label{dTL beta as}
\beta_k=1+hb_k+O(h^2)
\end{equation}
\end{theorem}

{\bf Remark.} The matrices $\mbB_{\pm}$ have the following expressions:
\begin{equation}\label{dTL B+}
\mbB_+(a,b,\lambda)=\Pi_+\Big(I+hT\Big)=
\sum_{k=1}^N\beta_kE_{k,k}+h\lambda\sum_{k=1}^{N}E_{k+1,k}
\end{equation}
\begin{equation}\label{dTL B-}
\mbB_-(a,b,\lambda)=\Pi_-\Big(I+hT\Big)=
I+h\lambda^{-1}\sum_{k=1}^{N}\frac{a_k}{\beta_k}E_{k,k+1}
\end{equation}
\vspace{1.5mm}

\noindent
{\bf Proof.} The bi--diagonal structure of the factors $\mbB_{\pm}$, as well
as the expressions for the entries of $\mbB_-$, follow from the tri--diagonal
structure of the matrix $T$. The recurrent relation (\ref{dTL beta}) for the
entries of $\mbB_+$ is equivalent to $\mbB_+\mbB_-=I+hT$. The equations of 
motion (\ref{dTL}) are now nothimg but the componentwise form of the matrix
equation $\mbB_+\wT=T\mbB_+$. \qed
\vspace{2mm}

The map (\ref{dTL}) will be denoted dTL. Due to the asymptotics 
(\ref{dTL beta as}) it is easy to see that the equations of motion (\ref{dTL}) 
of the dTL serve as a difference approximation to the Toda flow TL (\ref{TL}). 
The construction assures numerous positive properties of this discretization: 
the map dTL is Poisson with respect to each one of the Poisson brackets 
(\ref{TL l br}), (\ref{TL q br}), (\ref{TL c br}), it has the same integrals of 
motion as the flow TL, the Lax representation for the dTL lies in its very
definition, etc. The only unpleasant property of the equations of motion 
(\ref{dTL}), when compared with their continuous time counterparts (\ref{TL}),
is the nonlocality. The source of nonlocality are the functions $\beta_k$.
In the open--end case they have explicit expressions in terms of finite 
continued fractions:
\[
\beta_k=1+hb_k-\frac{h^2a_{k-1}}{1+hb_{k-1}-\;
\raisebox{-3mm}{$\ddots$}
\raisebox{-4.5mm}{$\;-\displaystyle\frac{h^2a_1}{1+hb_1}$}}
\]
In the periodic case for small $h$ the $\beta_k$'s may be expressed as analogous
infinite (periodic) continued fractions.

\subsection{Local equations of motion for dTL}
Fortunately, there exist different ways to bring these equations of motion into
a local form connected with a localizing change of variables
in the sense of Sect. \ref{Sect local}. Consider another copy of the 
phase space $\cT$.  The coordinates in this another copy will be denoted by 
$\ba_k$, $\bb_k$. The localizing change of variables for dTL is
the map $\cT(\ba,\bb)\mapsto\cT(a,b)$  defined by the following formulas:
\begin{equation}\label{dTL loc map}
a_k=\ba_k(1+h\bb_k),\qquad b_k=\bb_k+h\ba_{k-1}
\end{equation}
The implicit functions theorem assures that this is a local diffeomorphism
between the two copies of $\cT$ by small enough values of $h$. This change
of variables appeared in \cite{K1} as a Miura map in connection with the
problem of deformation of integrable systems, but without any relation to
integrable discretizations.
\begin{theorem}\label{local dTL}
The change of variables {\rm(\ref{dTL loc map})} conjugates the map {\rm dTL} 
with the map \newline $(\ba,\bb)\mapsto(\widetilde{\ba},\widetilde{\bb})$
described by the following equations of motion:
\begin{equation}\label{dTL loc}
\widetilde{\ba}_k(1+h\widetilde{\bb}_k)=\ba_k(1+h\bb_{k+1}),\qquad
\widetilde{\bb}_k=\bb_k+h(\ba_k-\widetilde{\ba}_{k-1})
\end{equation}
\end{theorem}
{\bf Proof.} The key point is the following observation: the auxiliary
functions $\beta_k$ aquire in the new coordinates local expressions, namely,
\begin{equation}\label{dTL loc beta}
\beta_k=1+h\bb_k
\end{equation}
To demonstrate this, it is enough to notice that from (\ref{dTL loc map})
there follows that the quantities $1+h\bb_k$ satisfy the same recurrent 
relation as the quantities $\beta_k$, namely
\[
1+h\bb_k=1+hb_k-\frac{h^2a_{k-1}}{1+h\bb_{k-1}}
\]
Due to the uniqueness of the solution  with the asymptotics $1+O(h)$ the
formula (\ref{dTL loc beta}) is proved. The statement of the theorem follows
now immediately from (\ref{dTL}), (\ref{dTL loc beta}). \qed
\vspace{1.5mm}

Of course, the map (\ref{dTL loc}) is Poisson with respect to pull--backs of 
the three invariant Poisson structures of the Toda lattice. These pull--backs
are described by highly nonlocal and non--polynomial formulas. However, there 
exist certain linear combinations of the basic Poisson structures in 
coordinates $(a,b)$ whose pull--backs to the coordinates $(\ba,\bb)$ are local. 
\begin{theorem}\label{dTL loc invariant PB}
{\rm a)} The pull--back of the bracket
\begin{equation}\label{dTL loc m1 br}
\{\cdot,\cdot\}_1+h\{\cdot,\cdot\}_2
\end{equation}
on $\cT(a,b)$ under the change of variables {\rm(\ref{dTL loc map})} is the
following bracket on $\cT(\ba,\bb)$:
\begin{equation}\label{dTL loc br 1}
\{\bb_k,\ba_k\}=-\ba_k(1+h\bb_k),\qquad \{\ba_k,\bb_{k+1}\}=-\ba_k(1+h\bb_{k+1})
\end{equation}
{\rm b)} The pull--back of the bracket 
\begin{equation}\label{dTL loc m2 br}
\{\cdot,\cdot\}_2+h\{\cdot,\cdot\}_3
\end{equation}
on $\cT(a,b)$ under the change of variables {\rm(\ref{dTL loc map})} is the
following bracket on $\cT(\ba,\bb)$:
\begin{eqnarray}
\{\bb_k,\ba_k\}=-\ba_k(\bb_k+h\ba_k)(1+h\bb_k),&\quad&
\{\ba_k,\bb_{k+1}\}=-\ba_k(\bb_{k+1}+h\ba_k)(1+h\bb_{k+1})\nonumber\\ 
\{\ba_k,\ba_{k+1}\}=-\ba_k\ba_{k+1}(1+h\bb_{k+1}),&\quad&
\{\bb_k,\bb_{k+1}\}=-\ba_k(1+h\bb_k)(1+h\bb_{k+1})\nonumber\\
\label{dTL loc br 2}
\end{eqnarray}
{\rm c)} The brackets {\rm(\ref{dTL loc br 1}), (\ref{dTL loc br 2})}
are compatible. The map {\rm(\ref{dTL loc})} is Poisson with respect
to both of them.
\end{theorem}
{\bf Proof.} To prove the theorem, one has, for example, in the (less 
laborious) case a) to verify the following statement: the formulas 
(\ref{dTL loc br 1}) imply that the nonvanishing pairwise Poisson brackets 
of the functions (\ref{dTL loc map}) are 
\begin{eqnarray*}
\{b_k,a_k\}-a_k(1+hb_k), & \qquad & \{a_k,b_{k+1}\}=-a_k(1+hb_{k+1})\\
\{a_k,a_{k+1}\}= -ha_ka_{k+1}, &\qquad & \{b_k,b_{k+1}\}=-ha_k
\end{eqnarray*}
This verification consists of straightforward calculations. In what follows
we do not repeat analogous arguments in the similar situations. \qed
\vspace{2mm}

The map (\ref{dTL loc}) was first found in \cite{HTI}, along with the Lax 
representation. In \cite{So} it was stressed that this map is nothing other 
than the so--called $qd$ algorithm well known in the numerical analysis.
Its Poisson structure and its place in the continuous time Toda hierarchy 
were not discussed in \cite{HTI}. The previous theorem provides 
a bi--Hamiltonian structure of the $qd$ algorithm. This result in a slightly 
different form was found in \cite{S6}. 
\vspace{2mm}

\begin{theorem} {\rm \cite{K1}}. 
The pull--back of the flow {\rm TL} under the change of variables {\rm(\ref{dTL 
loc map})} is described by the following differential equations:
\begin{equation}\label{TL in loc map}
\dot{\ba}_k=\ba_k(\bb_{k+1}-\bb_k),\qquad 
\dot{\bb}_k=(\ba_k-\ba_{k-1})(1+h\bb_k)
\end{equation}
\end{theorem}
{\bf Proof.} To determine the pull--back of the flow TL, we can use the
Hamiltonian formalism. An opportunity to apply it is given by the Theorem
\ref{dTL loc invariant PB}. We shall use the statement a) only.
Consider the function $h^{-1}\rH_1(a,b)=h^{-1}\sum_{k=1}^N b_k$.
It is a Casimir of the bracket $\{\cdot,\cdot\}_1$, and generates exactly
the flow TL in the bracket $h\{\cdot,\cdot\}_2$. Hence it generates the flow
TL also in the bracket (\ref{dTL loc m1 br}). The pull--back of this Hamilton
function is equal to $h^{-1}\sum_{k=1}^N(\bb_k+h\ba_{k-1})$. It remains only 
to calculate the flow generated by this function in the Poisson brackets 
(\ref{dTL loc br 1}). This results in the equations of motion 
(\ref{TL in loc map}). \qed

\setcounter{equation}{0}
\section{Second flow of the Toda hierarchy}
\label{Sect Toda 2}

Now we want to demonstrate that our method for finding integrable
discretizations and local equations of motion for them works not only for
the flow TL, but equally well for the higher flows of the Toda
hierarchy. We consider here the second flow (called hereafter TL2).

\subsection{Equations of motion and tri--Hamiltonian structure}
This is the flow on $\cT$ governed by the differential equations
\begin{equation}\label{TL2}
\dot{a}_k=a_k(b_{k+1}^2-b_k^2+a_{k+1}-a_{k-1}),\quad
\dot{b}_k=a_k(b_{k+1}+b_k)-a_{k-1}(b_k+b_{k-1})
\end{equation}
This flow is Hamiltonian with respect to all three brackets {\rm (\ref{TL l br}),
(\ref{TL q br}), (\ref{TL c br})}. The corresponding Hamilton functions are:
\begin{equation}\label{TL2 l H}
\rH_3(a,b)=\frac{1}{3}\sum_{k=1}^N b_k^3+\sum_{k=1}^N b_k(a_k+a_{k-1})=
\frac{1}{3}\Big({\rm tr}\, T^3(a,b,\lambda)\Big)_0
\end{equation}
for the bracket $\{\cdot,\cdot\}_1$, $\rH_2(a,b)$ for the bracket 
$\{\cdot,\cdot\}_2$, and $\rH_1(a,b)$ for the bracket $\{\cdot,\cdot\}_3$.

\subsection{Lax representation}
Naturally, the Lax matrix for the flow TL2 is the same as for the flow TL.
The difference lies in the auxiliary matrices $B_{\pm}$ taking part in the
Lax representation (since in this section we are dealing only with the flow 
TL2,  using the same notations $B_{\pm}$ as in Sect. \ref{Chapter Toda} 
will not lead to confusions; the same holds also for some other notations in 
this section).

The equations of motion {\rm(\ref{TL2})} are equivalent to the Lax equations
in $\g$
\begin{equation}\label{TL2 Lax}
\dot{T}=[T,B_+]=-[T,B_-]\quad{\rm with}\quad B_{\pm}=\pi_{\pm}(T^2)
\end{equation}
so that
\begin{eqnarray}
B_+(a,b,\lambda) & = & 
\sum_{k=1}^N(b_k^2+a_k+a_{k-1})E_{k,k}+\lambda\sum_{k=1}^N
(b_{k+1}+b_k)E_{k+1,k}+\lambda^2\sum_{k=1}^N E_{k+2,k} \label{TL2 B+}\\
B_-(a,b,\lambda) & = & 
\lambda^{-1}\sum_{k=1}^N (b_{k+1}+b_k)a_kE_{k,k+1}+
\lambda^{-2}\sum_{k=1}^N a_{k+1}a_kE_{k,k+2} \label{TL2 B-}
\end{eqnarray}

\subsection{Discretization}
In order to obtain an integrable discretization of the flow TL2
we can apply the recipe of Sect. \ref{Sect recipe} with $F(T)=I+hT^2$,
i.e. consider the map described by the discrete time Lax equation
\begin{equation}\label{dTL2 Lax}
\widetilde{T}=\mbB_+^{-1}T\mbB_+=\mbB_-T\mbB_-^{-1}\quad
{\rm with}\quad \mbB_{\pm}=\Pi_{\pm}(I+hT^2)
\end{equation}

\begin{theorem}\label{discrete Toda 2}
The discrete time Lax equation {\rm(\ref{dTL2 Lax})}
is equivalent to the map $(a,b)\mapsto(\wa,\wb)$ described by the following
equations:
\begin{equation}\label{dTL2}
\wa_k=a_k\,\frac{\beta_{k+1}}{\beta_k}, \qquad 
\wb_k=b_k+h\left(\frac{a_k\delta_k}{\beta_k}-
\frac{a_{k-1}\delta_{k-1}}{\beta_{k-1}}\right)
\end{equation}
where the auxiliary functions $\delta_k=\delta_k(a,b)=O(1)$ and 
$\beta_k=\beta_k(a,b)=1+O(h)$ are uniquely defined for $h$ small enough by the 
recurrent relations
\begin{equation}\label{dTL2 delta}
\delta_k=b_{k+1}+b_k-\frac{ha_{k-1}\delta_{k-1}}{\beta_{k-1}}
\end{equation}
\begin{equation}\label{dTL2 beta}
\beta_k=1+h(b_k^2+a_k+a_{k-1})-\frac{h^2a_{k-1}\delta_{k-1}^2}{\beta_{k-1}}
-\frac{h^2a_{k-1}a_{k-2}}{\beta_{k-2}}
\end{equation}
and have the asymptotics
\begin{equation}\label{dTL2 delta as}
\delta_k=b_{k+1}+b_k+O(h)
\end{equation}
\begin{equation}\label{dTL2 beta as}
\beta_k=1+h(b_k^2+a_k+a_{k-1})+O(h^2)
\end{equation}
\end{theorem}

{\bf Remark.} The matrices $\mbB_{\pm}$ from this theorem have the following
expressions:
\begin{equation}\label{dTL2 B+}
\mbB_+(a,b,\lambda)=\sum_{k=1}^N\beta_kE_{k,k}+
h\lambda\sum_{k=1}^N\delta_kE_{k+1,k}+h\lambda^2\sum_{k=1}^N E_{k+2,k}
\end{equation}
\begin{equation}\label{dTL2 B-}
\mbB_-(a,b,\lambda)=
I+h\lambda^{-1}\sum_{k=1}^N\frac{a_k\delta_k}{\beta_k}E_{k,k+1}
+h\lambda^{-2}\sum_{k=1}^N\frac{a_{k+1}a_k}{\beta_k}E_{k,k+2}
\end{equation}
\vspace{2mm}

\noindent
{\bf Proof.} The tri--diagonal structure of the matrix $T$ assures that the
factors $\mbB_{\pm}$ have the structure as in (\ref{dTL2 B+}), (\ref{dTL2 B-}).
The expressions for the entries of $\mbB_-$ and the recurrent relations for
the entries of $\mbB_+$ follow easily from the equality $\mbB_+\mbB_-=I+hT^2$.
After that the equations of motion follow from the equation $\mbB_+\wT=T\mbB_+$.
\qed

\subsection{Local equations of motion for dTL2}
For the map (\ref{dTL2}), called hereafter dTL2, there exists a localizing 
change of variables, however different from the one used for the map dTL.
It is given by the formulas
\begin{equation}\label{dTL2 loc map}
a_k=\ba_k(1+h\ba_{k-1})(1+h\bb_k^2),\qquad 
b_k=\bb_k(1+h\ba_{k-1})+h\bb_{k-1}\ba_{k-1}
\end{equation}
(Actually, discretizations of all higher flows of the Toda hierarchy should
possess their own localizing changes of variables).

The name ''localizing change of variables'' is justified by the 
following theorems.
\begin{theorem}\label{local dTL2}
The change of variables {\rm(\ref{dTL2 loc map})} conjugates the map {\rm dTL2}
with the map \newline 
$(\ba,\bb)\mapsto(\widetilde{\ba},\widetilde{\bb})$ governed
by the following {\rm local} equations of motion:
\begin{eqnarray}\label{dTL2 loc}
&\widetilde{\ba}_k(1+h\widetilde{\ba}_{k-1})(1+h\widetilde{\bb}_k^2)
=\ba_k(1+h\ba_{k+1})(1+h\bb_{k+1}^2)&\nonumber\\ \\
&\widetilde{\bb}_k-\bb_k=h\ba_k(\bb_k+\bb_{k+1})
-h\widetilde{\ba}_{k-1}(\widetilde{\bb}_{k-1}+\widetilde{\bb}_k)&
\nonumber
\end{eqnarray}
\end{theorem}
{\bf Proof.} This time the key point of the proof is obtaining the following 
local expressions for the coefficients of the factor $\Pi_+(I+hT^2)$:
\begin{eqnarray}
\beta_k  & = & (1+h\ba_k)(1+h\ba_{k-1})(1+h\bb_k^2)
\label{beta in dTL2 loc map}\\
\delta_k & = & (1+h\ba_k)(\bb_k+\bb_{k+1})
\label{delta in dTL2 loc map}
\end{eqnarray}
Indeed, the equations of motion (\ref{dTL2 loc}) follow directly from 
(\ref{dTL2}) and the latter formulas. To prove the latter formulas, 
{\it define} the quantities $\beta_k$, $\delta_k$ 
by the equations (\ref{beta in dTL2 loc map}), (\ref{delta in dTL2 loc map}). 
A straightforward calculation convinces that then the recurrent relations
(\ref{dTL2 beta}), (\ref{dTL2 delta}) are satisfied. The uniqueness of solution
to these equations proves the desired expressions for $\beta_k$, $\delta_k$.
\qed
\vspace{2mm}

There still exists a linear combination of invariant Poisson brackets of
the Toda hierarchy, whose pull--back with respect to the above change of
variables is local.

\begin{theorem}\label{dTL2 loc invariant PB}
The pull--back of the bracket
\begin{equation}\label{dTL2 loc m br}
\{\cdot,\cdot\}_1+h\{\cdot,\cdot\}_3
\end{equation}
on $\cT(a,b)$ under the change of variables {\rm(\ref{dTL2 loc map})}
is the following bracket on $\cT(\ba,\bb)$:
\begin{eqnarray}
\{\bb_k,\ba_k\} & = & -\ba_k(1+h\ba_k)(1+h\bb_k^2)\nonumber\\ 
\label{dTL2 loc br}\\
\{\ba_k,\bb_{k+1}\} & = & -\ba_k(1+h\ba_k)(1+h\bb_{k+1}^2)
\nonumber
\end{eqnarray}
The map {\rm(\ref{dTL2 loc})} is Poisson with respect to {\rm
(\ref{dTL2 loc br})}.
\end{theorem}
{\bf Proof} -- by straightforward verification. \qed
\vspace{2mm}

\begin{theorem}
The pull--back of the flow {\rm TL} under the map {\rm(\ref{dTL2 loc map})} is
described by the following differential equations:
\begin{equation}\label{TL in dTL2 loc map}
\dot{\ba}_k=\ba_k(\bb_{k+1}-\bb_k)(1+h\ba_k),\qquad 
\dot{\bb}_k=(\ba_k-\ba_{k-1})(1+h\bb_k^2)
\end{equation}
\end{theorem}
{\bf Proof.} The equations of motion we are looking for describe the flow TL
as a Hamiltonian system in the Poisson bracket (\ref{dTL2 loc m br}). The
corresponding Hamilton function is, obviously, given by
\[
(I+hT^2)\nabla\rH(T)=T
\]
Hence we have $\rH(T)=(2h)^{-1}\Big(\log\det(I+hT^2)\Big)_0$. From the
expressions for the factors of the generalized $LU$ factorization of the 
matrix $I+hT^2$ given in the formulas (\ref{dTL2 B+}), (\ref{dTL2 B-}) we 
conclude that $\rH(T)=(2h)^{-1}\sum_{k=1}^N\log(\beta_k)$. Taking into account 
the expressions (\ref{beta in dTL2 loc map}), we find finally:
\begin{equation}\label{TL in dTL2 loc map Ham}
\rH=\frac{1}{2h}\sum_{k=1}^N\log(1+h\bb_k^2)
+\frac{1}{h}\sum_{k=1}^N\log(1+h\ba_k)
\end{equation}
The Hamiltonian equations of motion generated by this Hamilton function and
the Poisson brackets (\ref{dTL2 loc br}), coincide with (\ref{TL in dTL2 loc 
map}).  \qed
\vspace{1.5mm}

The differential equations (\ref{TL in dTL2 loc map}) are a particular case
of a two--parameter deformation of TL found in \cite{K1}.

\newpage
\setcounter{equation}{0}
\section{Volterra lattice}\label{Sect Volterra in a}

\subsection{Equations of motion and bi--Hamiltonian structure}
The second flow (\ref{TL2}) of the Toda hierarchy, unlike the first one, allows
an important reduction $b_k=0$, and it results in the famous {\it Volterra 
lattice} (hereafter VL), known also under the names of Lotka--Volterra system,
discrete KdV equation, Langmuir lattice, Kac--van Moerbecke lattice, etc. 
\cite{M}, \cite{KM}:
\begin{equation}\label{VL in a}
\dot{a}_k=a_k(a_{k+1}-a_{k-1})
\end{equation} 
The phase space of VL in the case of periodic boundary conditions is
\begin{equation}\label{VL in a phase sp}
\cV={\Bbb R}^N(a_1,\ldots,a_N)
\end{equation}
Clearly, it is the subspace $b_k=0$ of $\cT(a,b)$.
Unfortunately, neither of three invariant Poison brackets 
$\{\cdot,\cdot\}_1$, $\{\cdot,\cdot\}_2$, and $\{\cdot,\cdot\}_3$ of the Toda
hierarchy can be {\it properly restricted} to the set $\cV$.
However, it is easy to see that the quadratic bracket $\{\cdot,\cdot\}_2$
allows a {\it Dirac reduction} to this set, the reduced bracket being defined
by the relations
\begin{equation}\label{VL in a q br}
\{a_k,a_{k+1}\}_{2}=-a_ka_{k+1}
\end{equation}
The system VL is Hamiltonian with respect to this bracket with the Hamilton
function
\[
\rH_1(a)=\sum_{k=1}^N a_k
\]
There exists one more local Poisson bracket on $\cV$ invariant with respect to 
VL and compatible with (\ref{VL in a q br}) \cite{K1}, \cite{FT1}. It is given  
by the relations
\begin{equation}\label{VL in a c br}
\{a_k,a_{k+1}\}_3=-a_ka_{k+1}(a_k+a_{k+1}), \qquad 
\{a_k,a_{k+2}\}_3=-a_ka_{k+1}a_{k+2}
\end{equation}
the corresponding Hamilton function being equal to
\[
\rH_0(a)=\frac{1}{2}\sum_{k=1}^N \log(a_k)
\]

\subsection{Lax representation}
The Lax representation of the previous section survives in the process 
of restriction to $\cV$. So, we obtain the Lax representation of the VL 
with the matrices from $\g$ \cite{M}, \cite{KM}:
\begin{equation}\label{VL in a Lax}
\dot{T}=[T,B_+]=-[T,B_-]
\end{equation}
where
\begin{eqnarray}
T(a,\lambda) & = & 
\lambda^{-1}\sum_{k=1}^N a_kE_{k,k+1}+\lambda\sum_{k=1}^NE_{k+1,k}
\label{VL in a T}\\ 
B_+(a,\lambda) & = & \pi_+(T^2)\;=\; 
\sum_{k=1}^N(a_k+a_{k-1})E_{k,k}+\lambda^2\sum_{k=1}^N E_{k+2,k}
\label{VL in a B+}\\ 
B_-(a,\lambda) & = & \pi_-(T^2)\;=\; 
\lambda^{-2}\sum_{k=1}^N a_{k+1}a_kE_{k,k+2}
\label{VL in a B-}
\end{eqnarray}
These equations allow an $r$--matrix interpretation in case of the quadratic
bracket $\{\cdot,\cdot\}_2$, see \cite{S4}. The Volterra hierarchy consists
of the flows allowing Lax representations of the form (\ref{VL in a Lax}) with
$B_{\pm}=\pi_{\pm}(f(T^2))$, where $f:\g\mapsto\g$ is ${\rm Ad}^*$--covariant.

\subsection{Discretization}

The discretization of the second Toda flow (\ref{dTL2}) also may be restricted 
to the set $\cV$. It is easy to see from (\ref{dTL2 delta}), (\ref{dTL2})
that in this situation we have with necessity $\delta_k=0$, and the recurrent
relations (\ref{dTL2 beta}) turn into
\begin{equation}\label{dVL in a beta}
\beta_k=1+h(a_k+a_{k-1})-\frac{h^2a_{k-1}a_{k-2}}{\beta_{k-2}}
\end{equation}

\begin{theorem} \label{discrete VL in a} {\rm \cite{S8} (see also \cite{K2}).}
The discrete time Lax equation
\begin{equation}\label{dVL in a Lax}
\widetilde{T}=\mbB_+^{-1}T\mbB_+=\mbB_-T\mbB_-^{-1}\quad {\rm with}\quad
\mbB_{\pm}=\Pi_{\pm}(I+hT^2)
\end{equation}
is equivalent to the following map $a\mapsto\wa$:
\begin{equation}\label{dVL in a}
\wa_k=a_k\,\frac{\beta_{k+1}}{\beta_k}
\end{equation}
where the quantities $\beta_k=\beta_k(a)=1+O(h)$ are uniquely defined by the 
recurrent relations
\begin{equation}\label{dVL in a beta alt}
\beta_k-ha_k=\frac{\beta_{k-1}}{\beta_{k-1}-ha_{k-1}}
=1+\frac{ha_{k-1}}{\beta_{k-1}-ha_{k-1}}
\end{equation}
and have the asymptotics 
\begin{equation}\label{dVL in a beta as}
\beta_k=1+h(a_k+a_{k-1})+O(h^2)
\end{equation}
\end{theorem}

{\bf Remark.} The matrices $\mbB_{\pm}$ have the following expressions:
\begin{eqnarray}
\mbB_+(a,\lambda) & = & \Pi_+(I+hT^2)\;=\;\sum_{k=1}^N\beta_kE_{k,k}
+h\lambda^2\sum_{k=1}^N E_{k+2,k}
\label{dVL in a fact+}\\
\mbB_-(a,\lambda) & = & \Pi_-(I+hT^2)\;=\;I+h\lambda^{-2}\sum_{k=1}^N
\frac{a_ka_{k+1}}{\beta_k}E_{k,k+2}
\label{dVL in a fact-}
\end{eqnarray}
\vspace{1.5mm}

{\bf Proof.} Referring to Theorem \ref{discrete Toda 2}, we need only to prove 
that the auxiliary quantities $\beta_k$ defined by the recurrent relations 
(\ref{dVL in a beta}), may be alternatively characterized as solutions to 
(\ref{dVL in a beta alt}). To do this, consider the matrix equation
$\mbB_+\wT=T\mbB_+$. In coordinates it is equivalent to the set of two 
scalar equations: (\ref{dVL}) and 
\[
\beta_{k+1}-ha_{k+1}=\frac{\beta_k}{\beta_{k-1}}\,(\beta_{k-1}-ha_{k-1})
\]
The last equation means that the following quantity does not depend on $k$:
\begin{equation}\label{dVL aux}
\frac{(\beta_k-ha_k)(\beta_{k-1}-ha_{k-1})}{\beta_{k-1}}=c={\rm const}
\end{equation}
and we need only to prove that $c=1$. But applying (\ref{dVL aux}) twice, we
have:
\[
\beta_k-ha_k=c+\frac{hca_{k-1}}{\beta_{k-1}-ha_{k-1}}=
c+\frac{ha_{k-1}(\beta_{k-2}-ha_{k-2})}{\beta_{k-2}}=
c+ha_{k-1}-\frac{h^2a_{k-1}a_{k-2}}{\beta_{k-2}}
\]
Comparing this with (\ref{dVL in a beta}), we conclude that $c=1$. \qed
\vspace{2mm}

The map (\ref{dVL in a}), (\ref{dVL in a beta alt}) will be denoted dVL. 
It is Poisson with respect to the brackets (\ref{VL in a q br}), (\ref{VL in
a c br}), but nonlocal, as opposed to the flow VL. In the open--end case we
have terminating continued fractions for the nonlocal quantities $\beta_k$
(in the periodic case for small $h$ the $\beta_k$ may be expressed
as analogous infinite periodic continued fractions).  From (\ref{dVL in a 
beta}) we derive, for example, for $\beta_k$ with odd indices:
\begin{equation}\label{dVL in a beta cf}
\beta_{2k+1}=1+h(a_{2k+1}+a_{2k})-\frac{h^2a_{2k}a_{2k-1}}
{1+h(a_{2k-1}+a_{2k-2})-\;\raisebox{-3mm}{$\ddots$}
 \raisebox{-4.5mm}{$\;-\displaystyle\frac{h^2a_2a_1}{1+ha_1}$}}
\end{equation}
(with obvious changes for $\beta_{2k}$). 
Analogously, from (\ref{dVL in a beta alt}) we derive alternative continued
fractions:
\begin{equation}\label{dVL in a beta alt cf}
\beta_k-ha_k=1+\frac{ha_{k-1}}
{1+\displaystyle\frac{ha_{k-2}}
 {1+\;\raisebox{-3mm}{$\ddots$}
  \raisebox{-4.5mm}{$\;+\displaystyle\frac{ha_2}{1+ha_1}$}}}
\end{equation}
One says that the continued fractions (\ref{dVL in a beta cf}) are obtained 
from (\ref{dVL in a beta alt cf}) with the help of {\it compression}.

\subsection{Local equations of motion for dVL}

The localizing change of variables $\cV(\ba)\mapsto\cV(a)$ for dVL follows 
from (\ref{dTL2 loc map}) upon setting $\bb_k=0$:
\begin{equation}\label{dVL in a loc map}
a_k=\ba_k(1+h\ba_{k-1})
\end{equation}
Again, this change of variables was found in \cite{K1} in the context of
integrable deformations, but with no relation to integrable disacretizations.

As it follows from the proof of Theorem \ref{local dTL2}, in the variables 
$\ba_k$ the auxiliary quantities $\beta_k$ become local:
\begin{equation}
\beta_k=(1+h\ba_k)(1+h\ba_{k-1})
\end{equation}
and we obtain the following result.
\begin{theorem}\label{local dVL in a}
The change of variables {\rm(\ref{dVL in a loc map})} conjugates the map 
{\rm dVL} with the map $\ba\mapsto\widetilde{\ba}$ governed by the following 
{\rm local} equations of motion:
\begin{equation}\label{dVL loc in a}
\widetilde{\ba}_k(1+h\widetilde{\ba}_{k-1})=\ba_k(1+h\ba_{k+1})
\end{equation}
\end{theorem}

The only known local invariant Poisson bracket (\ref{dTL2 loc br}) for 
the local form of dTL2 does not admit a proper restriction or a reduction to
the subset $\cV(\ba)$ of $\cT(\ba,\bb)$ characterized by $\bb_k=0$. 
Nevertheless, there appears to exist a local bracket on $\cV(\ba)$ invariant 
under the local form of dVL.

\begin{theorem} The pull--back of the bracket
\[
\{\cdot,\cdot\}_2+h\{\cdot,\cdot\}_3
\]
on $\cV(a)$ under the change of variables {\rm(\ref{dVL in a loc map})} is 
the following bracket on $\cV(\ba)$:
\begin{equation}\label{dVL in a loc PB}
\{\ba_k,\ba_{k+1}\}=-\ba_k\ba_{k+1}(1+h\ba_k)(1+h\ba_{k+1})
\end{equation}
This bracket is invariant under the map {\rm(\ref{dVL loc in a})}.
\end{theorem}
{\bf Proof} -- by direct calculation. \qed
\vspace{2mm}

The local equations of motion (\ref{dVL loc in a}) together with Lax 
representation were found by different methods in 
\cite{THO}, \cite{PN}, \cite{S8}. The invariant Poisson structure 
and the relation to factorization problem were pointed out only in the latter 
of these references.

\begin{theorem} {\rm \cite{K1}}.
The pull--back of the flow {\rm VL} under the map {\rm(\ref{dVL 
in a loc map})} is described by the following equations of motion:
\begin{equation}\label{VL in a in loc map}
\dot{\ba}_k=\ba_k(1+h\ba_k)(\ba_{k+1}-\ba_{k-1})
\end{equation}
\end{theorem}
{\bf Proof.} Since the flow VL has a Hamilton function
$(2h)^{-1}\,\sum_{k=1}^N\log(a_k)$ in the Poisson bracket $h\{\cdot,\cdot\}_3$, 
and this function is a Casimir of the bracket $\{\cdot,\cdot\}_2$, we conclude
that this function generates the flow VL also in the bracket $\{\cdot,\cdot\}_2
+h\{\cdot,\cdot\}_3$. That means that the pull--back of the flow VL is a 
Hamiltonian flow  in the
bracket (\ref{dVL in a loc PB}) with the Hamilton function
\[
(2h)^{-1}\sum_{k=1}^N\log\Big(\ba_k(1+h\ba_{k-1})\Big)
\]
Calculating the corresponding equations of motion, we arrive at (\ref{VL
in a in loc map}). \qed
\vspace{1.5mm}

The system (\ref{VL in a in loc map}) is known under the name of the 
{\it modified Volterra lattice}.
\vspace{1.5mm}

{\bf Remark.} An additional map
\[
\ba_k\mapsto\mba_k=\frac{\ba_k}{1+h\ba_k}
\]
delivers an alternative version of the localizing change of variables
\begin{equation}\label{dVL in a loc map alt}
a_k=\frac{\mba_k}{(1-h\mba_k)(1-h\mba_{k-1})}
\end{equation}
The corresponding local version of the dVL reads:
\begin{equation}\label{dVL in a loc alt}
\frac{\widetilde{\mba}_k}{(1-h\widetilde{\mba}_k)(1-h\widetilde{\mba}_{k-1})}
=\frac{\mba_k}{(1-h\mba_k)(1-h\mba_{k+1})}
\end{equation}
This version was introduced in \cite{HT} under the name ''discrete 
Lotka--Volterra equation of type II'' (while the type I was assigned to
(\ref{dVL loc in a})). The modified Volterra lattice in the variables $\mba_k$
takes the following form:
\begin{equation}
\dot{\mba}_k=\mba_k\left(\frac{\mba_{k+1}}{1-h\mba_{k+1}}-
\frac{\mba_{k-1}}{1-h\mba_{k-1}}\right)
\end{equation}
A remarkable feature of the variables $\mba_k$ (not mentioned in \cite{HT})
is the formal coincidence of the invariant Poisson bracket (\ref{dVL in a 
loc PB}) with the quadratic invariant Poisson bracket of the continuous time 
flow VL. Indeed, in the variables $\mba_k$ the bracket (\ref{dVL in a loc PB}) 
takes the form
\begin{equation}\label{dVL in a loc PB alt}
\{\mba_k,\mba_{k+1}\}=-\mba_k\mba_{k+1}
\end{equation}


\setcounter{equation}{0}
\section{Second flow of the Volterra hierarchy}\label{Sect VL2 in a}

Now we apply the general procedure of integrable discretization to the second 
flow of the Volterra hierarchy. We use the notations $\beta_k$, $B_{\pm}$ etc. 
for objects analogous to those of the previous section without danger of 
confusion.

\subsection{Equations of motion and bi--Hamiltonian structure}
The second flow of the Volterra hierarchy (hereafter VL2) is described by the 
following differential equations on $\cV$:
\begin{equation}\label{VL2 in a}
\dot{a}_k=a_k\Big(a_{k+1}(a_{k+2}+a_{k+1}+a_k)-
a_{k-1}(a_k+a_{k-1}+a_{k-2})\Big)
\end{equation}
This flow is Hamiltonian with respect to both Poisson brackets 
(\ref{VL in a q br}), (\ref{VL in a c br}). The corresponding Hamilton 
functions are:
\[
\rH_2(a)=\frac{1}{2}\sum_{k=1}^N a_k^2+\sum_{k=1}^Na_{k+1}a_k
\]
for the quadratic bracket $\{\cdot,\cdot\}_2$, and $\rH_1(a)$ for the cubic
bracket $\{\cdot,\cdot\}_3$.

\subsection{Lax representation}
The Lax representation for the flow VL2 is of the type (\ref{Lax in recipe}) 
with $f(T)=T^4$. 
\begin{theorem}\label{Lax for VL2}  
The flow {\rm(\ref{VL2 in a})} admits the following Lax representation in $\g$:
\begin{equation} \label{VL2 Lax}
\dot{T}=[T,B_+]=-[T,B_-]
\end{equation}
with the matrices
\begin{eqnarray*}
B_+(a,\lambda)\;=\;\pi_+(T^4) & = &
\sum_{k=1}^N \Big(a_{k+1}a_k+(a_k+a_{k-1})^2+a_{k-1}a_{k-2}\Big)E_{kk}\\
&&+\lambda^2\sum_{k=1}^N(a_{k+2}+a_{k+1}+a_k+a_{k-1})E_{k+2,k}+
\lambda^4\sum_{k=1}^N E_{k+4,k}\\
B_-(a,\lambda)\;=\;\pi_-(T^4) & = & 
 \lambda^{-2}\sum_{k=1}^N(a_{k+2}+a_{k+1}+a_k+a_{k-1})a_{k+1}a_kE_{k,k+2}\\
&&+\lambda^{-4}\sum_{k=1}^Na_{k+3}a_{k+2}a_{k+1}a_kE_{k,k+4}
\end{eqnarray*}
\end{theorem}

\subsection{Discretization}
Applying the recipe of Sect. \ref{Sect recipe} with $F(T)=I+hT^4$,
we take as a discretization of the flow VL2 the map described by the discrete 
time Lax equation
\begin{equation}\label{dVL2 Lax}
\wT=\mbB_+^{-1}T\mbB_+=\mbB_-T\mbB_-^{-1} \quad {\rm with} \quad
\mbB_{\pm}=\Pi_{\pm}\Big(I+hT^4\Big)
\end{equation}
\begin{theorem}\label{discrete VL2 in a}
The discrete time Lax equations  {\rm(\ref{dVL2 Lax})}
are equivalent to the map $a\mapsto\wa$ described by the following
equations:
\begin{equation}\label{dVL2 in a}
\wa_k=a_k\;\frac{\beta_{k+1}}{\beta_k}
\end{equation}
where the functions $\beta_k=\beta_k(a)=1+O(h)$ are uniquely defined for $h$ 
small enough simultaneously with the functions $\delta_k=O(1)$ by the system 
of recurrent relations
\begin{equation} \label{dVL2 in a beta}
\beta_k-h(\delta_k-a_{k+2})a_k =
\frac{\beta_{k-1}}{\beta_{k-1}-h(\delta_{k-1}-a_{k+1})a_{k-1}}
\end{equation}
\begin{equation} \label{dVL2 in a delta}
\delta_k = 
a_{k+2}+a_{k+1}+a_k+a_{k-1}-\frac{ha_{k-1}a_{k-2}\delta_{k-2}}{\beta_{k-2}}
\end{equation}
The auxiliary functions $\beta_k$ have the asymptotics
\begin{equation} \label{dVL2 in a beta as}
\beta_k = 1+h\Big(a_{k+1}a_k+(a_k+a_{k-1})^2+a_{k-1}a_{k-2}\Big)+O(h^2)
\end{equation}
\end{theorem}
\noindent

{\bf Remark.} The matrices $\mbB_{\pm}$ have the following 
expressions:
\begin{eqnarray*}
\mbB_+(a,\lambda) & = & 
\sum_{k=1}^N\beta_kE_{k,k}+h\lambda^2\sum_{k=1}^{N}\delta_kE_{k+2,k}
+h\lambda^4\sum_{k=1}^N E_{k+4,k}\\
\mbB_-(a,\lambda) & = & 
I+h\lambda^{-2}\sum_{k=1}^N\frac{a_{k+1}a_k\delta_k}{\beta_k}E_{k,k+2}
+h\lambda^{-4}\sum_{k=1}^N \frac{a_{k+3}a_{k+2}a_{k+1}a_k}{\beta_k}E_{k,k+4}
\end{eqnarray*}
\vspace{2mm}

\noindent
{\bf Proof.} The scheme of the proof is standard. First of all, the general
structure of the factors $\mbB_{\pm}$ is clear from the bi--diagonal 
structure of the matrix $T$. The expressions for the entries
of $\mbB_-$ and the recurrent relations for the entries of $\mbB_+$ follow
from the equality $\mbB_+\mbB_-=I+hT^4$. However, in this way we come to
the recurrent relation (\ref{dVL2 in a delta}) for $\delta_k$ and the following 
recurrent relation for $\beta_k$:
\begin{equation}\label{dVL2 in a beta alt}
\beta_k=1+h\Big(a_{k+1}a_k+(a_k+a_{k-1})^2+a_{k-1}a_{k-2}\Big)-
\frac{h^2a_{k-1}a_{k-2}\delta_{k-2}^2}{\beta_{k-2}}-
\frac{h^2a_{k-1}a_{k-2}a_{k-3}a_{k-4}}{\beta_{k-4}}
\end{equation}
different from (\ref{dVL2 in a beta}). In order to ''decompress'' this 
recurrent relation for $\beta_k$ and to bring it into the form 
(\ref{dVL2 in a beta}), we proceed as follows. The matrix equation
$\mbB_+\wT=T\mbB_+$ is equivalent to the set of three scalar ones: the equation
of motion (\ref{dVL2 in a}) and 
\begin{eqnarray}
\beta_{k+1}-h\delta_ka_{k+1} & = & \frac{\beta_k}{\beta_{k-1}}\,
(\beta_{k-1}-h\delta_{k-1}a_{k-1}) \label{dVL2 aux1}\\
\nonumber\\
\delta_{k+1}-a_{k+3} & = & \delta_k-a_{k-1}\,\frac{\beta_k}{\beta_{k-1}}
\label{dVL2 aux2}
\end{eqnarray}
From the last two equations it is easy to derive:
\[
\beta_{k+1}-h(\delta_{k+1}-a_{k+3})a_{k+1}=\frac{\beta_k}{\beta_{k-1}}\,
\Big(\beta_{k-1}-h(\delta_{k-1}-a_{k+1})a_{k-1}\Big)
\]
which means that the following quantity does not depend on $k$:
\begin{equation}\label{dVL2 aux3}
\frac{\Big(\beta_k-h(\delta_k-a_{k+2})a_k\Big)
\Big(\beta_{k-1}-h(\delta_{k-1}-a_{k+1})a_{k-1}\Big)}{\beta_{k-1}}=c
={\rm const}
\end{equation}
It remains to prove that this constant is equal to 1. To this end we apply
(\ref{dVL2 aux3}) twice (as in the proof of Theorem \ref{discrete VL in a})
to derive
\[
\beta_k=c+h(\delta_k-a_{k+2})a_k+h(\delta_{k-1}-a_{k+1})a_{k-1}-
\frac{h^2(\delta_{k-1}-a_{k+1})(\delta_{k-2}-a_k)a_{k-1}a_{k-2}}{\beta_{k-2}}
\]
Transforming the second and the third terms on right--hand side 
with the help of (\ref{dVL2 in a delta}), and the fourth one with
the help of (\ref{dVL2 aux2}), we see that the $O(h)$-terms of the last
formula exactly coincide with the $O(h)$--terms on the right--hand side of 
(\ref{dVL2 in a beta alt}). Equating the free terms, we get $c=1$. \qed

\subsection{Local equations of motion for dVL2}
The localizing change of variables for the map dVL2 is given by the formulas
\begin{equation}\label{dVL2 in a loc map}
a_k=\ba_k\,\frac{(1+h\ba_{k-1}^2)}{(1-h\ba_k\ba_{k-1})(1-h\ba_{k-1}\ba_{k-2})}
\end{equation}
\begin{theorem} \label{local discrete VL2 in a}
The change of variables {\rm(\ref{dVL2 in a loc map})} conjugates
the map {\rm dVL2} with the map $\ba\mapsto\widetilde{\ba}$
described by the following local equations of motion:
\begin{equation}\label{dVL2 in a loc}
\widetilde{\ba}_k\,\displaystyle\frac{(1+h\widetilde{\ba}_{k-1}^2)}
{(1-h\widetilde{\ba}_k\widetilde{\ba}_{k-1})(1-h\widetilde{\ba}_{k-1}
\widetilde{\ba}_{k-2})} = \ba_k\,\displaystyle\frac{(1+h\ba_{k+1}^2)}
{(1-h\ba_{k+2}\ba_{k+1})(1-h\ba_{k+1}\ba_k)}
\end{equation}
\end{theorem}
{\bf Proof.} The statement will follow immediately, if we prove the
following expressions for the auxiliary quantities $\beta_k$:
\begin{equation}\label{dVL2 in a beta in loc map}
\beta_k= \frac{(1+h\ba_k^2)(1+h\ba_{k-1}^2)}
{(1-h\ba_{k+1}\ba_k)(1-h\ba_k\ba_{k-1})^2(1-h\ba_{k-1}\ba_{k-2})}
\end{equation}
To do this, it is sufficient to verify that the recurrent relations 
(\ref{dVL2 in a beta}), (\ref{dVL2 in a delta}) are satisfied by the quantities
(\ref{dVL2 in a beta in loc map}) and
\begin{equation}\label{dVL2 in a delta in loc map}
\delta_k=\frac{\ba_{k+2}+\ba_{k+1}}{(1-h\ba_{k+2}\ba_{k+1})(1-h\ba_{k+1}\ba_k)}+
\frac{\ba_k+\ba_{k-1}}{(1-h\ba_{k+1}\ba_k)(1-h\ba_k\ba_{k-1})}
\end{equation}
Such a verification is a matter of straightforward, though somewhat tedious
algebra. We omit it, giving only an important intermediate formula:
\begin{equation}\label{dVL2 in a aux}
\beta_k-h(\delta_k-a_{k+2})a_k =
\frac{(1+h\ba_{k-1}^2)}{(1-h\ba_k\ba_{k-1})(1-h\ba_{k-1}\ba_{k-2})}
\end{equation}
(it implies immediately (\ref{dVL2 in a beta})). \qed
%
%
\vspace{1.5mm}

Unlike the situation with the previously encountered localizing changes of 
variables, we did not find a linear combination of invariant Poisson brackets
of the Volterra hierarchy, which would be pulled back into a local bracket
on $\cV(\ba)$ (presumably, such a linear combination does not exist).
Therefore, to find a pull back of the flow VL under the change of variables
(\ref{dVL2 in a loc map}), we cannot use the Hamiltonian formalism and are 
forced to turn to a direct analysis of equations of motion.

\begin{theorem} The pull--back of the flow {\rm VL} under the change of 
variables {\rm(\ref{dVL2 in a loc map})} is described by the following 
equations of motion:
\begin{equation}\label{VL in a in dVL2 loc map}
\dot{\ba}_k = 
\ba_k(1+h\ba_k^2)\left(\displaystyle\frac{\ba_{k+1}}{1-h\ba_{k+1}\ba_k}-
\displaystyle\frac{\ba_{k-1}}{1-h\ba_k\ba_{k-1}}\right)
\end{equation}
\end{theorem}
{\bf Proof.} It is a matter of straightforward calculations to verify that 
the equations (\ref{VL in a in dVL2 loc map}) are sent into (\ref{VL in a}) 
by the change of variables (\ref{dVL2 in a loc map}). \qed

\subsection{Local discretization of the KdV}
Probably the most interesting feature of the flow VL2 is that 
a certain linear combination of this flow with the original VL gives
a direct spatial discretization of the famous KdV \cite{AL1}. Indeed, the 
equations of motion of the linear combination $\alpha\cdot {\rm VL}+{\rm VL2}$ 
read:
\begin{equation}\label{VL2 lin comb}
\dot{a}_k=a_k\Big(a_{k+1}(\alpha+a_{k+2}+a_{k+1}+a_k)-
a_{k-1}(\alpha+a_k+a_{k-1}+a_{k-2})\Big)
\end{equation}
Setting $\alpha=-6$ and
\[
a_k=1+\varepsilon^2 p_k
\]
we get the equations of motion
\[
\dot{p}_k=(1+\varepsilon^2p_k)\Big(p_{k+2}-2p_{k+1}+2p_{k-1}-p_{k-2}+
\varepsilon^2p_{k+1}(p_{k+2}+p_{k+1}+p_k)-\varepsilon^2p_{k-1}(p_k+p_{k-1}+
p_{k-2})\Big)
\]
Assuming that $p_k(t)\approx p(t,k\varepsilon)$ with a smooth $p(t,x)$ and
rescaling the time $t\mapsto t/(2\varepsilon^3)$, we see that the previous
equation approximates 
\[
p_t=p_{xxx}+3p_xp
\]
which is the KdV. Therefore a local discretization of the linear combinations
of the flows VL and VL2 will result (for $\alpha=-6$) in a local 
spatio--temporal discretization of the KdV (see \cite{TA} for a highly nonlocal 
integrable discretization).

\begin{theorem} A localizing change of variables for the discretization of
the above flow corresponding to $F(T)=I+h(\alpha T^2+T^4)$ is given by the 
formulas
\begin{equation}
a_k=\ba_k\,\frac{(1+h\alpha\ba_{k-1}+h\ba_{k-1}^2)}
{(1-h\ba_k\ba_{k-1})(1-h\ba_{k-1}\ba_{k-2})}
\end{equation}
This change of variables conjugates the above discretization with the map
$\ba\mapsto\widetilde{\ba}$ described by the local equations of motion
\begin{equation}
\widetilde{\ba}_k\,
\frac{(1+h\alpha\widetilde{\ba}_{k-1}+h\widetilde{\ba}_{k-1}^2)}
{(1-h\widetilde{\ba}_k\widetilde{\ba}_{k-1})
(1-h\widetilde{\ba}_{k-1}\widetilde{\ba}_{k-2})}=
\ba_k\,\frac{(1+h\alpha\ba_{k+1}+h\ba_{k+1}^2)}
{(1-h\ba_{k+2}\ba_{k+1})(1-h\ba_{k+1}\ba_k)}
\end{equation}
\end{theorem}
{\bf Proof} is completely analogous to that of Theorem \ref{local discrete VL2 
in a}, and will not be repeated here. \qed


\setcounter{equation}{0}
\section{Modified Volterra lattice}\label{Sect modified Volterra}

\subsection{Equations of motion and Hamiltonian structure}
As we have seen many times in the preceding exposition, the localizing changes
of variables are interesting not only in their connection with the problem
of integrable discretization, but  already on the level of continuous time 
systems. Namely, they may be viewed as {\it Miura transformations} and used to 
find the so called {\it modified equations of motion}. These modified systems
are often interesting in their own rights -- and one can also wish to 
discretize them as well. We consider in the present section one example
of modified systems, namely, the {\it modified Volterra lattice} (MVL).

To this end we use the change of variables (\ref{dVL in a loc map}) in 
slightly different notations. Namely, we define a Miura change of variables 
$\cM:{\cal MV}(q)\mapsto\cV(a)$ by the formula
\begin{equation}\label{MVL Miura}
a_k=q_k(1+\alpha q_{k-1})
\end{equation}
We denote the parameter $\alpha$ instead of $h$ ($\alpha$ is not related
to the time step of discretizations and is not supposed to be small). The 
MVL is the pull--back of the VL (\ref{VL in a}) under the change of variables 
(\ref{MVL Miura}). Its equations of motion were already determined in Sect.
\ref{Sect Volterra in a} to be
\begin{equation}\label{MVL}
\dot{q}_k=q_k(1+\alpha q_k)(q_{k+1}-q_{k-1})
\end{equation}
As usual, this system may be considered under periodic or open--end boundary
conditions, the phase space in case of the periodic ones being
\[
{\cal MV}={\Bbb R}^N(q_1,\ldots,q_N)
\]
As pointed out in Sect. \ref{Sect Volterra in a}, this system is Hamiltonian 
with respect to the following Poisson bracket on ${\cal MV}$:
\begin{equation}\label{MVL PB}
\{q_k,q_{k+1}\}=-q_kq_{k+1}(1+\alpha q_k)(1+\alpha q_{k+1})
\end{equation}
and with one of the Hamilton functions
\begin{equation}
\rH_0(q)=\sum_{k=1}^N \log(q_k)\quad{\rm or}\quad
\rH_0(q)=\alpha^{-1}\sum_{k=1}^N\log(1+\alpha q_k)
\end{equation}
(their difference is a Casimir of the bracket (\ref{MVL PB})).
The bracket (\ref{MVL PB}) is the pull--back under the Miura change
of variables (\ref{MVL Miura}) of the invariant Poisson bracket
$\{\cdot,\cdot\}_2+\alpha\{\cdot,\cdot\}_3$ of the VL (see (\ref{VL in a q 
br}), (\ref{VL in a c br}) for the relevant definitions).

\subsection{Discretization}
We define the discretization dMVL of MVL (\ref{MVL}) 
as a pull--back of the dVL under the Miura change of variables 
(\ref{MVL Miura}). 
\begin{theorem} The equations of motion of the map {\rm dMVL} read:
\begin{equation}\label{dMVL}
\wq_k=q_k\;\frac{\gamma_{k+1}}{\gamma_k}
\end{equation}
where the quantities $\gamma_k=\gamma_k(q)=1+O(h)$ are uniquely defined by
$h$ small enough by the system of recurrent relations
\begin{equation}\label{dMVL gamma}
\gamma_k-hq_k=\frac{\gamma_{k-1}+h\alpha q_kq_{k-1}}{\gamma_{k-1}-hq_{k-1}}
\quad\Leftrightarrow \quad\gamma_k-hq_k+\alpha q_k 
=(1+\alpha q_k)\,\frac{\gamma_{k-1}}{\gamma_{k-1}-hq_{k-1}}
\end{equation}
and have the asymptotics
\begin{equation}\label{dMVL gamma as}
\gamma_k=1+h(q_k+q_{k-1}+\alpha q_kq_{k-1})+O(h^2)
\end{equation}
\end{theorem}
{\bf Proof.} From the equations of motion (\ref{dMVL}) and the definition 
(\ref{dMVL gamma}) we derive:
\begin{equation}\label{dMVL aux1}
1+\alpha\wq_k=
\frac{1}{\gamma_k}\,(\gamma_k+h\alpha q_{k+1}q_k)\left(1+\frac{\alpha q_k}
{\gamma_k-hq_k}\right)=
(1+\alpha q_k)\;\frac{\gamma_{k-1}\,(\gamma_{k+1}-hq_{k+1})}
{\gamma_k\,(\gamma_{k-1}-hq_{k-1})}
\end{equation}
Hence the variables $a_k$ defined as in (\ref{MVL Miura}) satisfy the
equations of motion (\ref{dVL in a}) with
\begin{equation}\label{dMVL beta}
\beta_k=\frac{\gamma_k}{\gamma_{k-2}}\,(\gamma_{k-1}-hq_{k-1})
(\gamma_{k-2}-hq_{k-2})
\end{equation}
It remains to prove that these quantities $\beta_k$ satisfy the recurrent
relations (\ref{dVL in a beta alt}). But this follows immediately from the
formula
\[
\beta_k-ha_k=\frac{\gamma_{k-1}}{\gamma_{k-2}}\,(\gamma_{k-2}-hq_{k-2})
\]
which is an easy consequence of (\ref{dMVL beta}), (\ref{MVL Miura}),
and (\ref{dMVL gamma}). \qed
\vspace{1.5mm}

Clearly, by definition the map dMVL is Poisson with respect to the 
bracket (\ref{MVL PB}), but the above theorem renders dMVL nonlocal. 
In particular, in the open--end case we have terminating continued fractions
\begin{equation}\label{dMVL gamma cf}
\gamma_k-hq_k=1+\frac{hq_{k-1}(1+\alpha q_k)}
{1+\displaystyle\frac{hq_{k-2}(1+\alpha q_{k-1})}
 {1+\;\raisebox{-3.5mm}{$\ddots$}
  \raisebox{-5mm}{$\;+\displaystyle
   \frac{hq_2(1+\alpha q_3)}{1+hq_1(1+\alpha q_2)}$}}}
\end{equation}
In the periodic case these continued fractions are also periodic.

\subsection{Local equations of motion for dMVL}

It is possible to find a localizing change of variables 
${\cal MV}(\bq)\mapsto{\cal MV}(q)$ for the map dMVL:
\begin{equation}\label{dMVL loc map}
q_k=\frac{\bq_k(1+h\bq_{k-1})}{1-h\alpha\bq_k\bq_{k-1}}
\end{equation}
As usual, this map is a local diffeomorphism for $h$ small enough.
\begin{theorem}
The change of variables {\rm(\ref{dMVL loc map})} conjugates the map {\rm dMVL}
with the map $\bq\mapsto\widetilde{\bq}$ described by the following {\em local} 
equations of motion:
\begin{equation}\label{dMVL loc}
\frac{\widetilde{\bq}_k}{1+\alpha\widetilde{\bq}_k}\,(1+h\widetilde{\bq}_{k-1})
=\frac{\bq_k}{1+\alpha\bq_k}\,(1+h\bq_{k+1})
\end{equation}
or, equivalently,
\begin{equation}\label{dMVL loc alt}
\frac{1+\alpha\widetilde{\bq}_k}
{1-h\alpha\widetilde{\bq}_k\widetilde{\bq}_{k-1}}
=\frac{1+\alpha\bq_k}{1-h\alpha\bq_{k+1}\bq_k}
\end{equation}
\end{theorem}
{\bf Proof.} The first step is to prove the following local expressions for 
the quantities $\gamma_k$ in the variables $\bq_k$: 
\begin{equation}\label{dMVL gamma in loc map}
\gamma_k=\frac{(1+h\bq_k)(1+h\bq_{k-1})}
{1-h\alpha\bq_k\bq_{k-1}}
\end{equation}
Indeed, from (\ref{dMVL}), (\ref{dMVL loc map}) and (\ref{dMVL gamma in loc
map}) we derive immediately the equations of motion in the form
\[
\frac{\widetilde{\bq}_k(1+h\widetilde{\bq}_{k-1})}
{1-h\alpha\widetilde{\bq}_k\widetilde{\bq}_{k-1}}
=\frac{\bq_k(1+h\bq_{k+1})}{1-h\alpha\bq_{k+1}\bq_k}
\]
which is equivalent to either of (\ref{dMVL loc}) and (\ref{dMVL loc alt}).
To prove (\ref{dMVL gamma in loc map}), it is as usual enough to verify that 
the quantities defined by this formula satisfy the recurrent relations 
(\ref{dMVL gamma}). This verification is the matter of a simple algebra. \qed
\vspace{1.5mm}

The equations of motion (\ref{dMVL loc}) were introduced in \cite{HT}.
\vspace{1.5mm}

In general, the only expression for a Poisson bracket invariant with
respect to the map (\ref{dMVL loc}) is non--local and non--polynomial 
(it may be characterized as the pull--back of the bracket (\ref{MVL PB}) 
under the localizing change of variables (\ref{dMVL loc map})). However, a 
direct analysis of equations of motion allows to prove the following statement.
\begin{theorem}
The pull--back of the flow {\rm(\ref{MVL})} under the change of
variables {\rm(\ref{dMVL loc map})} is described by the following 
equations of motion:
\begin{equation}\label{MVL in loc map}
\dot{\bq}_k=\bq_k(1+\alpha\bq_k)(1+h\bq_k)\left(
\frac{{\bq}_{k+1}}{1-h\alpha\bq_{k+1}\bq_k}-
\frac{\bq_{k-1}}{1-h\alpha\bq_k\bq_{k-1}}\right)
\end{equation}
\end{theorem}
This system may be called the {\it second modification of the Volterra lattice}.
Notice its remarkable symmetry with respect to the interchange $\alpha
\leftrightarrow h$. Also the expressions for $a_k$ in terms of $\bq_k$ enjoy
this symmetry. Indeed, the composition of two changes of variables (\ref{MVL
Miura}) and (\ref{dMVL loc map}) resilts in
\begin{equation}\label{dMVL a in loc map}
a_k=\frac{\bq_k(1+h\bq_{k-1})(1+\alpha\bq_{k-1})}
{(1-h\alpha\bq_k\bq_{k-1})(1-h\alpha\bq_{k-1}\bq_{k-2})}
\end{equation}
However, in the discrete equations of motion (\ref{dMVL loc}) this symmetry 
goes lost.

The formula (\ref{dMVL a in loc map}) allows also to translate the Miura map
$\cM:{\cal MV}(q)\mapsto\cV(a)$ into the language of localizing variables.
Namely, if we define the map $\bM:{\cal MV}(\bq)\mapsto\cV(\ba)$ by the
formula
\begin{equation}\label{MVL Miura in loc map}
\ba_k=\frac{\bq_k(1+\alpha\bq_{k-1})}{1-h\alpha\bq_k\bq_{k-1}}
\end{equation}
then an easy calculation shows the commutativity of the following diagram:
\begin{center}
\unitlength1cm
\begin{picture}(9,6.5)
\put(3.5,1.1){\vector(1,0){2}}
\put(3.5,5.1){\vector(1,0){2}}
\put(2,4.1){\vector(0,-1){2}}
\put(7,4.1){\vector(0,-1){2}}
\put(1,0.6){\makebox(2,1){${\cal MV}(q)$}} 
\put(1,4.6){\makebox(2,1){${\cal MV}(\bq)$}}
\put(6,4.6){\makebox(2,1){$\cV(\ba)$}}
\put(6,0.6){\makebox(2,1){$\cV(a)$}}
\put(0,2.6){\makebox(2,1){(\ref{dMVL loc map})}}
\put(7,2.6){\makebox(2,1){(\ref{dVL in a loc map})}}
\put(3.8,-0.2){\makebox(1.4,1.4){$\cM$}}
\put(3.8,5.0){\makebox(1.4,1.4){$\bM$}}
\end{picture}
\end{center}
\vspace{1.5mm}

{\bf Remark.} Sometimes another form of MVL is more convenient:
\begin{equation}\label{MVL in c}
\dot{c}_k=(c_k^2-\varepsilon^2)(c_{k+1}-c_{k-1})
\end{equation}
It is related to (\ref{MVL}) with $\alpha=(2\varepsilon)^{-1}$ by means of 
a linear change of variables:
\begin{equation}
q_k=c_k-\varepsilon
\end{equation}
and time rescaling $t\mapsto t/\alpha$.  Analogous linear change of localizing 
variables $\bq_k=\bc_k-\varepsilon$ accompanied by the rescaling of the time 
step $h\mapsto h/\alpha$ allows to derive from (\ref{dMVL loc}) the following
local integrable discretization of the flow (\ref{MVL in c}):
\begin{equation}\label{dMVL in c loc}
\frac{\widetilde{\bc}_k-\varepsilon}{\widetilde{\bc}_k+\varepsilon}\,
(1+2h\varepsilon
\widetilde{\bc}_{k-1})=
\frac{\bc_k-\varepsilon}{\bc_k+\varepsilon}\,(1+2h\varepsilon\bc_{k+1})
\end{equation}

\subsection{Particular case $\alpha\to\infty$}\label{Modified Volterra spec}
There exists a particular case of MVL, for which a more detailed information
is available. It corresponds to the $\alpha\to\infty$ limit of the previous
constructions, which has to be accompanied by the time rescaling $t\mapsto
t/\alpha$. (Alternatively, one could consider the $\varepsilon\to 0$ limit of 
the system (\ref{MVL in c})). Let us list the corresponding results. The 
equations of motion read:
\begin{equation}\label{MVL part}
\dot{q}_k=q_k^2(q_{k+1}-q_{k-1})
\end{equation}
and the Miura map $\cM: \cM\cV(q)\mapsto\cV(a)$ relating this system to the 
VL, takes the form
\begin{equation}\label{MVL part Miura}
a_k=q_kq_{k-1}
\end{equation}
An invariant Poisson bracket (\ref{MVL PB}) in the limit $\alpha\to\infty$ 
turns (being suitably rescaled) into 
\begin{equation}\label{MVL part PB}
\{q_k,q_{k+1}\}_3=-q_k^2q_{k+1}^2
\end{equation}
(see below for the reason for assuming the index ''3'' to this bracket).
The Hamilton function of the flow (\ref{MVL part}) in this bracket is equal to
\[
\rH_0(q)=\sum_{k=1}^N \log(q_k)
\]
The localizing change of variables (\ref{dMVL loc map}) after rescaling
$h\mapsto h/\alpha$ and going to the limit $\alpha\to\infty$ becomes
\begin{equation}\label{dMVL part loc map}
q_k=\frac{\bq_k}{1-h\bq_k\bq_{k-1}}
\end{equation}
and the discrete time equations of motion (\ref{dMVL loc alt}) become
\begin{equation}\label{dMVL part loc alt}
\frac{\widetilde{\bq}_k}{1-h\widetilde{\bq}_k\widetilde{\bq}_{k-1}}
=\frac{\bq_k}{1-h\bq_k\bq_{k+1}}
\end{equation}
or, equivalently,
\begin{equation}\label{dMVL part loc}
\widetilde{\bq}_k-\bq_k=h\widetilde{\bq}_k\bq_k(\bq_{k+1}-\widetilde{\bq}_{k-1})
\end{equation}
Finally, the pull--back of the system (\ref{MVL part}) under the localizing
change of variables (\ref{dMVL part loc map}) reads:
\begin{equation}\label{MVL part in loc map}
\dot{\bq}_k=\bq_k^2\left(\frac{\bq_{k+1}}{1-h\bq_{k+1}\bq_k}
-\frac{\bq_{k-1}}{1-h\bq_k\bq_{k-1}}\right)
\end{equation}
\vspace{1.5mm}

Now notice that the bracket (\ref{MVL part PB}) is the pull--back with respect
to the Miura map $\cM$ (\ref{MVL part Miura}) of the {\it cubic} invariant
bracket $\{\cdot,\cdot\}_3$ of the Volterra hierarchy. It turns out that in
this particular case the pull--back of the quadratic bracket $\{\cdot,\cdot\}_2$
is given by more elegant formulas than in the general case. Namely, the 
corresponding bracket on $\cM\cV(q)$ is {\it quadratic} albeit still nonlocal:
\begin{equation}\label{MVL part q br}
\{q_k,q_j\}_2=\pi_{kj}q_kq_j
\end{equation}
with the coefficients 
\begin{equation}\label{MVL pi}
\pi_{kj}=\left\{\begin{array}{rcl}
0 & \quad & k-j=0\\ 
-1 & \quad & k-j>0\;\;{\rm and\;\;odd\;\;or}\;\;k-j<0\;\;{\rm and\;\;even}\\
1 & \quad & k-j>0\;\;{\rm and\;\;even\;\;or}\;\;k-j<0\;\;{\rm and\;\;odd}
\end{array}\right\}
\end{equation}
The corresponding Hamilton function of the flow (\ref{MVL part}) is given by
\[
\rH(q)=\sum_{k=1}^N q_kq_{k+1}
\]
So, in this case we know a simple formula for an additional invariant Poisson 
bracket. This allows to find a nice formula for an invariant Poisson
structure of the map (\ref{dMVL part loc}) and of the flow (\ref{MVL part in 
loc map}) -- the result lacking in the general case. 
\begin{theorem} \label{MVL part local bracket}
The pull--back of the bracket
\begin{equation}\label{MVL spec mixed br}
\{\cdot,\cdot\}_2+h\{\cdot,\cdot\}_3
\end{equation}
on $\cM\cV(q)$ under the map {\rm(\ref{dMVL part loc map})} is the following
bracket on $\cM\cV(\bq)$:
\begin{equation}\label{dMVL part loc br}
\{\bq_k,\bq_j\}=\pi_{kj}\bq_k\bq_j
\end{equation}
with the coefficients {\rm(\ref{MVL pi})}. The map {\rm(\ref{dMVL part loc})} 
and the flow {\rm(\ref{MVL part in loc map})} are Poisson  with respect to this 
bracket.
\end{theorem}
{\bf Proof.} A direct calculation shows that the brackets (\ref{dMVL part
loc br}) for the variables $\bq_k$ are sent by the map (\ref{dMVL part loc map}) 
into the brackets (\ref{MVL spec mixed br}) for the variables $q_k$. \qed
\vspace{2mm}

We close this section by noticing that the translation of the Miura map
$\cM$ (\ref{MVL part Miura}) into the localizing variables is the map
$\bM:\cM\cV(\bq)\mapsto\cV(\ba)$ given by the formula
\begin{equation}\label{dMVL part Miura in loc map}
\ba_k=\frac{\bq_k\bq_{k-1}}{1-h\bq_k\bq_{k-1}}
\end{equation}
For the localizing variables $\mba_k$ of VL introduced by {\rm(\ref{dVL in a 
loc map alt})}, we have:
\begin{equation}\label{dMVL to dVL loc map}
\mba_k=\bq_k\bq_{k-1}
\end{equation}


\setcounter{equation}{0}
\section{Bogoyavlensky lattices}\label{Chapter Bogoyavlensky}

\subsection{Equations of motion and Hamiltonian structure}
\label{Sect Bog introduction}

There are three basic families of integrable lattice systems carrying the name
of Bogoyavlensky \cite{B} (although some of them were found earlier in
\cite{N}, \cite{K1}, \cite{I}). These systems are 
enumerated by integer parameters $m,p\ge 1$ ($p>1$ for the third one) and read:
\begin{equation}\label{BL1}
\dot{a}_k=a_k\left(\sum_{j=1}^m a_{k+j}-\sum_{j=1}^m a_{k-j}\right)
\end{equation}
\begin{equation}\label{BL2}
\dot{a}_k=a_k\left(\prod_{j=1}^p a_{k+j}-\prod_{j=1}^p a_{k-j}\right)
\end{equation}
\begin{equation}\label{BL3}
\dot{a}_k=\prod_{j=1}^{p-1}a_{k+j}^{-1}-\prod_{j=1}^{p-1}a_{k-j}^{-1}
\end{equation}

We shall call these systems BL1$(m)$, BL2$(p)$, and BL3$(p)$, respectively.

The lattices BL1$(m)$ and BL2$(p)$ serve as generalizations of the Volterra 
lattice, which arises from them by $m=1$ and $p=1$, respectively. 
The lattice BL3$(p)$ after the change of variables $a_k\mapsto q_k=a_k^{-1}$ 
and $t\mapsto -t$ turns into
\begin{equation}\label{mBL3}
\dot{q}_k=q_k^2\left(\prod_{j=1}^{p-1}q_{k+j}-\prod_{j=1}^{p-1}q_{k-j}\right)
\end{equation} 
We call the latter system {\it modified} BL2$(p)$. It serves as a 
generalization of the modified Volterra lattice (\ref{MVL part}), which is 
the $p=2$ particular case of (\ref{mBL3}).

These systems may be considered on an infinite lattice (all the subscripts 
belong to ${\Bbb Z}$), and admit also periodic finite--dimensional reductions 
(all indices belong to ${\Bbb Z}/N{\Bbb Z}$, where $N$ is the number of 
particles). The lattices BL1 and BL2 admit also finite--dimensional versions 
with the open--end boundary conditions. The phase space of the periodic BL's 
is
\begin{equation}\label{BL phase sp}
\cB={\Bbb R}^N(a_1,\ldots,a_N)
\end{equation}

The Hamiltonian structure of the BL1 was determined in \cite{K1} and later
in \cite{B}; for the BL2 and BL3 in the infinite setting this was done in 
\cite{ZTOF}; in the open--end and the periodic setting (where some subtleties 
come out) the Hamiltonian structures were determined in \cite{S4}. We reproduce 
here the corresponding result for the periodic boundary conditions.

The invariant quadratic Poisson brackets for Bogoyavlensky lattices are 
given by the formula (we set $p=1$ for BL1$(m)$, $m=1$ for BL2$(p)$, 
and $m=-1$ for BL3$(p)$):
\begin{equation}\label{BL PB aa}
\{a_k,a_j\}_2=\pi_{kj}a_ka_j
\end{equation}
with the coefficients
\begin{equation}\label{BL PB pi}
\pi_{kj}=\frac{1}{2}
\Big(\delta_{k,j+m}-\delta_{k+m,j}
+w_{k+m,j+m}^{(p)}-w_{k,j+m}^{(p)}-w_{k+m,j}^{(p)}+w_{k,j}^{(p)}\,\Big)
\end{equation}
Here, in turn, the coefficients $w_{kj}^{(p)}$ are given in the $N$--periodic 
case with ${\rm g.c.d.}(N,p)=1$ by the formula
\begin{equation}\label{BL Wp per}
w_{kj}^{(p)}=\left\{\begin{array}{lcl} 
{\rm sgn}(k-j) & \quad & k\equiv j \pmod N\\ \\
2n/p-1 & \quad & k-j\equiv nN\pmod p,\;\;1\le n\le p-1
\end{array}\right.
\end{equation}

The Poisson structure (\ref{BL PB aa}) is nonlocal unless $p=1$ 
(the case of BL1$(m)$), when it is given by the following brackets:
\begin{equation}\label{BL1 PB}
\{a_k,a_{k+1}\}=-a_ka_{k+1}\,,\quad\ldots\quad \{a_k,a_{k+m}\}=-a_ka_{k+m}
\end{equation}
In particular, for $m=1$ we obtain the quadratic Poisson bracket for the 
Volterra lattice. 

The corresponding Hamiltonians are:
\begin{eqnarray}
\rH(a) & = & \sum_{k=1}^N a_k\quad{\rm for\;\;BL1}(m)
\label{BL1 Ham}\\
\rH(a) & = & \sum_{k=1}^N a_ka_{k+1}\ldots a_{k+p-1}\quad{\rm for\;\;BL2}(p)
\label{BL2 Ham}\\
\rH(a) & = & \sum_{k=1}^N a_k^{-1}a_{k+1}^{-1}\ldots a_{k+p-1}^{-1}\quad 
{\rm for\;\;BL3}(p)  \label{BL3 Ham}
\end{eqnarray}

\subsection{Lax representations}

The Lax representations of the Bogoyavlensky lattices fall into the class
considered in \cite{K1}, and were also specified in \cite{B}. They have 
the form
\begin{equation}\label{BL Lax}
\dot{T}=\left[T,B\right]
\end{equation}
with the spectral--dependent (in the periodic case) matrices $T=T(a,\lambda)$,
$B=B(a,\lambda)$. Actually, the natural phase space of all these Lax 
representations is the same algebra $\g$ as before. For BL1$(m)$ the matrices 
$T$, $B$ are given by
\begin{equation}\label{BL1 T}
T(a,\lambda)=\lambda\sum_{k=1}^N E_{k+1,k}+\lambda^{-m}\sum_{k=1}^N a_kE_{k,k+m}
\end{equation}
\begin{equation}\label{BL1 B}
B(a,\lambda)=\pi_+\Big(T^{m+1}(a,\lambda)\Big)
=\sum_{k=1}^N (a_k+a_{k-1}+\ldots+a_{k-m})E_{k,k}+
\lambda^{m+1}\sum_{k=1}^N E_{k+m+1,k}
\end{equation}
for BL2$(p)$ they are given by
\begin{equation}\label{BL2 T}
T(a,\lambda)=\lambda^p\sum_{k=1}^N E_{k+p,k}+
\lambda^{-1}\sum_{k=1}^N a_kE_{k,k+1}
\end{equation}
\begin{equation}\label{BL2 B}
B(a,\lambda)=-\pi_-\Big(T^{p+1}(a,\lambda)\Big)
=-\lambda^{-p-1}\sum_{k=1}^N a_ka_{k+1}\ldots a_{k+p}E_{k,k+p+1}
\end{equation}
and for BL3$(p)$ they are given by
\begin{equation}\label{BL3 T}
T(a,\lambda)=\lambda^{p}\sum_{k=1}^N E_{k+p,k}+
\lambda\sum_{k=1}^N a_{k+1}E_{k+1,k}
\end{equation}
\begin{equation}\label{BL3 B}
B(a,\lambda)=-\pi_-\Big(T^{-p+1}(a,\lambda)\Big)
=-\lambda^{-p+1}\sum_{k=1}^N a_{k+1}^{-1}a_{k+2}^{-1}\ldots 
a_{k+p-1}^{-1}E_{k,k+p-1}
\end{equation}

All three Lax representations above may be seen as having the form
\[
\dot{T}=\Big[\,T,\,\pm \pi_{\pm}(d\varphi(T))\,\Big]
\]
with appropriate $Ad$--invariant functions $\varphi$ on $\g$, namely,
\begin{eqnarray}
\varphi(T)& = & \frac{1}{m+1}\Big({\rm tr}(T^{m+1})\Big)_0\quad
{\rm for\;\;BL1}(m)  \label{BL1 Ham in T}\\
\varphi(T) & = & \frac{1}{p+1}\Big({\rm tr}(T^{p+1})\Big)_0\quad
{\rm for\;\;BL2}(p)  \label{BL2 Ham in T}\\
\varphi(T) & = & -\frac{1}{p-1}\Big({\rm tr}(T^{-p+1})\Big)_0\quad 
{\rm for\;\;BL3}(p)  \label{BL3 Ham in T}
\end{eqnarray}
(it is easy to see that the values of these functions in coordinates $a_k$ 
coincide with (\ref{BL1 Ham})--(\ref{BL3 Ham}), respectively).

These Lax equations allow an $r$--matrix interpretation \cite{S4}, in which a 
certain quadratic bracket on $\g$ and its Dirac reduction to the manifold of Lax
matrices are involved. It may be demonstrated that in general the Dirac
correction to the Lax equations of motion vanishes for the Hamilton functions
of the form $\varphi(T)=\Big({\rm tr}(T^n)\Big)_0$, where $n=(p_1+m_1)n_1$ 
with some $n_1\in{\Bbb Z}$ and $m=m_1d$, $p=p_1d$, $d={\rm g.c.d.}(m,p)$.
In particular, this explains the Lax equations for the Hamilton functions
(\ref{BL1 Ham in T})--(\ref{BL3 Ham in T}).

\subsection{Discretization of BL1}

We are now in a position to apply the recipe of Sect. \ref{Sect recipe}
to the problem of discretizing the Bogoyavlensky lattices. This was done for the
first time in \cite{S8}. Closely related results were obtained in \cite{THO}, \cite{PN} by 
different methods. We reproduce here the results of \cite{S8} without proofs. 
As usual, the construction gives automatically for all discrete time systems 
(called hereafter dBL1, dBL2, dBL3, respectively) the invariant Poisson 
structure, the Lax representation, the integrals of motion, the interpolating 
Hamiltonian flows etc. The maps of all three families are nonlocal, but we 
demonstrate how to bring them to a local form by means of a suitable change 
of variables. Of course, the local forms of the maps dBL1--dBL3 are Poisson 
with respect to the Poisson brackets on $\cB(\ba)$ which are the pull--backs 
of the corresponding brackets on $\cB(a)$. However, we did not succeed in 
finding more or less nice formulas for such pull--backs (the only 
exceptions -- the Volterra and the modified Volterra lattices).

As dBL1$(m)$ we take the map described by the discrete time Lax equation
\begin{equation}\label{dBL1 Lax}
\wT=\mbB_+^{-1} T\mbB_+, \qquad \mbB_+=\Pi_+\Big(I+hT^{m+1}\Big)
\end{equation}

\begin{theorem} \label{discrete BL1} {\rm\cite{S8}}. 
The discrete time Lax equation {\rm(\ref{dBL1 Lax})} is equivalent to the map 
$a\mapsto\wa$ described by the equations
\begin{equation}\label{dBL1}
\wa_k=\frac{\beta_{k+m}}{\beta_k}\,a_k
\end{equation}
where the functions $\beta_k=\beta_k(a)=1+O(h)$ are uniquely defined for $h$
small enough by the recurrent relations
\begin{equation}\label{dBL1 beta}
\beta_k-ha_k=\prod_{j=1}^m
\left(1+\frac{ha_{k-j}}{\beta_{k-j}-ha_{k-j}}\right)
\end{equation}
and have the following asymptotics:
\begin{equation}\label{dBL1 asymp beta}
\beta_k=1+h\sum_{j=0}^m a_{k-j}+O(h^2)
\end{equation}
\end{theorem}
{\bf Remark.} The factor $\mbB_+$ is of the form
\begin{equation}\label{dBL1 B}
\mbB_+(a,\lambda)=\Pi_+\Big(I+hT^{m+1}\Big)=
\sum_{k=1}^N\beta_k E_{k,k}+h\lambda^{m+1}\sum_{k=1}^N E_{k+m+1,k}
\end{equation}
\vspace{2mm}

The map (\ref{dBL1}) is nonlocal due to the functions $\beta_k$. In the simplest
case $m=1$ one has the continued fractions, terminating for the open--end
boundary conditions:
\begin{equation}\label{cont fr}
\beta_k-ha_k=1+\displaystyle\frac{ha_{k-1}}
{1+\;\raisebox{-3mm}{$\ddots$}
  \raisebox{-4.5mm}{$\;+\displaystyle\frac{ha_2}{1+ha_1}$}}
\end{equation}
and periodic for the periodic boundary conditions. For $m>1$ there lacks
even such an expressive mean as continued fractions to represent these nonlocal
functions.

The localizing change of variables $\cB(\ba)\mapsto\cB(a)$ for dBL1$(m)$ is:
\begin{equation}\label{dBL1 loc map}
a_k=\ba_k\prod_{j=1}^m(1+h\ba_{k-j})
\end{equation}
Obviously, the map (\ref{dBL1 loc map}) for $h$ small enough is a local 
diffeomorphism.

\begin{theorem}\label{discrete BL1 local} 
The change of variables {\rm(\ref{dBL1 loc map})} conjugates the map 
{\rm dBL1}$(m)$ with the following one:
\begin{equation}\label{dBL1 loc}
\widetilde{\ba}_k\prod_{j=1}^m(1+h\widetilde{\ba}_{k-j})
=\ba_k\prod_{j=1}^m(1+h\ba_{k+j})
\end{equation}
\end{theorem}
{\bf Proof.} Let us {\it define} the quantities $\beta_k$ by the relation
\begin{equation}\label{dBL1 loc beta}
\beta_k=\prod_{j=0}^m(1+h\ba_{k-j})
\end{equation}
and prove that they satisfy the recurrent relations (\ref{dBL1 beta}).
Indeed, from (\ref{dBL1 loc map}), (\ref{dBL1 loc beta}) it follows:
\[
\beta_k-ha_k=\prod_{j=1}^m(1+h\ba_{k-j})=\frac{a_k}{\ba_k}
\]
Hence $\ba_k=a_k/(\beta_k-ha_k)$, which, being substituted in the previous 
formula, implies (\ref{dBL1 beta}). Now the uniqueness of solution of this 
latter recurrent system yields (\ref{dBL1 loc beta}). Plugging (\ref{dBL1 
loc beta}), (\ref{dBL1 loc map}) into the equations of motion (\ref{dBL1}) 
results in (\ref{dBL1 loc}). \qed
\vspace{2mm}

\begin{theorem}
The pull--back of equations of motion {\rm(\ref{BL1})} under the change of 
variables {\rm(\ref{dBL1 loc map})} is given by the following formula:
\begin{equation}\label{BL1 in dBL1 loc map}
\dot{\ba}_k=\ba_k(1+h\ba_k)\frac{\prod_{j=1}^m(1+h\ba_{k+j})-\prod_{j=1}^m
(1+h\ba_{k-j})}{h}
\end{equation}
\end{theorem}
{\bf Proof.} By a direct calculation one checks that the equations
(\ref{BL1 in dBL1 loc map}) under the change of variables (\ref{dBL1 loc map}) 
are mapped to the equations of motion (\ref{BL1}). \qed

\subsection{Discretization of BL2}
As dBL2$(p)$ we take the map described by the discrete time Lax equation
\begin{equation}\label{dBL2 Lax}
\wT=\mbC_- T\mbC_-^{-1},\qquad \mbC_-=\Pi_-\Big(I+hT^{p+1}\Big)
\end{equation}

\begin{theorem} \label{discrete BL2} {\rm\cite{S8}}. 
The discrete time Lax equation {\rm(\ref{dBL2 Lax})} is equivalent to the 
map $a\mapsto\wa$ described by the equations
\begin{equation}\label{dBL2}
\wa_k=\frac{a_k-h\gamma_{k-p}}{a_{k+p+1}-h\gamma_{k+1}}\,a_{k+p+1}
\end{equation}
where the functions $\gamma_k=\gamma_k(a)=O(1)$ are uniquely defined for $h$
small enough by the recurrent relations
\begin{equation}\label{dBL2 gamma}
a_k-h\gamma_{k-p}=\frac{a_k}
{1+h\displaystyle\prod_{j=1}^p(a_{k-j}-h\gamma_{k-p-j})}
\end{equation}
and have the asymptotics
\begin{equation}\label{dBL2 gamma as}
\gamma_k=\prod_{j=0}^p a_{k+j}\,(1+O(h))
\end{equation}
\end{theorem}
{\bf Remark. }
The factor $\mbC_-$ is of the form
\begin{equation}\label{dBL2 C}
\mbC_-(a,\lambda)=\Pi_-\Big(I+hT^{p+1}\Big)=
I+h\lambda^{-(p+1)}\sum_{k=1}^N\gamma_kE_{k,k+p+1}
\end{equation}
\vspace{2mm}

The quantities $\gamma_k$ render the equations of motion (\ref{dBL2})
nonlocal.  
The localizing change of variables for dBL2$(p)$ reads:
\begin{equation}\label{dBL2 loc map}
a_k=\ba_k\left(1+h\prod_{j=1}^p \ba_{k-j}\right)
\end{equation}
As usual, its bijectivity follows from the implicit functions theorem.

\begin{theorem}\label{discrete BL2 local} 
The change of variables {\rm(\ref{dBL2 loc map})} conjugates the map
{\rm dBL2}$(p)$ with the following one:
\begin{equation}\label{dBL2 loc}
\widetilde{\ba}_k\left(1+h\prod_{j=1}^p\widetilde{\ba}_{k-j}\right)=
\ba_k\left(1+h\prod_{j=1}^p \ba_{k+j}\right)
\end{equation}
\end{theorem}
{\bf Proof.} We proceed according to the by now already usual scheme. 
{\it Define} the quantities $\gamma_k$ by the formula 
\begin{equation}\label{dBL2 loc gamma}
\gamma_k=\prod_{j=0}^p \ba_{k+j}
\end{equation}
Then we immediately derive:
\begin{equation}\label{dBL2 loc aux}
a_k-h\gamma_{k-p}=\ba_k
\end{equation}
and plugging this expression for $\ba_k$ into (\ref{dBL2 loc map}) shows
that the recurrent relations (\ref{dBL2 gamma}) are satisfied. The uniqueness
of the system of functions $a_k-h\gamma_{k-p}$ satisfying these relations 
justifies the expressions (\ref{dBL2 loc gamma}). Now putting (\ref{dBL2 
loc map}), (\ref{dBL2 loc aux}) into the equations of motion (\ref{dBL2}) 
allows to rewrite them as (\ref{dBL2 loc}). \qed
\vspace{2mm}

\begin{theorem} The pull--back of the equations of motion {\rm(\ref{BL2})}
under the change of variables {\rm(\ref{dBL2 loc map})} is given by the 
following formula:
\begin{equation}\label{BL2 in dBL2 loc map}
\dot{\ba}_k=\ba_k
\left(\prod_{j=1}^p \ba_{k+j}-\prod_{j=1}^p \ba_{k-j}\right)
\prod_{n=1}^p\bigg(1+h\prod_{i=1}^p\ba_{k-n+i}\bigg)
\end{equation}
\end{theorem}
{\bf Proof.} A direct, though tedious calculation shows that the equations 
of motion (\ref{BL2 in dBL2 loc map}) are mapped on (\ref{BL2}) by means of
the map (\ref{dBL2 loc map}). \qed

\subsection{Discretization of BL3}

As dBL3$(p)$ we define the map with the discrete time Lax representation
\begin{equation}\label{dBL3 Lax}
\wT=\mbA_- T\mbA_-^{-1},\qquad \mbA_-=\Pi_-\Big(I+hT^{-p+1}\Big)
\end{equation}

\begin{theorem}\label{discrete BL3} {\rm\cite{S8}}. 
The discrete time Lax equation {\rm(\ref{dBL3 Lax})} is equivalent to the 
map $a\mapsto\wa$ described by the equations
\begin{equation}\label{dBL3}
\wa_k=\frac{a_k-h\alpha_{k-p}}{a_{k+p-1}-h\alpha_{k-1}}\,a_{k+p-1}
\end{equation}
where the functions $\alpha_k=\alpha_k(a)=O(1)$ are uniquely defined for $h$
small enough by the recurrent relations
\begin{equation}\label{dBL3 alpha}
\alpha_k=\prod_{j=1}^{p-1}\frac{1}{a_{k+j}-h\alpha_{k+j-p}}
\end{equation}
and have the asymptotics
\begin{equation}\label{dBL3 alpha as}
\alpha_k=\prod_{j=1}^{p-1}a_{k+j}^{-1}\,(1+O(h))
\end{equation}
\end{theorem}
{\bf Remark. } The factor $\mbA_-$ is of the form
\begin{equation}\label{dBL3 A}
\mbA_-(a,\lambda)=\Pi_-\Big(I+hT^{-p+1}\Big)=
I+h\lambda^{-p+1}\sum_{k=1}^N\alpha_kE_{k,k+p-1}
\end{equation}
\vspace{2mm}

The nonlocality of the equations of motion (\ref{dBL3}) is due to the functions
$\alpha_k(a)$. For example, for $p=2$ they can be expressed as periodic 
continued fractions of the following structure:
\[
h\alpha_k=\frac{h}{a_{k+1}-\displaystyle\frac{h}{a_k-
\displaystyle\frac{h}{a_{k-1}-\;
\raisebox{-3mm}{$\ddots$}}}}
\]
The localizing change of variables for dBL3$(p)$ is given by:
\begin{equation}\label{dBL3 loc map}
a_k=\ba_k\left(1+h\prod_{j=0}^{p-1} \ba_{k-j}^{-1}\right)
\end{equation}
The bijectivity of this map is assured by the implicit functions theorem.

\begin{theorem}\label{discrete BL3 local}
The change of variables {\rm(\ref{dBL3 loc map})} conjugates the map 
{\rm dBL3}$(p)$ with the following one:
\begin{equation}\label{dBL3 loc}
\widetilde{\ba}_k\left(1+h\prod_{j=0}^{p-1}\widetilde{\ba}_{k-j}^{-1}\right)=
\ba_k\left(1+h\prod_{j=0}^{p-1} \ba_{k+j}^{-1}\right)
\end{equation}
\end{theorem}
{\bf Proof.} Defining the quantities $\alpha_k$ by the formula 
\begin{equation}\label{dBL3 loc alpha}
\alpha_k=\prod_{j=1}^{p-1}\ba_{k+j}^{-1}
\end{equation}
we obtain with the help of (\ref{dBL3 loc map}):
\begin{equation}\label{dBL3 loc aux}
a_k-h\alpha_{k-p}=\ba_k
\end{equation}
Substituting this expression into (\ref{dBL3 loc alpha}), we see that the
recurrent relations (\ref{dBL3 alpha}) are satisfied, which proves (\ref{dBL3
loc alpha}). Substituting (\ref{dBL3 loc map}), (\ref{dBL3 loc aux}) into 
(\ref{dBL3}), we immediately arrive at the equations of motion (\ref{dBL3 loc}). 
\qed
\vspace{2mm}

\begin{theorem} The pull--back of the equations of motion {\rm(\ref{BL3})}
under the change of variables {\rm(\ref{dBL3 loc map})} is given by the 
following formula:
\begin{equation}\label{BL3 in dBL3 loc map}
\dot{\ba}_k=\left(\prod_{j=1}^{p-1} \ba_{k+j}^{-1}-
\prod_{j=1}^{p-1} \ba_{k-j}^{-1}\right)
\prod_{n=0}^{p-1}\bigg(1+h\prod_{i=0}^{p-1}\ba_{k-n+i}^{-1}\bigg)^{-1}
\end{equation}
\end{theorem}
{\bf Proof} -- by direct (but tiresome) calculations. \qed
\vspace{2mm}

We close the discussion of the local equations of motion for the dBL3$(p)$
by noticing that under the change of variables $\ba_k\mapsto\bq_k=\ba_k^{-1}$, 
$h\mapsto -h$ the map (\ref{dBL3 loc})  turns into
\begin{equation}\label{dmBL3 loc}
\widetilde{\bq}_k\left(1-h\prod_{j=0}^{p-1}\widetilde{\bq}_{k-j}\right)^{-1}=
\bq_k\left(1-h\prod_{j=0}^{p-1} \bq_{k+j}\right)^{-1}
\end{equation}
which is a local integrable discretization of the system (\ref{mBL3}), while
the differential equations (\ref{BL3 in dBL3 loc map}) under the change of 
variables $\ba_k\mapsto\bq_k=\bq_k^{-1}$, $h\mapsto -h$, $t\mapsto -t$ go into
\begin{equation}\label{mBL3 in dBL3 loc map}
\dot{\bq}_k=\bq_k^2\left(\prod_{j=1}^{p-1} \bq_{k+j}-
\prod_{j=1}^{p-1} \bq_{k-j}\right)
\prod_{n=0}^{p-1}\bigg(1-h\prod_{i=0}^{p-1}\bq_{k-n+i}\bigg)^{-1}
\end{equation}
which is an integrable one--parameter deformation of (\ref{mBL3}).

\subsection{Particular case $p=2$}\label{Modified Volterra in a}

The case $p=2$ of the Bogoyavlensky lattice BL3 is  remarkable in several
respects. The equations of motion for this case read:
\begin{equation}\label{BL3 spec}
\dot{a}_k=\frac{1}{a_{k+1}}-\frac{1}{a_{k-1}}
\end{equation}
The localizing change of variables for the discretization of the system 
(\ref{BL3 spec}) obtained in the previous subsection, is:
\begin{equation}\label{dBL3 spec loc map}
a_k=\ba_k\left(1+\frac{h}{\ba_k\ba_{k-1}}\right)=\ba_k+\frac{h}{\ba_{k-1}}
\end{equation}
In the variables $\ba_k$ we have the following discretization:
\begin{equation}\label{dBL3 spec loc}
\widetilde{\ba}_k-\ba_k=h\left(\frac{1}
{\ba_{k+1}}-\frac{1}{\widetilde{\ba}_{k-1}}\right)
\end{equation}
(When considered on the lattice $(t,k)$, the equation (\ref{dBL3 spec loc}) 
is equivalent to the so called {\it lattice KdV}, which is a very popular
object nowadays, cf. \cite{NC}, \cite{NS}, and references therein). 
The localizing change of variables (\ref{dBL3 spec loc map}) brings the system 
(\ref{BL3 spec}) itself into the form
\begin{eqnarray}
\dot{\ba}_k & = &\left(\frac{1}{\ba_{k+1}}-\frac{1}{\ba_{k-1}}\right)
\left(1+\frac{h}{\ba_{k-1}\ba_k}\right)^{-1}
\left(1+\frac{h}{\ba_k\ba_{k+1}}\right)^{-1}
\nonumber\\ \nonumber\\
& = & \ba_k\left(\frac{1}{\ba_{k+1}\ba_k+h}-\frac{1}{\ba_k\ba_{k-1}+h}\right)
\label{BL3 spec in loc map}
\end{eqnarray} 
\vspace{1.5mm}

The special properties of this case begin with the following observation.
The subset of $\g$ consisting of the Lax matrices
\begin{equation}\label{BL3 spec space}
T(a,\lambda)=\lambda^2\sum_{k=1}^N E_{k+2,k}+\lambda\sum_{k=1}^Na_{k+1}E_{k+1,k}
\end{equation}
is a Poisson submanifold for the corresponding quadratic $r$--matrix bracket,
so that the Dirac reduction is not needed in giving an $r$--matrix 
interpretation to the bracket 
\begin{equation}\label{BL3 spec q br}
\{a_k,a_j\}_2=\pi_{kj}a_ka_j
\end{equation}
The coefficients $\pi_{kj}$ are given (in the case of odd $N$) by the formula 
(\ref{MVL pi}).

Moreover, not only the quadratic $r$--matrix bracket, but also the linear 
one may be properly restricted to the set of the matrices (\ref{BL3 spec space}). 
The coordinate representation of the induced bracket on this set is given by:
\begin{equation}\label{BL3 spec l br}
\{a_k,a_{k+1}\}_1=-1
\end{equation}

The two Poisson brackets (\ref{BL3 spec q br}) and (\ref{BL3 spec l br}) are 
compatible, hence the system (\ref{BL3 spec}) and its discretization given by 
Theorem \ref{discrete BL3} with $p=2$ are bi--Hamiltonian. Obviously, the 
Hamilton function of the system (\ref{BL3 spec}) in the Poisson bracket 
(\ref{BL3 spec l br}) may be taken as
\begin{equation}\label{BL3 spec l Ham}
\rH_0(a)=\sum_{k=1}^N \log(a_k)
\end{equation}
It is easy to check that this function is a Casimir of the quadratic bracket.

Being bi--Hamiltonian, the system (\ref{BL3 spec}) and its discretization
admit also an arbitrary linear combination of the brackets (\ref{BL3 spec q br})
and (\ref{BL3 spec l br}) as an invariant Poisson structure. A further 
remarkable feature is the following: there exists a linear combination of 
these two brackets whose pull--back under the map (\ref{dBL3 spec loc map}) 
allows a nice representation in the variables $\ba_k$.
\begin{theorem} \label{BL3 local bracket}
The pull--back of the bracket
\begin{equation}\label{BL3 spec mixed br}
\{\cdot,\cdot\}_2+h\{\cdot,\cdot\}_1
\end{equation}
on $\cB(a)$ under the map {\rm(\ref{dBL3 spec loc map})} is the following
bracket on $\cB(\ba)$:
\begin{equation}\label{dBL3 spec loc br}
\{\ba_k,\ba_j\}=\pi_{kj}\ba_k\ba_j
\end{equation}
with the coefficients {\rm(\ref{MVL pi})}. The map {\rm(\ref{dBL3 spec loc})} 
and the flow {\rm(\ref{BL3 spec in loc map})} are Poisson  with respect to this 
bracket.
\end{theorem}
{\bf Proof.} A direct calculation shows that the brackets (\ref{dBL3 spec 
loc br}) for the variables $\ba_k$ are sent by the map (\ref{dBL3 spec loc map}) 
into the brackets (\ref{BL3 spec mixed br}) for the variables $a_k$. \qed
\vspace{2mm}

This theorem allows also to derive the equations of motion (\ref{BL3 spec in 
loc map}) in a Hamiltonian manner. Indeed, these equations describe the 
Hamiltonian flow with the Hamilton function 
\[
h^{-1}\sum_{k=1}^N\log\Big(\ba_k(1+h\ba_k^{-1}\ba_{k-1}^{-1})\Big)
\]
in the Poisson brackets (\ref{dBL3 spec loc br}). Indeed, this function is a 
pull--back of $h^{-1}\sum_{k=1}^N\log(a_k)$, which generates (\ref{BL3 spec})
in the bracket (\ref{BL3 spec mixed br}). A direct calculation shows 
that this Hamiltonian system is governed by the differential equations
(\ref{BL3 spec in loc map}). 
\vspace{1.5mm}

{\bf Remark.} The results of this subsection agree with the results of the
subsection \ref{Modified Volterra spec} after the change of variables
$a_k\mapsto q_k=a_k^{-1}$, $\ba_k\mapsto\bq_k=\ba_k^{-1}$.


\setcounter{equation}{0}
\section{Alternative approach to Volterra lattice}\label{Sect Volterra}

Starting from this point, we consider the systems with Lax representations
in the direct products $\bg=\g^{\otimes\,m}$ rather than in $\g$ itself.
We start with an alternative approach to the Volterra lattice based on a
Lax representation in $\g\otimes\g$.

\subsection{Equations of motion and bi--Hamiltonian structure}
The version of VL we consider here is:
\begin{equation}\label{VL}
\dot{u}_k=u_k(v_k-v_{k-1}),\qquad \dot{v}_k=v_k(u_{k+1}-u_{k})
\end{equation}
Usually we let the subscript $k$ change in the interval
$1\le k \le N$ and consider either open--end boundary conditions 
($v_0=u_{N+1}=0$) or periodic ones (all indices are taken (mod $N$). 
We consider mainly the periodic case, because the open--end one is
similar and more simple. The relation to the form (\ref{VL in a}) is 
achieved by re-naming the variables according to 
\begin{equation}\label{VL renaming a}
u_k\mapsto a_{2k-1},\qquad v_k\mapsto a_{2k}
\end{equation}
So, in the present setting the $N$--periodic VL consists of $2N$ particles. 
The phase space of VL in the case of periodic boundary conditions is
the space
\begin{equation}\label{VL phase sp}
\cW={\Bbb R}^{2N}(u_1,v_1,\ldots,u_N,v_N)
\end{equation}

Two compatible local Poisson brackets on $\cW$ invariant under the flow
VL are given by the relations
\begin{equation}\label{VL q br}
\{u_k,v_k\}_2=-u_kv_k,\qquad \{v_k,u_{k+1}\}_2=-v_ku_{k+1}
\end{equation}
and
\begin{equation}\label{VL c br}
\begin{array}{cclcccl}
\{u_k,v_k\}_3 & = & -u_kv_k(u_k+v_k), & \quad &
\{v_k,u_{k+1}\}_3 & = & -v_ku_{k+1}(v_k+u_{k+1})\\ \\
\{u_k,u_{k+1}\}_3 & = & -u_kv_ku_{k+1}, & \quad &
\{v_k,v_{k+1}\}_3 & = & -v_ku_{k+1}v_{k+1}
\end{array}
\end{equation}
respectively.
The corresponding Hamilton functions for the flow VL are equal to
\begin{equation}\label{VL q Ham}
\rH_1(u,v)=\sum_{k=1}^N u_k+\sum_{k=1}^N v_k
\end{equation}
and
\begin{equation}\label{VL c Ham}
\rH_0(u,v)=\sum_{k=1}^N \log(u_k)\qquad{\rm or}\qquad
\rH_0(u,v)=\sum_{k=1}^N \log(v_k)
\end{equation}
(the second function makes sense only in the periodic case;
the difference of these two functions is a Casimir of $\{\cdot,\cdot\}_3$
whose value is fixed to $\infty$ in the open--end case).

\subsection{Lax representation}\label{Sect Volterra Lax}

Consider the following two matrices:
\begin{equation}\label{VL U}
U(u,v,\lambda)=\sum_{k=1}^N u_kE_{k,k}+\lambda\sum_{k=1}^N E_{k+1,k}, \qquad
V(u,v,\lambda)=I+\lambda^{-1}\sum_{k=1}^N v_kE_{k,k+1}
\end{equation}
These formulas define the ''Lax matrix'' $(U,V):\cW\mapsto\bg=\g\otimes\g$. 

\begin{theorem}\label{Lax for Volterra in g+g}  {\rm\cite{K1} (see also 
\cite{S12})}.
The flow {\rm(\ref{VL})} admits the following Lax representation in 
$\g\otimes\g$:
\begin{eqnarray}
\dot{U}=UC_+-B_+U =B_-U-UC_- \nonumber\\  \label{VL Lax in g+g}\\ 
\dot{V}=VB_+-C_+V=C_-V-VB_-  \nonumber
\end{eqnarray}
with the matrices
\begin{eqnarray}
B_+(u,v,\lambda) & = & \sum_{k=1}^N (u_k+v_{k-1})E_{kk}+
\lambda\sum_{k=1}^NE_{k+1,k}
\label{VL B+}\\
C_+(u,v,\lambda) & = & \sum_{k=1}^N (u_k+v_k)E_{kk}+
\lambda\sum_{k=1}^NE_{k+1,k}
\label{VL C+}\\
B_-(u,v,\lambda) & = & \lambda^{-1}\sum_{k=1}^N u_kv_kE_{k,k+1}
\label{VL B-}\\
C_-(u,v,\lambda) & = & \lambda^{-1}\sum_{k=1}^N u_{k+1}v_kE_{k,k+1}
\label{VL C-}
\end{eqnarray}
\end{theorem}

\noindent
{\bf Corollary.} {\it The matrices 
\begin{equation}\label{VL T}
T_+(u,v,\lambda)=U(u,v,\lambda)V(u,v,\lambda),\qquad
T_-(u,v,\lambda)=V(u,v,\lambda)U(u,v,\lambda)
\end{equation}
satisfy the usual Lax equations in $\g$:}
\begin{equation}\label{VL Lax in g}
\dot{T}_+=[T_+,B_+]=-[T_+,B_-],\qquad 
\dot{T}_-=[T_-,C_+]=-[T_-,C_-]
\end{equation}

The matrices $T_{\pm}$ are easy to calculate explicitly. From the corresponding
formulas one sees that the matrices $B_{\pm}$, $C_{\pm}$ allow the following 
representations:
\[
B_+=\pi_+(T_+)=\pi_+(UV),\qquad B_-=\pi_+(T_-)=\pi_+(VU)
\]
\[
C_+=\pi_-(T_+)=\pi_-(UV),\qquad C_-=\pi_-(T_-)=\pi_-(VU)
\]

The Lax equations (\ref{VL Lax in g+g}) may be given an $r$--matrix 
interpretation \cite{S12} in terms of a certain {\it quadratic} bracket on 
$\bg=\g\otimes\g$, the induced bracket on $\cW$ being $\{\cdot,\cdot\}_2$.

\subsection{Discretization}
\label{Sect discretization VL}

To find an integrable time discretization for the flow VL, we 
apply the recipe of Sect. \ref{Sect recipe} with $F(T)=I+hT$, 
i.e. we consider the map described by the
discrete time ''Lax triads''
\begin{equation}\label{dVL Lax in g+g}
\wU=\mbB_+^{-1}U\mbC_+=\mbB_-U\mbC_-^{-1}\,, \qquad 
\wV=\mbC_+^{-1}U\mbB_+=\mbC_-V\mbB_-^{-1}
\end{equation}
with
\[
\mbB_{\pm}=\Pi_{\pm}(I+hUV), \qquad \mbC_{\pm}=\Pi_{\pm}(I+hVU)
\]
\begin{theorem}\label{discrete VL}
The discrete time Lax equations  {\rm(\ref{dVL Lax in g+g})}
are equivalent to the map $(u,v)\mapsto(\wu,\wv)$ described by the following
equations:
\begin{equation}\label{dVL}
\wu_k=u_k\;\frac{\gamma_k}{\beta_k}, \qquad 
\wv_k=v_k\;\frac{\beta_{k+1}}{\gamma_k}
\end{equation}
where the functions $\beta_k=\beta_k(u,v)=1+O(h)$, $\gamma_k=\gamma_k(u,v)
=1+O(h)$ are uniquely defined for $h$ small enough by the recurrent relations
\begin{equation}\label{dVL beta alt}
\beta_k-hu_k=\frac{\gamma_{k-1}}{\gamma_{k-1}-hv_{k-1}}=
1+\frac{hv_{k-1}}{\gamma_{k-1}-hv_{k-1}}
\end{equation}
\begin{equation}\label{dVL gamma alt}
\gamma_k-hv_k=\frac{\beta_k}{\beta_k-hu_k}=
1+\frac{hu_k}{\beta_k-hu_k}
\end{equation}
and have the asymptotics
\begin{eqnarray}
\beta_k & = & 1+h(u_k+v_{k-1})+O(h^2)
\label{dVL beta as}\\
\gamma_k & = & 1+h(u_k+v_k)+O(h^2)
\label{dVL gamma as}
\end{eqnarray}
\end{theorem}
\noindent

{\bf Remark.} The matrices $\mbB_{\pm}$, $\mbC_{\pm}$ have the following 
expressions:
\begin{eqnarray}
\mbB_+(u,v,\lambda) & = & \Pi_+\Big(I+hUV\Big)\;=\;
\sum_{k=1}^N\beta_kE_{k,k}+h\lambda\sum_{k=1}^{N}E_{k+1,k}
\label{dVL B+}\\
\mbC_+(u,v,\lambda) & = & \Pi_+\Big(I+hVU\Big)\;=\;
\sum_{k=1}^N\gamma_kE_{k,k}+h\lambda\sum_{k=1}^{N}E_{k+1,k}
\label{dVL C+}\\
\mbB_-(u,v,\lambda) & = & \Pi_-\Big(I+hUV\Big)\;=\;
I+h\lambda^{-1}\sum_{k=1}^{N}\frac{u_kv_k}{\beta_k}E_{k,k+1}
\label{dVL B-}\\
\mbC_-(u,v,\lambda) & = & \Pi_-\Big(I+hVU\Big)\;=\;
I+h\lambda^{-1}\sum_{k=1}^{N}\frac{u_{k+1}v_k}{\gamma_k}E_{k,k+1}
\label{dVL C-}
\end{eqnarray}
\vspace{2mm}

\noindent
{\bf Proof.} The general structure of the factors $\mbB_{\pm}$, $\mbC_{\pm}$,
as given in (\ref{dVL B+})--(\ref{dVL C-}), as well as the expressions
for the entries of $\mbB_-$, $\mbC_-$, follow directly from the 
defining equalities $\mbB_+\mbB_-=I+hUV$, $\mbC_+\mbC_-=I+hVU$. For the
entries $\beta_k$, $\gamma_k$ of $\mbB_+$, $\mbC_+$ one obtains the following
recurrent relations:
\begin{equation}\label{dVL beta gamma}
\beta_k=1+h(u_k+v_{k-1})-\frac{h^2u_{k-1}v_{k-1}}{\beta_{k-1}}, \qquad
\gamma_k=1+h(u_k+v_k)-\frac{h^2u_kv_{k-1}}{\gamma_{k-1}}
\end{equation}
Now notice that these relations coincide with (\ref{dVL in a beta}) after
re--naming (\ref{VL renaming a}) and
\[
\beta_k\mapsto\beta_{2k-1}, \qquad \gamma_k\mapsto\beta_{2k}
\]
Hence we may use the proof of Theorem \ref{discrete VL in a} to establish
the alternative recurrent relations (\ref{dVL beta alt}), (\ref{dVL gamma alt}).
The equations of motion (\ref{dVL}) follow directly from $\mbB_+\wU=U\mbC_+$, 
$\mbC_+\wV=V\mbB_+$. It is important to notice that (\ref{dVL}) become identical
with (\ref{dVL in a}) after the above mentioned re--namings of variables.
This allows to denote consistently the map constructed in this theorem by dVL.
\qed

\subsection{Local equations of motion for dVL}
Now we can simply reformulate the results of Sect. \ref{Sect Volterra in a} in 
our new notations.

The localizing change of variables $\cW(\bu,\bv)\mapsto\cW(u,v)$ for dVL
is given by the following formulas:
\begin{equation}\label{dVL loc map}
u_k=\bu_k(1+h\bv_{k-1}), \qquad v_k=\bv_k(1+h\bu_k)
\end{equation}
Due to the implicit function theorem, these formulas define a local 
diffeomorphism for $h$ small enough. 
\begin{theorem}\label{local dVL}
The change of variables {\rm(\ref{dVL loc map})} conjugates the map {\rm dVL} 
with the map $(\bu,\bv)\mapsto(\widetilde{\bu},\widetilde{\bv})$ governed 
by the following {\em local} equations of motion:
\begin{equation}\label{dVL loc}
\widetilde{\bu}_k(1+h\widetilde{\bv}_{k-1})=\bu_k(1+h\bv_k),\qquad
\widetilde{\bv}_k(1+h\widetilde{\bu}_k)=\bv_k(1+h\bu_{k+1})
\end{equation}
\end{theorem}
\vspace{1.5mm}

Let us mention that the functions $\beta_k(u,v)$, $\gamma_k(u,v)$ in the 
localizing variables are given by the formulas
\begin{equation}\label{dVL loc beta gamma}
\beta_k=(1+h\bu_k)(1+h\bv_{k-1}), \qquad \gamma_k=(1+h\bv_k)(1+h\bu_k)
\end{equation}
so that
\begin{equation}\label{dVL loc aux}
\beta_k-hu_k=1+h\bv_{k-1},\qquad \gamma_k-hv_k=1+h\bu_k
\end{equation}
\vspace{1.5mm}

{\bf Remark.} Cosidering the equations (\ref{dVL loc}) as lattice equations on
the lattice $(t,k)$, and performing a linear change of independent variables,
one arrives at the {\it explicit} version of dVL, cf. \cite{V1}, \cite{FV}.
\vspace{2mm}

\begin{theorem}\label{local PB for local dVL}
The pull--back of the bracket
\begin{equation}\label{dVL loc PB}
\{\cdot,\cdot\}_2+h\{\cdot,\cdot\}_3
\end{equation} 
on $\cW(u,v)$ under the change of variables {\rm(\ref{dVL loc map})} is 
the following bracket on $\cW(\bu,\bv)$:
\begin{equation}\label{dVL loc PB loc}
\{\bu_k,\bv_k\}=-\bu_k\bv_k(1+h\bu_k)(1+h\bv_k),\qquad
\{\bv_k,\bu_{k+1}\}=-\bv_k\bu_{k+1}(1+h\bv_k)(1+h\bu_{k+1})
\end{equation}
The map {\rm(\ref{dVL loc})} is Poisson with respect to the bracket 
{\rm (\ref{dVL loc PB loc})}.
\end{theorem}

\begin{theorem}\label{VL in local dVL coords}
The pull--back of the flow {\rm VL} under the map {\rm(\ref{dVL loc map})} is
described by the following equations of motion:
\begin{equation}\label{VL in loc dVL coords}
\dot{\bu}_k=\bu_k(1+h\bu_k)(\bv_k-\bv_{k-1}),\qquad
\dot{\bv}_k=\bv_k(1+h\bv_k)(\bu_{k+1}-\bv_k)
\end{equation}
\end{theorem}

\subsection{Lax representation for VL2}

We consider now the $\g\otimes\g$ formulation of the second flow VL2 of the 
Volterra hierarchy. We use the notations $\beta_k$, $\gamma_k$, $B_{\pm}$, 
$C_{\pm}$ etc. for objects analogous to those relevant for VL without danger 
of confusion.

The flow  VL2 is described by the following differential equations on $\cW$:
\begin{eqnarray}
\dot{u}_k & = & u_k\Big(v_k(u_{k+1}+v_k+u_k)-v_{k-1}(u_k+v_{k-1}+u_{k-1})\Big)
\nonumber\\
\dot{v}_k & = & v_k\Big(u_{k+1}(v_{k+1}+u_{k+1}+v_k)-u_k(v_k+u_k+v_{k_1})\Big)
\label{VL2}
\end{eqnarray}
The Hamilton functions of this flow are:
\[
\rH_2(u,v)=\frac{1}{2}\sum_{k=1}^N(u_k^2+v_k^2)+\sum_{k=1}^N(u_{k+1}v_k+v_ku_k)
\]
in the quadratic bracket $\{\cdot,\cdot\}_2$, and $\rH_1(u,v)$ in the cubic
bracket $\{\cdot,\cdot\}_3$.

The Lax representation for the flow VL2 is of the type (\ref{Lax triads in 
recipe}) with $m=2$ and $f(T)=T^2$. 
\begin{theorem}\label{Lax for VL2 in g+g}  {\rm \cite{K1}}.
The flow {\rm(\ref{VL2})} admits the following Lax representation in 
$\g\otimes\g$:
\begin{eqnarray}
\dot{U}=UC_+-B_+U =B_-U-UC_- \nonumber\\  \label{VL2 Lax in g+g}\\ 
\dot{V}=VB_+-C_+V=C_-V-VB_-  \nonumber
\end{eqnarray}
with the matrices
\begin{eqnarray*}
B_+(u,v,\lambda) & = & 
\sum_{k=1}^N \Big((u_k+v_{k-1})^2+u_kv_k+u_{k-1}v_{k-1}\Big)E_{kk}\\
&&+\lambda\sum_{k=1}^N(u_{k+1}+v_k+u_k+v_{k-1})E_{k+1,k}+
\lambda^2\sum_{k=1}^N E_{k+2,k}\\
C_+(u,v,\lambda) & = & 
 \sum_{k=1}^N \Big((u_k+v_k)^2+u_{k+1}v_k+u_kv_{k-1}\Big)E_{kk}\\
&&+\lambda\sum_{k=1}^N(u_{k+1}+v_{k+1}+u_k+v_k)E_{k+1,k}+
\lambda^2\sum_{k=1}^N E_{k+2,k}\\
B_-(u,v,\lambda) & = & 
 \lambda^{-1}\sum_{k=1}^N(u_{k+1}+v_k+u_k+v_{k-1})u_kv_kE_{k,k+1}\\
&&+\lambda^{-2}\sum_{k=1}^Nu_{k+1}v_{k+1}u_kv_kE_{k,k+2}\\
C_-(u,v,\lambda) & = & 
\lambda^{-1}\sum_{k=1}^N (u_{k+1}+v_{k+1}+u_k+v_k)u_{k+1}v_kE_{k,k+1}\\
&&+\lambda^{-2}\sum_{k=1}^Nu_{k+2}v_{k+1}u_{k+1}v_kE_{k,k+2}
\end{eqnarray*}
\end{theorem}
Obviously, the expressions for $B_+$, $C_+$  may be obtained from (\ref{TL2 B+})
with the help of the substitutions (\ref{TL M+}), (\ref{TL M-}), respectively,
and $B_-$, $C_-$ follow analogously from (\ref{TL2 B-}).

\subsection{Discretization of VL2}
Applying the recipe of Sect. \ref{Sect recipe} with $F(T)=I+hT^2$,
we take as a discretization of the flow VL2 the map described by the discrete 
time Lax triads
\begin{equation}\label{dVL2 Lax in g+g}
\wU=\mbB_+^{-1}U\mbC_+=\mbB_-U\mbC_-^{-1}\,, \qquad 
\wV=\mbC_+^{-1}U\mbB_+=\mbC_-V\mbB_-^{-1}
\end{equation}
with
\[
\mbB_{\pm}=\Pi_{\pm}\Big(I+h(UV)^2\Big), \qquad 
\mbC_{\pm}=\Pi_{\pm}\Big(I+h(VU)^2\Big)
\]
\begin{theorem}\label{discrete VL2}
The discrete time Lax equations  {\rm(\ref{dVL2 Lax in g+g})}
are equivalent to the map $(u,v)\mapsto(\wu,\wv)$ described by the following
equations:
\begin{equation}\label{dVL2}
\wu_k=u_k\;\frac{\gamma_k}{\beta_k}, \qquad 
\wv_k=v_k\;\frac{\beta_{k+1}}{\gamma_k}
\end{equation}
where the functions $\beta_k=\beta_k(u,v)=1+O(h)$, $\gamma_k=\gamma_k(u,v)
=1+O(h)$ are uniquely defined for $h$ small enough simultaneously with the
functions $\delta_k=O(1)$, $\epsilon_k=O(1)$ by the system of recurrent 
relations
\begin{eqnarray}
\beta_k-h(\delta_k-u_{k+1})u_k & = &
\frac{\gamma_{k-1}}{\gamma_{k-1}-h(\epsilon_{k-1}-v_k)v_{k-1}}
\label{dVL2 beta}\\
\gamma_k-h(\epsilon_k-v_{k+1})v_k & = & 
\frac{\beta_k}{\beta_k-h(\delta_k-u_{k+1})u_k} \label{dVL2 gamma}
\end{eqnarray}
\begin{eqnarray}
\delta_k & = & 
u_{k+1}+v_k+u_k+v_{k-1}-\frac{hu_{k-1}v_{k-1}\delta_{k-1}}{\beta_{k-1}}
\label{dVL2 delta}\\
\epsilon_k & = & 
u_{k+1}+v_{k+1}+u_k+v_k-\frac{hu_kv_{k-1}\epsilon_{k-1}}{\gamma_{k-1}}
\label{dVL2 eps}
\end{eqnarray}
The auxiliary functions $\beta_k$, $\gamma_k$ have the asymptotics
\begin{eqnarray}
\beta_k & = & 1+h\Big((u_k+v_{k-1})^2+u_kv_k+u_{k-1}v_{k-1}\Big)+O(h^2)
\label{dVL2 beta as}\\
\gamma_k & = & 1+h\Big((u_k+v_k)^2+u_{k+1}v_k+u_kv_{k-1}\Big)+O(h^2)
\label{dVL2 gamma as}
\end{eqnarray}
\end{theorem}
\noindent

{\bf Remark.} The matrices $\mbB_+$, $\mbC_+$ have the following 
expressions:
\begin{eqnarray*}
\mbB_+(u,v,\lambda) & = & \Pi_+\Big(I+h(UV)^2\Big)\;=\;
\sum_{k=1}^N\beta_kE_{k,k}+h\lambda\sum_{k=1}^{N}\delta_kE_{k+1,k}
+h\lambda^2\sum_{k=1}^N E_{k+2,k}\\
\mbC_+(u,v,\lambda) & = & \Pi_+\Big(I+h(VU)^2\Big)\;=\;
\sum_{k=1}^N\gamma_kE_{k,k}+h\lambda\sum_{k=1}^{N}\epsilon_kE_{k+1,k}
+h\lambda^2\sum_{k=1}^N E_{k+2,k}
\end{eqnarray*}
\vspace{2mm}

\noindent
{\bf Proof.} The scheme of the proof is standard. First of all, the general
structure of the factors $\mbB_{\pm}$, $\mbC_{\pm}$ is clear from the 
bi--diagonal structure of the matrices $U$, $V$. The recurrent relations
for the entries of the matrices $\mbB_+$, $\mbC_+$ follow, in principle, from
(\ref{dTL2 delta}), (\ref{dTL2 beta}) with the help of substitutions
(\ref{TL M+}), (\ref{TL M-}). It is easy to see that these relations coincide
with (\ref{dVL2 in a beta alt}), (\ref{dVL2 in a delta}) after re--naming
(\ref{VL renaming a}) and
\[
\beta_k\mapsto\beta_{2k-1},\quad \gamma_k\mapsto\beta_{2k},\quad
\delta_k\mapsto\delta_{2k-1}, \quad \epsilon_k\mapsto\delta_{2k}
\]
The equations of motion (\ref{dVL2}), following  from $\mbB_+\wU=U\mbC_+$, 
$\mbC_+\wV=V\mbB_+$, also coincide with (\ref{dVL2 in a}) after the above 
re--namings. Thus the discretization introduced in this theorem agrees with 
the one from Sect. \ref{Sect VL2 in a} and may be consistently denoted by dVL2.
 \qed

\subsection{Local equations of motion for dVL2}
Here we translate the corresponding results from Sect. \ref{Sect VL2 in a}
into our present notations. The localizing change of variables for the map 
dVL2 is given by the formulas
\begin{equation}\label{dVL2 loc map}
u_k=\bu_k\,\frac{(1+h\bv_{k-1}^2)}{(1-h\bu_k\bv_{k-1})(1-h\bv_{k-1}\bu_{k-1})}\,,
\qquad
v_k=\bv_k\,\frac{(1+h\bu_k^2)}{(1-h\bv_k\bu_k)(1-h\bu_k\bv_{k-1})}
\end{equation}
\begin{theorem} \label{local discrete VL2}
The change of variables {\rm(\ref{dVL2 loc map})} conjugates the map 
{\rm dVL2} with the map $(\bu,\bv)\mapsto(\widetilde{\bu},\widetilde{\bv})$
described by the following local equations of motion:
\begin{eqnarray}
\widetilde{\bu}_k\,\displaystyle\frac{(1+h\widetilde{\bv}_{k-1}^2)}
{(1-h\widetilde{\bu}_k\widetilde{\bv}_{k-1})(1-h\widetilde{\bv}_{k-1}
\widetilde{\bu}_{k-1})} & = & \bu_k\,\displaystyle\frac{(1+h\bv_k^2)}
{(1-h\bu_{k+1}\bv_k)(1-h\bv_k\bu_k)}\nonumber\\
\label{dVL2 loc}\\
\widetilde{\bv}_k\,\displaystyle\frac{(1+h\widetilde{\bu}_k^2)}
{(1-h\widetilde{\bv}_k\widetilde{\bu}_k)(1-h\widetilde{\bu}_k
\widetilde{\bv}_{k-1})} & = & \bv_k\,\displaystyle\frac{(1+h\bu_{k+1}^2)}
{(1-h\bv_{k+1}\bu_{k+1})(1-h\bu_{k+1}\bv_k)}\nonumber
\end{eqnarray}
\end{theorem}

\begin{theorem} The pull--back of the flow {\rm VL} under the change of 
variables {\rm(\ref{dVL2 loc map})} is described by the following equations of 
motion:
\begin{eqnarray}
\dot{\bu}_k & = & 
\bu_k(1+h\bu_k^2)\left(\displaystyle\frac{\bv_k}{1-h\bv_k\bu_k}-
\displaystyle\frac{\bv_{k-1}}{1-h\bu_k\bv_{k-1}}\right)
\nonumber\\
\label{VL in dVL2 loc map}\\
\dot{\bv}_k & = & 
\bv_k(1+h\bv_k^2)\left(\displaystyle\frac{\bu_{k+1}}{1-h\bu_{k+1}\bv_k}
-\displaystyle\frac{{\bu}_k}{1-h\bv_k\bu_k}\right)
\nonumber
\end{eqnarray}
\end{theorem}

\subsection{Miura relations to the Toda hierarchy}

We have seen in Sect. \ref{Sect Volterra in a} that the flow VL may be 
considered as a restriction of the {\it second} flow TL2 of the Toda hierarchy. 
There exists a relation of a completely different nature with
the flow TL. Namely, the flow VL is Miura related to the {\it first} flow TL
of the Toda hierarchy, while the flow VL2 is Miura related to TL2. To see this 
notice that the matrices $T_{\pm}(u,v,\lambda)$ from (\ref{VL T}) have the 
same tri--diagonal structure as the Toda lattice Lax matrix (\ref{TL T}) with 
\begin{equation}\label{TL M+}
a_k=u_kv_k,\qquad b_k=u_k+v_{k-1}
\end{equation}
or
\begin{equation}\label{TL M-}
a_k=u_{k+1}v_k,\qquad b_k=u_k+v_k
\end{equation}
respectively. These two pair of formulas may be considered as two Miura maps
$\cM_{\pm}:\cW(u,v)\mapsto\cT(a,b)$. The following holds (see \cite{K1}):
\begin{enumerate}
\item The both maps $\cM_{\pm}$ are Poisson, if $\cW$ is 
equipped with the bracket (\ref{VL q br}), and $\cT$ is equipped with 
the bracket (\ref{TL q br}).
\item The both maps $\cM_{\pm}$ are Poisson, if $\cW$ is 
equipped with the bracket (\ref{VL c br}), and $\cT$ is equipped with 
the bracket (\ref{TL c br}).
\item The flow VL (\ref{VL}) is conjugated with the flow TL (\ref{TL})
and the flow VL2 (\ref{VL2}) is conjugated with the flow TL2 (\ref{TL2})
by either of the maps $\cM_{\pm}$.
\end{enumerate}

We now translate these statements into the language of localizing variables.
Since we have two different localizing changes of variables (for dVL and
dVL2), two different translations are necessary.

We start with the case of the localizing change of variables (\ref{dVL loc 
map}).
\begin{theorem} {\rm \cite{K1}}
\begin{itemize}
\item[\rm a)] Define two maps $\bM_{\pm}:\cW(\bu,\bv)\mapsto\cT(\ba,\bb)$ 
by the formulas
\begin{equation}\label{loc dTL M+}
\ba_k=\bu_k\bv_k,\qquad 1+h\bb_k=(1+h\bu_k)(1+h\bv_{k-1})
\end{equation}
and
\begin{equation}\label{loc dTL M-}
\ba_k=\bu_{k+1}\bv_k,\qquad 1+h\bb_k=(1+h\bu_k)(1+h\bv_k)
\end{equation}
respectively. Then the following diagram is commutative:
\begin{center}
\unitlength1cm
\begin{picture}(9,6.5)
\put(3.5,1.1){\vector(1,0){2}}
\put(3.5,5.1){\vector(1,0){2}}
\put(2,4.1){\vector(0,-1){2}}
\put(7,4.1){\vector(0,-1){2}}
\put(1,0.6){\makebox(2,1){$\cW(u,v)$}} 
\put(1,4.6){\makebox(2,1){$\cW(\bu,\bv)$}}
\put(6,4.6){\makebox(2,1){$\cT(\ba,\bb)$}}
\put(6,0.6){\makebox(2,1){$\cT(a,b)$}}
\put(0,2.6){\makebox(2,1){{\rm(\ref{dVL loc map})}}}
\put(7,2.6){\makebox(2,1){{\rm(\ref{dTL loc map})}}}
\put(3.8,-0.2){\makebox(1.4,1.4){$\cM_{\pm}$}}
\put(3.8,5.0){\makebox(1.4,1.4){$\bM_{\pm}$}}
\end{picture}
\end{center}
 \item[\rm b)] The both maps $\bM_{\pm}$ are Poisson, if $\cW(\bu,\bv)$ is 
equipped with the bracket {\rm(\ref{dVL loc PB loc})}, and $\cT(\ba,\bb)$ is 
equipped with the bracket {\rm(\ref{dTL loc br 2})}.
\item[\rm c)] The local form of the {\rm dVL (\ref{dVL loc})} is conjugated 
with the local form of the {\rm dTL (\ref{dTL loc})} by either of the maps 
$\bM_{\pm}$. 
\end{itemize}
\end{theorem}
{\bf Proof.} The first statement is verified by a direct check, the second 
and the third ones are its consequences. \qed
\vspace{1.5mm}

In \cite{K1} this theorem was formulated without any relation to the problem 
of integrable discretization. In the context of discrete time systems the 
formulas (\ref{loc dTL M+}), (\ref{loc dTL M-}) were found also in \cite{HT},  
however, without discussing Poisson properties of these maps.
\vspace{1.5mm}

Concerning the localizing change of variables (\ref{dVL2 loc map}) for dVL2,
we come to the following results. 
\begin{theorem} Define the maps $\bN_{\pm}:\,\cW(\bu,\bv)\mapsto\cT(\ba,\bb)$
by the formulas
\begin{equation}
\ba_k=\frac{\bu_k\bv_k}{1-h\bu_k\bv_k}\,,\qquad
\bb_k=\frac{\bu_k+\bv_{k-1}}{1-h\bu_k\bv_{k-1}}
\end{equation}
and
\begin{equation}
\ba_k=\frac{\bu_{k+1}\bv_k}{1-h\bu_{k+1}\bv_k}\,,\qquad
\bb_k=\frac{\bu_k+\bv_k}{1-h\bu_k\bv_k}
\end{equation} 
respectively. Then the following diagram is commutative:
\begin{center}
\unitlength1cm
\begin{picture}(9,6.5)
\put(3.5,1.1){\vector(1,0){2}}
\put(3.5,5.1){\vector(1,0){2}}
\put(2,4.1){\vector(0,-1){2}}
\put(7,4.1){\vector(0,-1){2}}
\put(1,0.6){\makebox(2,1){$\cW(u,v)$}} 
\put(1,4.6){\makebox(2,1){$\cW(\bu,\bv)$}}
\put(6,4.6){\makebox(2,1){$\cT(\ba,\bb)$}}
\put(6,0.6){\makebox(2,1){$\cT(a,b)$}}
\put(0,2.6){\makebox(2,1){{\rm(\ref{dVL2 loc map})}}}
\put(7,2.6){\makebox(2,1){{\rm(\ref{dTL2 loc map})}}}
\put(3.8,-0.2){\makebox(1.4,1.4){$\cM_{\pm}$}}
\put(3.8,5.0){\makebox(1.4,1.4){$\bN_{\pm}$}}
\end{picture}
\end{center}
and the local form of the {\rm dVL2 (\ref{dVL2 loc})} is conjugated with 
the local form of the {\rm dTL2 (\ref{dTL2 loc})} by either of the maps 
$\bN_{\pm}$. 
\end{theorem}
{\bf Proof} -- by a direct check. \qed


\setcounter{equation}{0}
\section{Relativistic Toda lattice}\label{Chapter rel Toda}

\subsection{Equations of motion and tri--Hamiltonian structure}
\label{Sect rel Toda motivation}

The relativistic Toda lattice was invented by Ruijsenaars \cite{Ruij}, and
further studied in \cite{BR1}, \cite{OFZR} and numerous other papers. In
particular, the tri--Hamiltonian structure was elaborated in the latter 
reference.

We consider here two flows of the relativistic Toda hierarchy:
\begin{equation}\label{RTL+}
\dot{d}_k=d_k(c_k-c_{k-1}), \qquad
\dot{c}_k=c_k(d_{k+1}+c_{k+1}-d_k-c_{k-1})
\end{equation}
and
\begin{equation}\label{RTL-}
\dot{d}_k=d_k\left(\frac{c_k}{d_kd_{k+1}}-
\frac{c_{k-1}}{d_{k-1}d_k}\right), \quad
\dot{c}_k=c_k\left(\frac{1}{d_k}-\frac{1}{d_{k+1}}\right)
\end{equation}
The both systems may be considered either under open--end boundary conditions
($c_0=c_N=0$), or under periodic ones (all the subscripts are
taken $({\rm mod}\;N)$, so that $d_{N+1}\equiv d_1$, $d_0\equiv d_N$, 
$c_{N+1}\equiv c_1$, $c_0\equiv c_N$).
We shall denote the first flow by RTL+, and the second one by RTL$-$. 
The phase space of the flows RTL$\pm$ in the case of the periodic boundary 
conditions may be defined as
\begin{equation}\label{RTL phase space}
\cR={\Bbb R}^{2N}(d_1,c_1,\ldots,d_N,c_N)
\end{equation}
This space carries three compatible local Poisson bracket, with respect to
which the flows RTL$\pm$ are Hamiltonian.

The linear Poisson structure on $\cR$ is defined as
\begin{equation}\label{RTL l br}
\{d_k,c_k\}_1=-c_k, \qquad \{c_k,d_{k+1}\}_1=-c_k,
\qquad \{d_k,d_{k+1}\}_1=c_k
\end{equation}
The quadratic invariant Poisson structure on $\cR$ is given by the brackets 
\begin{equation}\label{RTL q br}
\{d_k,c_k\}_2=-d_kc_k \qquad \{c_k,d_{k+1}\}_2=-c_kd_{k+1}, \qquad 
\{c_k,c_{k+1}\}_2=-c_kc_{k+1}
\end{equation}
Finally, the cubic Poisson bracket on $\cR$ is given by the relations
\begin{equation}\label{RTL c br}
\begin{array}{cclcccl}
\{d_k,c_k\}_3     & = & -d_kc_k(d_k+c_k), & \quad &
\{c_k,d_{k+1}\}_3 & = & -c_kd_{k+1}(c_k+d_{k+1}), \\ \\
\{d_k,d_{k+1}\}_3 & = & -d_kc_kd_{k+1} ,  & \quad &
\{c_k,c_{k+1}\}_3 & = & -c_kc_{k+1}(c_k+2d_{k+1}+c_{k+1}) ,\\ \\
\{d_k,c_{k+1}\}_3 & = & -d_kc_kc_{k+1},  & \quad &
\{c_k,d_{k+2}\}_3 & = & -c_kc_{k+1}d_{k+2},\\ \\
\{c_k,c_{k+2}\}_3 & = & -c_kc_{k+1}c_{k+2}& & & & 
\end{array}
\end{equation}
The Hamilton functions for the flow RTL+ in the brackets (\ref{RTL l br}),
(\ref{RTL q br}), (\ref{RTL c br}) are equal to $\rH_2(c,d)$, $\rH_1(c,d)$, and
$\rH_0(c,d)$, respectively, where
\begin{eqnarray}
\rH_2(c,d) & = &
\frac{1}{2}\sum_{k=1}^N (d_k+c_k)^2+\sum_{k=1}^N (d_k+c_k)c_{k-1}\label{RTL H2}\\
\rH_1(c,d) & = & \sum_{k=1}^N(d_k+c_k) \label{RTL H1}\\
\rH_0(c,d) & = & \sum_{k=1}^N\log(d_k) \label{RTL H0}
\end{eqnarray}
Similarly, the Hamilton functions for the flow RTL$-$ in these three brackets
are equal to $-\rH_0(c,d)$, $\rH_{-1}(c,d)$, $\rH_{-2}(c,d)$, respectively, 
where
\begin{eqnarray}
\rH_{-1}(c,d) & = & \sum_{k=1}^N\frac{d_k+c_k}{d_kd_{k+1}} \label{RTL H-1}\\
\rH_{-2}(c,d) & = & \frac{1}{2}\sum_{k=1}^N \frac{(d_k+c_k)^2}{d_{k+1}^2d_k^2}+
\sum_{k=1}^N \frac{(d_{k-1}+c_{k-1})c_k}{d_{k+1}d_k^2d_{k-1}}
\label{RTL H-2}
\end{eqnarray}

\subsection{Lax representation}\label{Sect rel Toda Lax}
The most natural Lax representation for the relativistic Toda hierarchy
is the one living in $\bg=\g\otimes\g$ \cite{S3}, which is in many respects 
analogous to the Lax representation for the Volterra hierarchy considered in 
the previous section.

Introduce the matrices
\begin{equation}\label{RTL UV}
U(c,d,\lambda)=\sum_{k=1}^N d_kE_{kk}+\lambda\sum_{k=1}^N E_{k+1,k},\qquad
V(c,d,\lambda)=I-\lambda^{-1}\sum_{k=1}^N c_kE_{k,k+1}
\end{equation}

\begin{theorem}\label{Lax triads for rel Toda} {\rm\cite{S3}.} 
The equations of motion {\rm(\ref{RTL+})} are equivalent to the 
following Lax equations in $\g\otimes\g$ (Lax triads):
\begin{equation}
\dot{U}=UB-AU, \qquad \dot{V}=VB-AV
\end{equation}
where
\begin{eqnarray}
A(c,d,\lambda) & = & \sum_{k=1}^N(d_k+c_{k-1})E_{kk}+\lambda
\sum_{k=1}^N E_{k+1,k} \label{RTL+ A}\\
B(c,d,\lambda) & = & \sum_{k=1}^N(d_k+c_k)E_{kk}+\lambda
\sum_{k=1}^N E_{k+1,k} \label{RTL+ B}
\end{eqnarray}
The equations of motion {\rm(\ref{RTL-})} are equivalent to
the following Lax triads:
\begin{equation}
\dot{U}=UD-CU, \qquad \dot{V}=VD-CV
\end{equation}
where
\begin{eqnarray}
C(c,d,\lambda) & = & -\lambda^{-1}\sum_{k=1}^N 
\frac{c_k}{d_{k+1}}E_{k,k+1} \label{RTL- C}\\
D(c,d,\lambda) & = & -\lambda^{-1}\sum_{k=1}^N 
\frac{c_k}{d_k}E_{k,k+1} \label{RTL- D}
\end{eqnarray}
\end{theorem}

\noindent
{\bf Corollary.} {\it The matrices
\begin{equation}\label{RTL T}
T_+(c,d,\lambda)=U(c,d,\lambda)V^{-1}(c,d,\lambda),
 \quad T_-(c,d,\lambda)=V^{-1}(c,d,\lambda)U(c,d,\lambda)
\end{equation}
satisfy the usual Lax equations in $\g$. Namely, for the flow {\rm(\ref{RTL+})}
\begin{equation}\label{RTL+ Lax}
\dot{T}_+=\left[ T_+,A\right], \qquad \dot{T}_-=\left[ T_-,B\right] 
\end{equation}
and for the flow} {\rm(\ref{RTL-})}
\begin{equation}\label{RTL- Lax}
\dot{T}_+=\left[ T_+,C\right], \qquad \dot{T}_-=\left[ T_-,D\right] 
\end{equation}

The following formulas are easy to establish by a direct calculation:
\begin{equation}\label{RTL AB}
A=\pi_+(T_+),\qquad B=\pi_+(T_-)
\end{equation} 
\begin{equation}\label{RTL CD}
C=\pi_-(T_+^{-1}),\qquad D=\pi_-(T_-^{-1})
\end{equation} 
Hence the Lax equations (\ref{RTL+ Lax}) and (\ref{RTL- Lax}) may be presented 
as
\[
\dot{T}=\Big[T,\pi_+(T)\Big]\qquad{\rm and}\qquad
\dot{T}=\Big[T,\pi_-(T^{-1})\Big]
\]
respectively. These Lax equations in $\g$ may be given an $r$--matrix 
interpretation in the cases of the linear and of the quadratic Poisson brackets;
for the Lax triads in $\g\otimes\g$ an $r$--matrix interpretation is known
in the case of the quadratic bracket only \cite{S3}.

\subsection{Discretization of the relativistic Toda hierarchy}
\label{Sect discretizataion rel Toda}
We see that actually the pairs $(U,V^{-1})$ satisfy the Lax equations of the
form (\ref{Lax triads in recipe}) (with $m=2$). This allows to apply our 
general recipe to find integrable discretizations of the flows RTL$\pm$.
This results in considering the following discrete time Lax triads:
\begin{eqnarray*}
\wU & = & \Pi_+^{-1}(F(T_+))\cdot U\cdot\Pi_+(F(T_-))\;=\;
\Pi_-(F(T_+))\cdot U\cdot\Pi_-^{-1}(F(T_-))\\
\wV & = & \Pi_+^{-1}(F(T_+))\cdot V\cdot\Pi_+(F(T_-))\;=\;
\Pi_-(F(T_+))\cdot V\cdot\Pi_-^{-1}(F(T_-))
\end{eqnarray*}
with
\[
F(T)=I+hT\qquad{\rm and}\qquad F(T)=I-hT^{-1}
\]
respectively. It turns out that for the flow RTL+ the version with the $\Pi_+$ 
factors is more suitable, while for the RTL$-$ flow the version with the 
$\Pi_-$ factors is preferable.  

\subsection{Discretization of the flow RTL+}
Consider the discrete time Lax triad
\begin{equation}\label{dRTL+ Lax triads}
\wU=\mbA^{-1}\,U\,\mbB,\qquad \wV=\mbA^{-1}\,V\,\mbB
\end{equation}
with
\[
\mbA=\Pi_+(I+hT_+),\qquad \mbB=\Pi_+(I+hT_-)
\]
implying also either of the two equivalent forms of a convenient Lax equation:
\begin{equation}\label{dRTL+ Lax}
\wT_+=\mbA^{-1}\,T_+\,\mbA,\qquad 
\wT_-=\mbB^{-1}\,T_-\,\mbB
\end{equation}
\begin{theorem} \label{discrete RTL+} {\rm\cite{S7}.}
The equations {\rm(\ref{dRTL+ Lax triads})} are equivalent 
to the map $(c,d)\mapsto(\wc,\wid)$ described by the following equations:
\begin{equation}\label{dRTL+}
\wid_k=d_k\,\frac{\gb_k}{\ga_k}, \qquad \wc_k=c_k\,\frac{\gb_{k+1}}{\ga_k} 
\end{equation}
where the functions $\ga_k=\ga_k(c,d)=1+O(h)$ are uniquely defined by 
the recurrent relation 
\begin{equation}\label{dRTL+ a}
\ga_k=1+hd_k+\frac{hc_{k-1}}{\ga_{k-1}}
\end{equation}
and the coefficients $\gb_k=\gb_k(c,d)=1+O(h)$ are given by
\begin{equation}\label{dRTL+ b}
\gb_k=\ga_{k-1}\,\frac{\ga_k+hc_k}{\ga_{k-1}+hc_{k-1}}
=\ga_k\,\frac{\ga_{k+1}-hd_{k+1}}{\ga_k-hd_k}
\end{equation}
The following asymptotics hold:
\begin{equation}\label{dRTL+ a as}
\ga_k=1+h(d_k+c_{k-1})+O(h^2)
\end{equation}
\begin{equation}\label{dRTL+ b as}
\gb_k=1+h(d_k+c_k)+O(h^2)
\end{equation}
\end{theorem}
\vspace{2mm}

{\bf Remark.} The auxiliary matrices $\mbA$, $\mbB$ are bi--diagonal:
\begin{equation}\label{dRTL+ A}
{\mbA}(c,d,\lambda)=\Pi_+(I+hT_+)=
\sum_{k=1}^N\ga_kE_{kk}+h\lambda\sum_{k=1}^N E_{k+1,k}
\end{equation}
\begin{equation}\label{dRTL+ B}
{\mbB}(c,d,\lambda)=\Pi_+(I+hT_-)=
\sum_{k=1}^N\gb_kE_{kk}+h\lambda\sum_{k=1}^N E_{k+1,k}
\end{equation}
\vspace{2mm}

\noindent
{\bf Proof.} The general bi--diagonal structure of the factors $\mbA$, $\mbB$
follows from their definition. The simplest way to derive the recurrent
relations (\ref{dRTL+ a}) for the entries of $\mbA$ is to notice that
\[
\mbA=\Pi_+\Big(I+hUV^{-1}\Big)=\Pi_+(V+hU),\;\;{\rm because}\;\;V\in\G_-
\]
The equations of motion (\ref{dRTL+}) and the relations (\ref{dRTL+ b})
follow now from $\mbA\wU=U\mbB$, $\mbA\wV=V\mbB$. \qed
\vspace{2mm}

The map (\ref{dRTL+}) will be denoted dRTL+. Due to the asymptotic 
relations (\ref{dRTL+ a as}), (\ref{dRTL+ b as}) it clearly approximates the 
flow RTL+. As usual, it is tri--Poisson, 
allows the same integrals and the same Lax matrix as the flow RTL, but has a
drawback of nonlocality. Its source are the functions $\ga_k$, which in the
open--end case may be expressed as terminating continued fractions:
\[
\ga_k=1+hd_k+\frac{hc_{k-1}}{1+hd_{k-1}+\;
\raisebox{-3mm}{$\ddots$}
\raisebox{-4.5mm}{$\;+\displaystyle\frac{hc_1}{1+hd_1}$}}
\]
In the periodic case the $\ga_k$'s may be expressed as infinite periodic
continued fractions of an analogous structure.

\subsection{Local equations of motion for dRTL+}
Introduce another copy of the phase space $\cR$ parametrized by 
the variables $\bc_k$, $\bd_k$ and consider the change of variables
$\cR(\bc,\bd)\mapsto\cR(c,d)$ defined by the following formulas:
\begin{equation}\label{dRTL+ loc map}
d_k=\bd_k(1+h\bc_{k-1}),\qquad c_k=\bc_k(1+h\bd_k)(1+h\bc_{k-1})
\end{equation}
Obviously, due to the implicit function theorem, for $h$ small enough this map 
is locally a diffeomorphism.  

\begin{theorem}\label{dRTL+ in localization variables}
The change of variables {\rm(\ref{dRTL+ loc map})} conjugates {\rm dRTL+} 
with the map described by the following {\em local} equations of motion:
\begin{equation}\label{dRTL+ in loc map}
\begin{array}{l}
\widetilde{\bd}_k(1+h\widetilde{\bc}_{k-1})=\bd_k(1+h\bc_k)\\ \\
\widetilde{\bc}_k(1+h\widetilde{\bd}_k)(1+h\widetilde{\bc}_{k-1})
=\bc_k(1+h\bd_{k+1})(1+h\bc_{k+1})
\end{array}
\end{equation}
\end{theorem}
{\bf Proof.} The crucial point in the proof of this theorem is the following
observation: the parametrization of the variables $(c,d)$ according to
(\ref{dRTL+ loc map}) allows to find the coefficients $\ga_k$ (defined by the
recurrent relations (\ref{dRTL+ a})) in the closed form, namely:
\begin{equation}\label{dRTL+ loc a}
\ga_k=(1+h\bd_k)(1+h\bc_{k-1})
\end{equation}
Indeed, if we accept the last formula as the {\it definition} of the 
quantities $\ga_k$, then we obtain successively from 
(\ref{dRTL+ loc a}) and (\ref{dRTL+ loc map}): 
\[
\ga_k=1+h\bd_k(1+h\bc_{k-1})+h\bc_{k-1}=
1+hd_k+\frac{hc_{k-1}}{\ga_{k-1}}
\]
So, the quantities defined by (\ref{dRTL+ loc a}) satisfy the recurrent
relation (\ref{dRTL+ a}), and due to the uniqueness of solution our assertion
is demonstrated. Now (\ref{dRTL+ b}) and (\ref{dRTL+ loc a}) yield:
\begin{equation}\label{dRTL+ loc b}
\gb_k=(1+h\bd_k)(1+h\bc_k)
\end{equation}
and the equations of motion (\ref{dRTL+ in loc map}) follow directly from 
(\ref{dRTL+}), (\ref{dRTL+ loc a}), (\ref{dRTL+ loc b}). \qed
\vspace{2mm}

It turns out that the pull--back of either of the brackets (\ref{RTL l br}),
(\ref{RTL q br}), (\ref{RTL c br}) is highly nonlocal. Nevertheless, there 
exist certain linear combinations thereof, whose pull--backs are local. 

\begin{theorem}\label{local PB for local dRTL+}
\begin{itemize}
\item[{\rm a)}] The pull--back of the bracket 
\begin{equation}\label{dRTL+ loc PB1 target}
\{\cdot,\cdot\}_1+h\{\cdot,\cdot\}_2
\end{equation}
on $\cR(c,d)$ under the change of variables {\rm(\ref{dRTL+ loc map})}
is the following Poisson bracket on $\cR(\bc,\bd)$:
\begin{eqnarray}
\{\bd_k,\bc_k\} & = & -\bc_k(1+h\bd_k) \nonumber\\
\{\bc_k,\bd_{k+1}\} & = & -\bc_k(1+h\bd_{k+1})
\label{dRTL+ loc PB1}\\
\{\bd_k,\bd_{k+1}\} & = & \bc_k\,\displaystyle
\frac{(1+h\bd_k)(1+h\bd_{k+1})}{(1+h\bc_k)}
\nonumber
\end{eqnarray}

\item[{\rm b)}] The  pull--back of the bracket
\begin{equation}\label{dRTL+ loc PB2 target}
\{\cdot,\cdot\}_2+h\{\cdot,\cdot\}_3
\end{equation}
on $\cR(c,d)$ under the change of variables {\rm(\ref{dRTL+ loc map})}
is the following Poisson bracket on $\cR(\bc,\bd)$:
\begin{eqnarray}
\{\bd_k,\bc_k\} & = & -\bd_k\bc_k(1+h\bd_k)(1+h\bc_k) \nonumber\\
\{\bc_k,\bd_{k+1}\} & = & -\bc_k\bd_{k+1}(1+h\bc_k)(1+h\bd_{k+1})
\label{dRTL+ loc PB2}\\
\{\bc_k,\bc_{k+1}\} & = & -\bc_k\bc_{k+1}(1+h\bc_k)(1+h\bd_{k+1})(1+h\bc_{k+1})
\nonumber
\end{eqnarray}
\item[{\rm c)}] The brackets {\rm(\ref{dRTL+ loc PB1}), (\ref{dRTL+ loc PB2})}
are compatible. The map {\rm(\ref{dRTL+ in loc map})} is Poisson with respect 
to both of them.
\end{itemize}
\end{theorem}
{\bf Proof} -- by a direct check. Notice a curious feature of the bracket
(\ref{dRTL+ loc PB1}): it is nonpolynomial in coordinates (though still local).
\qed
\vspace{2mm}

\begin{theorem}\label{RTL+ in localization variables}
The pull--back of the flow {\rm RTL+} under the map {\rm(\ref{dRTL+ loc map})}
is described by the following equations of motion:
\begin{equation}\label{RTL+ in loc map}
\begin{array}{l}
\dot{\bd}_k=\bd_k(1+h\bd_k)(\bc_k-\bc_{k-1})\\ \\
\dot{\bc}_k=\bc_k(1+h\bc_k)\Big(\bd_{k+1}+\bc_{k+1}+h\bd_{k+1}\bc_{k+1}-\bd_k-
\bc_{k-1}-h\bd_k\bc_{k-1}\Big)
\end{array}
\end{equation}
\end{theorem}
{\bf Proof.}  We will use by the proof only the bracket of the part b) of the
previous theorem. Obviously, the pull--back we are looking for is a Hamiltonian 
system with the Hamilton function, which is a pull--back of 
$h^{-1}\rH_0(c,d)$ (indeed, this function is a Casimir function for 
$\{\cdot,\cdot\}_2$ and is the Hamilton function of the flow RTL+ in the 
bracket $h\{\cdot,\cdot\}_3$).
Calculating the equations of motion generated by the Hamilton function
\[
h^{-1}\sum_{k=1}^N \log\Big(\bd_k(1+h\bc_{k-1})\Big)
\]
in the Poisson brackets (\ref{dRTL+ loc PB2}), we obtain (\ref{RTL+ in loc map}). 
\qed

\subsection{Discretization of the flow RTL$-$}

Consider the discrete time Lax triad
\begin{equation}\label{dRTL- Lax triads}
\wU=\mbC\,U\,\mbD^{-1},\qquad \wV=\mbC\,V\,\mbD^{-1}
\end{equation}
with
\[
\mbC=\Pi_-(I-hT_+^{-1}),\qquad \mbD=\Pi_-(I-hT_-^{-1})
\]
implying also  the convenient Lax equations:
\begin{equation}\label{dRTL- Lax}
\wT_+=\mbC\,T_+\,\mbC^{-1},\qquad 
\wT_-=\mbD\,T_-\,\mbD^{-1}
\end{equation}
\begin{theorem}\label{discrete RTL-} {\rm\cite{S7}.} 
The equations {\rm(\ref{dRTL- Lax triads})} are equivalent to 
the map $(c,d)\mapsto(\wc,\wid)$ described by the following equations:
\begin{equation}\label{dRTL-}
\wid_k=d_{k+1}\,\frac{\gc_k}{\gd_k}, \qquad 
\wc_k=c_{k+1}\,\frac{\gc_k}{\gd_{k+1}} 
\end{equation}
where the functions $\gd_k=\gd_k(c,d)=O(1)$ are uniquely defined by the 
recurrent relation
\begin{equation}\label{dRTL- d}
\gd_k=\frac{c_k}{d_k-h-h\gd_{k-1}}
\end{equation}
and the coefficients $\gc_k=\gc_k(c,d)=O(1)$ are given by
\begin{equation}\label{dRTL- c}
\gc_k=\gd_k\,\frac{d_k-h\gd_{k-1}}{d_{k+1}-h\gd_k}
=\gd_{k+1}\,\frac{c_k+h\gd_k}{c_{k+1}+h\gd_{k+1}}
\end{equation}
The following asymptotics hold:
\begin{equation}\label{dRTL- dc as}
\gd_k= \frac{c_k}{d_k}+O(h), \qquad \gc_k=\frac{c_k}{d_{k+1}}+O(h)
\end{equation}
\end{theorem}

{\bf Remark.} The auxiliary matrices $\mbC$, $\mbD$ in the discrete time
Lax equations admit the following expressions:
\begin{equation}\label{dRTL- C}
\mbC(c,d,\lambda)=I+h\lambda^{-1}\sum_{k=1}^N\gc_kE_{k,k+1}
\end{equation}
\begin{equation}\label{dRTL- D}
\mbD(c,d,\lambda)=I+h\lambda^{-1}\sum_{k=1}^N\gd_kE_{k,k+1}
\end{equation}
\vspace{1.5mm}

\noindent
{\bf Proof} -- analogous to that of Theorem \ref{discrete RTL+}. \qed
\vspace{2mm}

The map (\ref{dRTL-}) will be called hereafter dRTL$-$. Like dRTL+, it is 
tri--Poisson etc. It is nonlocal because of the presence of the functions
$\gd_k$, which in the open--end case have the following finite continued
fractions expressions:
\[
\gd_{k}=\frac{c_{k}}{d_{k}-h-\displaystyle
\frac{hc_{k-1}}{d_{k-1}-h-\;
\raisebox{-3mm}{$\ddots$}
\raisebox{-4.5mm}{$\;-\displaystyle\frac{hc_1}{d_1-h}$}}}
\]

\subsection{Local equations of motion for dRTL$-$}

To bring the map dRTL$-$ to the local form, another change of variables 
$\cR(\bc,\bd)\mapsto\cR(c,d)$ is necessary:
\begin{equation}\label{dRTL- loc map}
d_k=\bd_k\left(1+\frac{h\bc_{k-1}}{\bd_{k-1}\bd_k}\right),\qquad 
c_k=\bc_k\left(1-\frac{h}{\bd_k}\right)
\end{equation}
Again, for $h$ small enough this map is locally a diffeomorphism, due to the 
implicit function theorem. 
\begin{theorem}\label{dRTL- in localization variables}
The change of variables {\rm(\ref{dRTL- loc map})} conjugates {\rm dRTL}$-$ with
the map described by the following {\em local} equations of motion:
\begin{equation}\label{dRTL- in loc map}
\begin{array}{l}
\widetilde{\bd}_k\left(1+\displaystyle\frac{h\widetilde{\bc}_{k-1}}
{\raisebox{-1.5mm}{$\,\widetilde{\bd}_{k-1}\widetilde{\bd}_k$}}\right)=
\bd_k\left(1+\displaystyle\frac{h\bc_k}{\bd_k\bd_{k+1}}\right)\\ \\
\widetilde{\bc}_k\left(1-\displaystyle\frac{h}
{\raisebox{-1.5mm}{$\,\widetilde{\bd}_k$}}\right)=
\bc_k\left(1-\displaystyle\frac{h}{\bd_{k+1}}\right)
\end{array}
\end{equation}
\end{theorem}
{\bf Proof.} This time the crucial component of the proof is the following
remarkably simple local formula for the coefficients $\gd_k$ (defined by the
recurrent relations (\ref{dRTL- d})) in the coordinates $(\bc_k,\bd_k)$:
\begin{equation}\label{dRTL- loc d}
\gd_k=\frac{\bc_k}{\bd_k}
\end{equation}
Indeed, if we use {\it both} (\ref{dRTL- loc map}) and (\ref{dRTL- loc d}) as
definitions, then we obtain: 
\[
\displaystyle\frac{c_k}{\gd_k}=
\bd_k-h=d_k-h-h\gd_{k-1}
\]
Hence, the quantities defined by (\ref{dRTL- loc d}) satisfy the recurrent
relation (\ref{dRTL- d}), and due to the uniqueness of solution our assertion
is proved. From (\ref{dRTL- c}) and (\ref{dRTL- loc d}) we obtain also:
\begin{equation}\label{dRTL- loc c}
\gc_k=\frac{\bc_k}{\bd_{k+1}}
\end{equation}
Now the equations of motion (\ref{dRTL- in loc map}) follow directly from 
(\ref{dRTL-}), (\ref{dRTL- loc d}), (\ref{dRTL- loc c}). \qed
\vspace{2mm}

The Poisson properties of the change of variables (\ref{dRTL- loc map})
are similar to that of (\ref{dRTL+ loc map}). Namely, the pull--backs of 
either of the brackets (\ref{RTL l br}), (\ref{RTL q br}), (\ref{RTL c br}) 
are nonlocal, but there exist linear combinations thereof, whose pull--backs 
are local. 

\begin{theorem}\label{local PB for local dRTL-}
\begin{itemize}
\item[\rm a)] The pull--back of the Poisson bracket 
\begin{equation}\label{dRTL- loc PB1 target}
\{\cdot,\cdot\}_2-h\{\cdot,\cdot\}_1
\end{equation}
on $\cR(c,d)$ under the change of variables {\rm(\ref{dRTL- loc map})} is 
the following bracket on $\cR(\bc,\bd)$:
\begin{eqnarray}
&\{\bd_k,\bc_k\}=-\bd_k\bc_k\left(1-\displaystyle\frac{h}{\bd_k}\right), \qquad
\{\bc_k,\bd_{k+1}\}=-\bc_k\bd_{k+1}\left(1-\displaystyle\frac{h}{\bd_{k+1}}
\right),&\nonumber\\
&\{\bc_k,\bc_{k+1}\}=-\bc_k\bc_{k+1}\left(1-\displaystyle\frac{h}{\bd_{k+1}}
\right)&  \label{dRTL- loc PB1}
\end{eqnarray}
\item[\rm b)] The pull--back of the Poisson bracket 
\begin{equation}\label{dRTL- loc PB2 target}
\{\cdot,\cdot\}_3-h\{\cdot,\cdot\}_2
\end{equation}
on $\cR(c,d)$ under the change of variables {\rm(\ref{dRTL- loc map})} is 
the following bracket on $\cR(\bc,\bd)$:
\begin{eqnarray}
\{\bd_k,\bc_k\}     & = & -\bd_k\bc_k(\bd_k+\bc_k)
\left(1-\displaystyle\frac{h}{\bd_k}\right)\nonumber\\
\{\bc_k,\bd_{k+1}\} & = & -\bc_k\bd_{k+1}(\bc_k+\bd_{k+1})
\left(1-\displaystyle\frac{h}{\bd_{k+1}}\right)\nonumber\\
\{\bd_k,\bd_{k+1}\} & = & -\bd_k\bc_k\bd_{k+1}
\left(1-\displaystyle\frac{h}{\bd_k}\right)
\left(1-\displaystyle\frac{h}{\bd_{k+1}}\right)\nonumber\\
\{\bc_k,\bc_{k+1}\} & = & -\bc_k\bc_{k+1}(\bc_k+2\bd_{k+1}+\bc_{k+1})
\left(1-\displaystyle\frac{h}{\bd_{k+1}}\right)
\label{dRTL- loc PB2}\\
\{\bd_k,\bc_{k+1}\} & = & -\bd_k\bc_k\bc_{k+1}
\left(1-\displaystyle\frac{h}{\bd_k}\right)
\left(1-\displaystyle\frac{h}{\bd_{k+1}}\right)\nonumber\\
\{\bc_k,\bd_{k+2}\} & = & -\bc_k\bc_{k+1}\bd_{k+2}
\left(1-\displaystyle\frac{h}{\bd_{k+1}}\right)
\left(1-\displaystyle\frac{h}{\bd_{k+2}}\right)\nonumber\\
\{\bc_k,\bc_{k+2}\} & = & -\bc_k\bc_{k+1}\bc_{k+2}
\left(1-\displaystyle\frac{h}{\bd_{k+1}}\right)
\left(1-\displaystyle\frac{h}{\bd_{k+2}}\right)\nonumber
\end{eqnarray}
\item[\rm c)] The brackets {\rm(\ref{dRTL- loc PB1}), (\ref{dRTL- loc PB2})}
are compatible. The map {\rm(\ref{dRTL- in loc map})} is Poisson with respect 
to both of them.
\end{itemize}
\end{theorem}
{\bf Proof} consists of straightforward calculations. \qed 
\vspace{2mm}

\begin{theorem}\label{RTL- in localization variables}
The pull--back of the flow {\rm RTL}$-$ under the map {\rm(\ref{dRTL- loc map})}
is described by the following equations of motion:
\begin{eqnarray}
\dot{\bd}_k & = & (\bd_k-h)\left(\frac{\bc_k}{\bd_{k}\bd_{k+1}}\bigg(1+
\frac{h\bc_k}{\bd_{k}\bd_{k+1}}\bigg)^{-1}-\frac{\bc_{k-1}}{\bd_{k-1}\bd_{k}}
\bigg(1+\frac{h\bc_{k-1}}{\bd_{k-1}\bd_{k}}\bigg)^{-1}\right)\nonumber\\
\label{RTL- in loc map}\\
\dot{\bc}_k & = &\bc_k\left(\frac{1}{\bd_k}-\frac{1}{\bd_{k+1}}\right)
\bigg(1+\frac{h\bc_k}{\bd_{k}\bd_{k+1}}\bigg)^{-1}\nonumber
\end{eqnarray}
\end{theorem}
{\bf Proof.} We use the part a) of the previous Theorem. The flow under 
consideration is a Hamiltonian system with
the Hamilton function, which is a pull--back of $h^{-1}\rH_0(c,d)$
(indeed, this function is a Casimir function for $\{\cdot,\cdot\}_2$ and is
a Hamilton function of the flow RTL$-$ in the bracket $-h\{\cdot,\cdot\}_1$).
Calculating the equations of motion generated by the Hamilton function
\[
h^{-1}\sum_{k=1}^N \log(\bd_k)+h^{-1}\sum_{k=1}^N\log\left(1+\frac{h\bc_k}
{\bd_k\bd_{k+1}}\right)
\]
in the Poisson brackets (\ref{dRTL- loc PB1}), we arrive at (\ref{RTL- in 
loc map}). \qed

\subsection{Third appearance of the Volterra lattice}

It is interesting to remark that the flow RTL+ allows the reduction $d_k=0$,
in which it turns into the Volterra lattice
\[
\dot{c}_k=c_k(c_{k+1}-c_{k-1})
\]
The Lax representation of the RTL+ flow survives this reduction, delivering
a new (third) Lax representation for the VL. Also the quadratic and the
cubic Poisson brackets (\ref{RTL q br}) and (\ref{RTL c br}) allow this 
reduction and turn into the corresponding objects for the VL.

It may be verified that the map dRTL+ in the reduction $d_k=0$ turns into dVL,
although this discretization is based on a completely different Lax 
representation. Naturally, the same holds for the local forms of these maps.


\setcounter{equation}{0}
\section{Belov--Chaltikian lattice}\label{Sect BC}

\subsection{Equations of motion and bi--Hamiltonian structure}
In their studies of the lattice analogs of $W$--algebras, Belov and Chaltikian
\cite{BC}, \cite{ABC} found an interesting integrable lattice (hereafter BCL):
\begin{equation}\label{BCL}
\dot{b}_k=b_k(b_{k+1}-b_{k-1})-c_k+c_{k-1},\qquad \dot{c}_k=c_k(b_{k+2}-b_{k-1})
\end{equation}
This system may be viewed as an extension of the Volterra lattice (which
appears as a $c_k=0$ reduction of the above system). Belov and Chaltikian
established also the bi--Hamiltonian structure of this system. Namely, its
phase space, which in the periodic case is
\begin{equation}\label{BC phase sp}
{\cal BC}={\Bbb R}^{2N}(b_1,c_1,\ldots,b_N,c_N)
\end{equation}
carries two compatible local Poisson brackets, with respect to which the system
BCL is Hamiltonian. The first (''quadratic'') Poisson bracket is given by
\begin{equation}\label{BC q br}
\begin{array}{cclcccl}
\{b_k,b_{k+1}\}_2 & = & -b_kb_{k+1}+c_k, & \quad & 
\{c_k,c_{k+1}\}_2 & = & -c_kc_{k+1}\\
\{b_k,c_{k+1}\}_2 & = & -b_kc_{k+1}, & \quad & 
\{c_k,b_{k+2}\}_2 & = & -c_kb_{k+2}\\
\{c_k,c_{k+2}\}_2 & = & -c_kc_{k+2} & & & & 
\end{array}
\end{equation}
the corresponding Hamilton function being
\begin{equation}\label{BC H1}
\rH_1(b,c)=\sum_{k=1}^N b_k
\end{equation}
The second (''cubic'') Poisson bracket on ${\cal BC}$ is given by
\begin{eqnarray}\label{BC c br}
\{b_k,c_k\}_3 & = & -c_k(b_kb_{k+1}-c_k) \nonumber\\
\{c_k,b_{k+1}\}_3 & = & -c_k(b_kb_{k+1}-c_k) \nonumber\\ 
\{b_k,b_{k+1}\}_3 & = & -(b_k+b_{k+1})(b_kb_{k+1}-c_k) \nonumber\\
\{c_k,c_{k+1}\}_3 & = & -c_kc_{k+1}(b_k+b_{k+2}) \nonumber\\
\{b_k,c_{k+1}\}_3 & = & -b_kc_{k+1}(b_k+b_{k+1})+c_kc_{k+1} \nonumber\\ 
\{c_k,b_{k+2}\}_3 & = & -c_kb_{k+2}(b_{k+1}+b_{k+2})+c_kc_{k+1} \\ 
\{b_k,b_{k+2}\}_3 & = & -b_kb_{k+1}b_{k+2}+b_kc_{k+1}+c_kb_{k+2} \nonumber\\ 
\{c_k,c_{k+2}\}_3 & = & -c_kc_{k+2}(b_{k+1}+b_{k+2})\nonumber\\
\{b_k,c_{k+2}\}_3 & = & -c_{k+2}(b_kb_{k+1}-c_k)  \nonumber\\
\{c_k,b_{k+3}\}_3 & = & -c_k(b_{k+2}b_{k+3}-c_{k+2}) \nonumber\\
\{c_k,c_{k+3}\}_3 & = & -c_kb_{k+2}c_{k+3} \nonumber 
\end{eqnarray}
and the corresponding Hamilton function is equal to
\begin{equation}\label{BC H0}
\rH_0(b,c)=\frac{1}{3}\sum_{k=1}^N \log(c_k)
\end{equation}

\subsection{Lax representation}
The Lax matrix of the BCL found in \cite{BC}, \cite{ABC}, is given in terms
of the matrices
\begin{equation}\label{BC UV}
U(\lambda)=\lambda\sum_{k=1}^N E_{k+1,k},\qquad  V(b,c,\lambda)=I-\lambda^{-1}
\sum_{k=1}^N b_kE_{k,k+1}+\lambda^{-2}\sum_{k=1}^N c_kE_{k,k+2}
\end{equation}

\begin{theorem}
The equations of motion {\rm(\ref{BCL})} are equivalent to the following 
matrix differential equation:
\begin{equation}\label{BC Vdot}
\dot{V}=VB-AV
\end{equation}
where
\begin{eqnarray}\label{BC AB}
A(b,c,\lambda) & = & \pi_+(UV^{-1})\;=\;
\sum_{k=1}^N b_{k-1}E_{kk}+\lambda\sum_{k=1}^N E_{k+1,k}\nonumber\\
B(b,c,\lambda) & = & \pi_+(V^{-1}U)\;=\;
\sum_{k=1}^N b_kE_{kk}+\lambda\sum_{k=1}^N E_{k+1,k}
\end{eqnarray}
so that also the following equation holds identically:
\begin{equation}\label{BC Udot}
UB-AU=0
\end{equation}
\end{theorem}

\noindent
{\bf Corollary} \cite{BC},\cite{ABC}. {\it The matrix
\begin{equation}\label{BC T}
T(b,c,\lambda)=U\cdot V^{-1}(b,c,\lambda)
\end{equation}
satisfies the usual Lax equation in $\g$:}
\begin{equation}\label{BC Lax}
\dot{T}=[T,A]
\end{equation}
\vspace{1.5mm}

This Lax equation can be given an $r$--matrix interpretation in the case of
quadratic Poisson bracket, which can be lifted to an $r$--matrix interpretation 
of the Lax triads for the pairs $(U,V)\in\g\otimes\g$.

\subsection{Discretization}
Since the Lax equation (\ref{BC Lax}) has the form (\ref{Lax in recipe}),
and moreover, the pairs $(U,V^{-1})$ satisfy the Lax triads of the form
(\ref{Lax triads in recipe}), we can apply the recipe of Sect. \ref{Sect recipe}.
Taking, as usual, $F(T)=I+hT$, we come to the discrete time matrix equation
\begin{equation}\label{dBC Lax triad V}
\wV=\mbA^{-1}V\mbB
\end{equation}
with
\[
\mbA=\Pi_+(I+hUV^{-1}),\qquad \mbB=\Pi_+(I+hV^{-1}U)
\]
Moreover, since the equation 
\begin{equation}\label{dBC Lax triad U}
U=\mbA^{-1}U\mbB
\end{equation}
holds, we have also the usual Lax equation
\begin{equation}\label{dBC Lax}
\wT=\mbA^{-1}T\mbA
\end{equation}
\begin{theorem} The equations {\rm(\ref{dBC Lax triad V}), (\ref{dBC Lax triad 
U})} are equivalent to the following equations:
\begin{equation}\label{dBC}
\wb_k=b_k\,\frac{\alpha_{k+2}}{\alpha_k}-h\left(c_k\,\frac{1}{\alpha_k}
-c_{k-1}\,\frac{\alpha_{k+2}}{\alpha_k\alpha_{k-1}}\right),\qquad 
\wc_k=c_k\,\frac{\alpha_{k+3}}{\alpha_k}
\end{equation}
where the coefficients $\alpha_k=\alpha_k(b,c)=1+O(h)$ are uniquely defined by
the recurrent relation
\begin{equation}\label{dBC alpha}
\alpha_k=1+\frac{hb_{k-1}}{\alpha_{k-1}}+\frac{h^2c_{k-2}}
{\alpha_{k-1}\alpha_{k-2}}
\end{equation}
The following asymptotics hold:
\begin{equation}\label{dBC alpha as}
\alpha_k=1+hb_{k-1}+O(h^2)
\end{equation}
\end{theorem}
{\bf Remark.} The auxiliary matrices $\mbA$, $\mbB$ are bi--diagonal:
\begin{eqnarray}
\mbA(b,c,\lambda) & = & \sum_{k=1}^N \alpha_kE_{kk}+h\lambda\sum_{k=1}^N 
E_{k+1,k} \label{dBC A}\\
\mbB(b,c,\lambda) & = & \sum_{k=1}^N \alpha_{k+1}E_{kk}+h\lambda\sum_{k=1}^N 
E_{k+1,k} \label{dBC B}
\end{eqnarray}

\noindent
{\bf Proof} is by now standard. The general bi--diagonal structure of the 
factors $\mbA$, $\mbB$ follows from $\mbA=\Pi_+(I+hUV^{-1})=\Pi_+(V+hU)$, the
latter representation implies also the recurrent relation
for the entries $\alpha_k$ of the matrix $\mbA$. From (\ref{dBC Lax triad U})
one derives immediately $\beta_k=\alpha_{k+1}$. The equations of motion
(\ref{dBC}) follow then easily from $\mbA\wV=V\mbB$. \qed
\vspace{1.5mm}

Hereafter we call the map (\ref{dBC}) dBCL. By construction, it is bi--Poisson
with respect to the brackets (\ref{BC q br}) and (\ref{BC c br}), approximates
the flow BCL due to the asymptotics (\ref{dBC alpha as}), but is nonlocal due
to the nature of the auxiliary quantities $\alpha_k$.

\subsection{Local equations of motion for the dBCL}
The localizing change of variables for dBCL is the map
${\cal BC}(\bb,\bc)\mapsto{\cal BC}(b,c)$ given by the formulas:
\begin{equation}\label{dBC loc map}
b_k=\bb_k(1+h\bb_{k-1})-h\bc_{k-1},\qquad c_k=\bc_k(1+h\bb_{k-1})
\end{equation}
As usual this is a local diffeomorphism for $h$ small enough.
\begin{theorem}
The change of variables {\rm(\ref{dBC loc map})} conjugates the map {\rm dBCL}
with the following one:
\begin{equation}\label{dBC loc}
\begin{array}{c}
\widetilde{\bb}_k(1+h\widetilde{\bb}_{k-1})-h\widetilde{\bc}_{k-1}=
\bb_k(1+h\bb_{k+1})-h\bc_k\\ \\
\widetilde{\bc}_k(1+h\widetilde{\bb}_{k-1})=\bc_k(1+h\bb_{k+2})
\end{array}
\end{equation}
\end{theorem}
{\bf Proof.} Introducing the quantities 
\begin{equation}\label{dBC loc alpha}
\alpha_k=1+h\bb_{k-1}
\end{equation}
we immediately see via simple check that they satisfy the recurrent relations
(\ref{dBC alpha}). Hence they represent the unique solution of these 
recurrencies with the asymptotics $\alpha_k=1+O(h)$. Now the equations of
motion follow directly from (\ref{dBC}), (\ref{dBC loc map}), 
(\ref{dBC loc alpha}). \qed
\vspace{1.5mm}

\begin{theorem} The pull back of the bracket
\begin{equation}
\{\cdot,\cdot\}_2+h\{\cdot,\cdot\}_3
\end{equation}
on ${\cal BC}(b,c)$ under the change of variables {\rm(\ref{dBC loc map})}
is the following {\em local} Poisson bracket on ${\cal BC}(\bb,\bc)$:
\begin{eqnarray}\label{dBC loc PB}
\{\bb_k,\bc_k\} & = & -h\bc_k(\bb_k\bb_{k+1}-\bc_k)(1+h\bb_k) \nonumber\\
\{\bc_k,\bb_{k+1}\} & = & -h\bc_k(\bb_k\bb_{k+1}-\bc_k)(1+h\bb_{k+1}) 
\nonumber\\ 
\{\bb_k,\bb_{k+1}\} & = & -(\bb_k\bb_{k+1}-\bc_k)(1+h\bb_k)(1+h\bb_{k+1}) 
\\
\{\bc_k,\bc_{k+1}\} & = & -\bc_k\bc_{k+1}\Big(1+h\bb_k+h\bb_{k+2}
+h^2(\bb_k\bb_{k+1}-\bc_k)+h^2(\bb_{k+1}\bb_{k+2}-\bc_{k+1})\Big) 
\nonumber\\
\{\bb_k,\bc_{k+1}\} & = & -\bc_{k+1}\Big(\bb_k+h(\bb_k\bb_{k+1}-\bc_k)\Big)
(1+h\bb_k) \nonumber\\ 
\{\bc_k,\bb_{k+2}\} & = & -\bc_k\Big(\bb_{k+2}+h(\bb_{k+1}\bb_{k+2}-\bc_{k+1})\Big)
(1+h\bb_{k+2}) \nonumber\\ 
\{\bc_k,\bc_{k+2}\} & = & -\bc_k\bc_{k+2}\Big(1+h\bb_{k+1}+h\bb_{k+2}+
h^2(\bb_{k+1}\bb_{k+2}-\bc_{k+1})\Big)\nonumber
\end{eqnarray}
The map {\rm(\ref{dBC loc})} is Poisson with respect to this bracket.
\end{theorem}
{\bf Proof} -- by a straightforward but tiresome calculation. \qed
\vspace{2mm}

\begin{theorem} The pull--back of the flow {\rm BCL} under the change of 
variables {\rm(\ref{dBC loc map})} is described by the following equations of 
motion:
\begin{eqnarray}\label{BC in dBC loc map}
\dot{\bb}_k & = & (1+h\bb_k)\Big(\bb_k(\bb_{k+1}-\bb_{k-1})-\bc_k+\bc_{k-1}\Big)
\nonumber\\ \\
\dot{\bc}_k & = & \bc_k\Big(\bb_{k+2}(1+h\bb_{k+1})-\bb_{k-1}(1+h\bb_k)
-h\bc_{k+1}+h\bc_{k-1}\Big)  \nonumber
\end{eqnarray}
\end{theorem}
{\bf Proof.} We can use the Hamiltonian formalism. The pull--back we are looking
for, is a Hamiltonian system on ${\cal BC}(\bb,\bc)$ with the Poisson bracket
(\ref{dBC loc PB}) and the Hamilton function which is a pull--back of
$(3h)^{-1}\sum_{k=1}^N\log(c_k)$. Indeed, this function is a Casimir function
for $\{\cdot,\cdot\}_2$ and is the Hamilton function for BCL in the bracket
$h\{\cdot,\cdot\}_3$. Calculating the equations of motion generated by the
Hamilton function
\[
(3h)^{-1}\sum_{k=1}^N\log(\bc_k)+(3h)^{-1}\sum_{k=1}^N\log(1+h\bb_k)
\]
in the bracket (\ref{dBC loc PB}), we arrive at (\ref{BC in dBC loc map}). \qed
 

\setcounter{equation}{0}
\section{A perturbation of the Volterra lattice}
\label{Sect perturbed Volterra}

\subsection{Equations of motion and bi--Hamiltonian structure}
Consider the following lattice system:
\begin{equation}\label{PVL}
\begin{array}{l}
\dot{u}_k = u_k(w_k-w_{k-1}+u_kw_k-u_{k-1}w_{k-1})\\ \\
\dot{w}_k = w_k(u_{k+1}-u_k+u_{k+1}w_{k+1}-u_kw_k)
\end{array}
\end{equation}
It may be considered as a perturbation of the Volterra lattice, therefore we
adopt a name PVL for it. Its phase space in the periodic case is
\begin{equation}\label{PVL phase sp}
\cP\cV={\Bbb R}^{2N}(u_1,w_1,\ldots,u_N,w_N)
\end{equation}
The system PVL is bi--Hamiltonian. First of all, it is Hamiltonian with
respect to a quadratic Poisson bracket on $\cP\cV$ which is identical with 
the invariant quadratic Poisson bracket of the Volterra lattice:
\begin{equation}\label{PVL q br}
\{u_k,w_k\}_2=-u_kw_k,\qquad \{w_k,u_{k+1}\}_2=-w_ku_{k+1}
\end{equation}
The corresponding Hamilton function is equal to
\begin{equation}\label{PVL H1}
\rH_1(u,w)=\sum_{k=1}^N(u_k+w_k+u_kw_k)
\end{equation}
The second (''cubic'') invariant Poisson bracket, compatible with the previous
one, is different from the cubic bracket of the VL and is given by
\begin{eqnarray}\label{PVL c br}
\{u_k,w_k\}_3=-u_kw_k(u_k+w_k+u_kw_k),&\quad&
\{w_k,u_{k+1}\}_3=-w_ku_{k+1}(w_k+u_{k+1})
\nonumber\\
\{u_k,u_{k+1}\}_3=-u_ku_{k+1}(w_k+u_kw_k),&\quad&
\{w_k,w_{k+1}\}_3=-w_kw_{k+1}(u_{k+1}+u_{k+1}w_{k+1})
\nonumber\\
\{w_k,u_{k+2}\}_3=-w_ku_{k+1}w_{k+1}u_{k+2}&&
\end{eqnarray}
The corresponding Hamilton function may be taken as
\begin{equation}\label{PVL H0}
\rH_0(u,w)=\sum_{k=1}^N \log(u_k)\quad{\rm or}\quad 
\rH_0(u,w)=\sum_{k-1}^N \log(w_k)
\end{equation}
(the difference of these two functions is a Casimir of the bracket 
(\ref{PVL c br})).

\subsection{Lax representation}
The Lax representation for PVL is given in terms of three matrices from $\g$:
\begin{eqnarray}
U(u,w,\lambda) & = & \sum_{k=1}^N u_k E_{k,k}+\lambda\sum_{k=1}^N E_{k+1,k} 
\nonumber\\
V(u,w,\lambda) & = & I-\lambda^{-1}\sum_{k=1}^N u_kw_kE_{k,k+1} \\
W(u,w,\lambda) & = & I+\lambda^{-1}\sum_{k=1}^N w_kE_{k,k+1}\nonumber
\end{eqnarray}
\begin{theorem} The equations of motion {\rm(\ref{PVL})} are 
equivalent to the following matrix differential equations:
\begin{equation}\label{PVL Lax UW}
\dot{U}=UC-AU,\qquad \dot{W}=WB-CW
\end{equation}
and imply also the matrix differential equation
\begin{equation}\label{PVL Lax V}
\dot{V}=VB-AV
\end{equation}
with the auxiliary matrices
\begin{eqnarray}\label{PVL ABC}
A(u,w,\lambda) & = & \sum_{k=1}^N (u_k+u_{k-1}w_{k-1}+w_{k-1}) E_{kk}+
\lambda\sum_{k=1}^N E_{k+1,k} \nonumber\\
B(u,w,\lambda) & = & \sum_{k=1}^N (u_k+u_kw_k+w_{k-1}) E_{kk}+
\lambda\sum_{k=1}^N E_{k+1,k} \\
C(u,w,\lambda) & = & \sum_{k=1}^N (u_k+u_kw_k+w_k) E_{kk}+
\lambda\sum_{k=1}^N E_{k+1,k}\nonumber 
\end{eqnarray}
\end{theorem}
{\bf Proof} -- an elementary check. \qed
\vspace{1.5mm}

It is easy to establish the following fact:
\[
A=\pi_+(UWV^{-1}),\qquad B=\pi_+(V^{-1}UW),\qquad C=\pi_+(WV^{-1}U)
\]
so that the triples $(U,V^{-1},W)\in\g\otimes\g\otimes\g$ satisfy the Lax 
equations of the type (\ref{Lax triads in recipe}) with $m=3$ and $f(T)=T$.
These equations may be given an $r$--matrix interpretation, at least in the
case of the quadratic bracket $\{\cdot,\cdot\}_2$. The corresponding quadratic
bracket on $\g\otimes\g\otimes\g$ turns out to be identical with the one
introduced in \cite{S12}.

\subsection{Discretization}
To discretize the PVL, we can apply the recipe of Sect. \ref{Sect recipe}
with $F(T)=I+hT$. So, we have to consider the following discrete time Lax 
representation:
\begin{equation}\label{dPVL Lax}
\wU=\mbA^{-1}U\mbC,\qquad \wV=\mbA^{-1}V\mbB, \qquad
\wW=\mbC^{-1}W\mbB
\end{equation}
where
\[
\mbA=\Pi_+\Big(I+hUWV^{-1}\Big),\quad \mbB=\Pi_+\Big(I+hV^{-1}UW\Big),\quad
\mbC=\Pi_+\Big(I+hWV^{-1}U\Big)
\]
\begin{theorem} The discrete time Lax equations {\rm(\ref{dPVL Lax})}
are equivalent to the following equations of motion:
\begin{equation}\label{dPVL}
\wu_k=u_k\,\frac{\gc_k}{\ga_k},\qquad
\ww_k=w_k\,\frac{\gb_{k+1}}{\gc_k}
\end{equation}
where the functions $\ga_k=\ga_k(u,w)=1+O(h)$ are uniquely defined by the 
system of recurrent relations
\begin{equation}\label{dPVL alpha}
\ga_{k+1}=1+h(u_{k+1}+w_k)+\displaystyle\frac{h(1-h)u_kw_k}{\ga_k}
\end{equation}
and the coefficients $\gb_k=\gb_k(u,w)=1+O(h)$, $\gc_k=\gc_k(u,w)=1+O(h)$
are given by
\begin{equation}\label{dPVL betagamma}
\gb_k=\ga_{k-1}\,\frac{\ga_k+hu_kw_k}{\ga_{k-1}+hu_{k-1}w_{k-1}},\qquad
\gc_k=\ga_k\,\frac{\ga_{k+1}-hu_{k+1}}{\ga_k-hu_k}
\end{equation}
The following asymptotics hold:
\begin{eqnarray}
\ga_k & = & 1+h(u_k+u_{k-1}w_{k-1}+w_{k-1})+O(h^2)\\
\gb_k & = & 1+h(u_k+u_kw_k+w_{k-1})+O(h^2)\\
\gc_k & = & 1+h(u_k+u_kw_k+w_k)+O(h^2)
\end{eqnarray}
\end{theorem}
{\bf Remark.} The auxiliary matrices $\mbA$, $\mbB$, $\mbC$ are bi--diagonal:
\begin{equation}\label{dPVL A}
\mbA=\sum_{k=1}^N \ga_kE_{k,k}+h\lambda\sum_{k=1}^N E_{k+1,k}
\end{equation}
\begin{equation}\label{dPVL B}
\mbB=\sum_{k=1}^N \gb_kE_{k,k}+h\lambda\sum_{k=1}^N E_{k+1,k}
\end{equation}
\begin{equation}\label{dPVL C}
\mbC=\sum_{k=1}^N \gc_kE_{k,k}+h\lambda\sum_{k=1}^N E_{k+1,k}
\end{equation}

\noindent
{\bf Proof.} We have: $\mbA=\Pi_+(I+hUWV^{-1})=\Pi_+(V+hUW)$, since
$V\in\G_-$. Here
\[
V+hUW=\sum_{k=1}^N(1+hu_k+hw_{k-1})E_{k,k}-\lambda^{-1}\sum_{k=1}^N
(1-h)u_kw_kE_{k,k+1}+h\lambda\sum_{k=1}^N E_{k+1,k}
\]
and the recurrent relations (\ref{dPVL alpha}) for $\ga_k$, the entries 
of the $\Pi_+$ factor of this tri--diagonal matrix, follow immediately. 
The expressions for $\gb_k$, $\gc_k$ through $\ga_k$, as well as the equations
of motion (\ref{dPVL}), follow directly from the Lax equations
$\mbA\wU=U\mbC$, $\mbA\wV=V\mbB$, $\mbC\wW=W\mbB$. \qed

We denote the map defined in this theorem by dPVL. As usual, it shares
with the system PVL the bi--Hamiltonian structure, the integrals of motion,
and so on, but is highly nonlocal.

\subsection{Local equations of motion for dPVL}
The localizing change of variables for the map dPLV is given by the formulas
\begin{equation}\label{dPVL loc map}
u_k=\bu_k\,\frac{1+h\bw_{k-1}}{1-h\bu_{k-1}\bw_{k-1}},\qquad
w_k=\bw_k\,\frac{1+h\bu_k}{1-h\bu_k\bw_k}
\end{equation}
Indeed, the following statement holds.
\begin{theorem}
The change of variables {\rm(\ref{dPVL loc map})} conjugates the map
{\rm dPVL} with the following one:
\begin{eqnarray}
\widetilde{\bu}_k\,\frac{1+h\widetilde{\bw}_{k-1}}
{1-h\widetilde{\bu}_{k-1}\widetilde{\bw}_{k-1}} 
& = &
\bu_k\,\frac{1+h\bw_k}{1-h\bu_k\bw_k}  \nonumber\\
\label{dPVL loc} \\
\widetilde{\bw}_k\,\frac{1+h\widetilde{\bu}_k}{1-h\widetilde{\bu}_k
\widetilde{\bw}_k} &  = &
\bw_k\,\frac{1+h\bu_{k+1}}{1-h\bu_{k+1}\bw_{k+1}} \nonumber
\end{eqnarray}
\end{theorem}
{\bf Proof.} The statement will follow immediately, if we establish the
local expressions for the quantities $\ga_k$:
\begin{equation}\label{dPVL loc alpha}
\ga_k=\frac{(1+h\bu_k)(1+h\bw_{k-1})}{(1-h\bu_{k-1}\bw_{k-1})}
\end{equation}
Indeed, from (\ref{dPVL loc map}), (\ref{dPVL loc alpha}) and
the formulas (\ref{dPVL betagamma}) we derive immediately:
\begin{equation}\label{dPVL loc beta}
\gb_k=\frac{(1+h\bu_k)(1+h\bw_{k-1})}{(1-h\bu_k\bw_k)}
\end{equation}
\begin{equation}\label{dPVL loc gamma}
\gc_k=\frac{(1+h\bu_k)(1+h\bw_k)}{(1-h\bu_k\bw_k)}
\end{equation}
and then (\ref{dPVL}) imply (\ref{dPVL loc}). To prove
(\ref{dPVL loc alpha}), we take this formula as a {\it definition} of 
the quantities $\ga_k$ and by means of a simple algebra verify that then
the recurrent relations (\ref{dPVL alpha}) hold. The reference to
the uniqueness of solution to these recurrent relations finishes the proof.
\qed
\vspace{1.5mm}

\begin{theorem}
The pull--back of the bracket
\begin{equation}
\{\cdot,\cdot\}_2+h\{\cdot,\cdot\}_3
\end{equation}
on $\cP\cV(u,w)$ under the change of variables {\rm(\ref{dPVL loc map})}
is the following bracket on $\cP\cV(\bu,\bw)$:
\begin{equation}\label{dPVL loc PB loc}
\{\bu_k,\bw_k\}=-\bu_k\bw_k(1+h\bu_k)(1+h\bw_k),\qquad
\{\bw_k,\bu_{k+1}\}=-\bw_k\bu_{k+1}(1+h\bw_k)(1+h\bu_{k+1})
\end{equation}
The map {\rm(\ref{dPVL loc})} is Poisson with respect to the bracket 
{\rm (\ref{dPVL loc PB loc})}.
\end{theorem}
{\bf Proof} -- by a straightforward verification. \qed
\vspace{1.5mm} 

It is very interesting that the bracket (\ref{dPVL loc PB loc}) again turns out
to be identical with the invariant local bracket (\ref{dVL loc PB loc}) of the 
local version of dVL (so that the contributions of different cubic brackets for 
VL and PVL are somehow compensated by different localizing changes of variables).

\begin{theorem} The pull--back of the flow {\rm PVL} under the change of 
variables {\rm(\ref{dPVL loc map})} is described by the following differential
equations:
\begin{eqnarray}\label{PVL in dPVL loc map}
\dot{\bu}_k & = & \bu_k(1+h\bu_k)\left(\displaystyle\frac{\bw_k+\bu_k\bw_k}
{1-h\bu_k\bw_k}-\displaystyle\frac{\bw_{k-1}+\bu_{k-1}\bw_{k-1}}
{1-h\bu_{k-1}\bw_{k-1}}\right)\nonumber\\ \\
\dot{\bw}_k & = & \bw_k(1+h\bw_k)\left(\displaystyle\frac
{\bu_{k+1}+\bu_{k+1}\bw_{k+1}}{1-h\bu_{k+1}\bw_{k+1}}
-\displaystyle\frac{\bu_k+\bu_k\bw_k}{1-h\bu_k\bw_k}\right)\nonumber
\end{eqnarray}
\end{theorem}
{\bf Proof.} To obtain these differential equations, one has to calculate 
the Hamiltonian equations of motion generated by the Hamilton function
\[
h^{-1}\sum_{k=1}^N\Big(\log(\bw_k)+\log(1+h\bu_k)-\log(1-h\bu_k\bw_k)\Big)
\]
with respect to the Poisson brackets (\ref{dPVL loc PB loc}). \qed


\setcounter{equation}{0}
\section{Some constrained lattice KP systems}
\label{Sect lattice cKP}

In this section we introduce a large family of systems generalizing
simulatneously the Volterra lattice, the relativistic Toda lattice, 
the Belov--Chaltikian lattice, and the perturbed Volterra lattice. 
For some reasons it is convenient 
to call these systems constrained lattice KP systems.

\subsection{Equations of motion and Hamiltonian structure}
Each system of this family may be treated as consisting of $m$ sorts of 
particles. The phase space of such systems in the periodic case is 
described as:
\begin{equation}\label{cKP phase sp}
\cK_m={\Bbb R}^{mN}\Big(v^{(1)},\ldots,v^{(m)}\Big)
\end{equation}
where each vector
\begin{equation}\label{cKP sort}
v^{(j)}=\Big(v_1^{(j)},\ldots,v_N^{(j)}\Big)\in{\Bbb R}^N
\end{equation}
represents the set of particles of the $j$th sort. 
We introduce the notion of the {\it signature} of the constrained lattice KP
as an ordered $m$--tuple of numbers:
\begin{equation}\label{cKP sign}
\epsilon=(\epsilon_1,\ldots,\epsilon_m),\qquad \epsilon_j\in\{0,1\},\quad 
\epsilon_1=0
\end{equation}
The {\it constrained KP lattice} (hereafter cKPL$(m)$) with the signature 
$\epsilon$ is the following system of differential equations:
\begin{equation}\label{cKP}
\dot{v}_k^{(j)}=v_k^{(j)}\left(
\sum_{i=1}^{j-1}\Big(v_{k+1}^{(i)}-v_k^{(i)}\Big)
+\epsilon_j\Big(v_{k+1}^{(j)}-v_{k-1}^{(j)}\Big)
+\sum_{i=j+1}^{m}\Big(v_k^{(i)}-v_{k-1}^{(i)}\Big)\right)
\end{equation}

Obviously, the Volterra lattice VL belongs to this class
and is characterized by the signature $\epsilon=(0,0)$. The RTL+ flow of the
relativistic Toda hierarchy also belongs to this class and has the signature
$\epsilon=(0,1)$. We will see later on that BCL is Miura related to a simple 
reduction of cKPL(3) with $\epsilon=(0,1,1)$, while PVL itself is a less trivial
reduction of cKPL(3) with $\epsilon=(0,1,0)$. The system cKPL$(m)$ with 
the signature consisting only of zeros, $\epsilon=(0,\ldots,0)$ is nothing but 
the Bogoyavlensky lattice BL1$(m-1)$. Also the $v_k^{(1)}=0$ reduction of 
cKPL$(m)$ with the signature $\epsilon=(0,1,1,\ldots,1)$ coincides with 
BL1$(m-1)$.

\begin{theorem}\label{Hamiltonian structure for cKP}
The system {\rm(\ref{cKP})} is Hamiltonian with respect to the following Poisson 
bracket on $\cK_m$:
\begin{equation}\label{cKP PB j}
\{v_k^{(j)},v_{k+1}^{(j)}\}_2=-\epsilon_jv_k^{(j)}v_{k+1}^{(j)}
\end{equation}
\begin{equation}\label{cKP PB i<j}
\{v_k^{(i)},v_k^{(j)}\}_2=-v_k^{(i)}v_k^{(j)},\qquad 
\{v_k^{(j)},v_{k+1}^{(i)}\}_2=-v_k^{(j)}v_{k+1}^{(i)}\quad
{\rm for}\quad 1\leq i<j\leq m
\end{equation}
with the Hamilton function
\begin{equation}\label{cKP H}
\rH_1(v)=\rH_1(v^{(1)},\ldots,v^{(m)})=\sum_{j=1}^{m}\sum_{k=1}^N v_k^{(j)}
\end{equation}
(independent of the signature).
\end{theorem} 

This statement can be easily checked and generalizes the quadratic brackets
(\ref{VL q br}) and (\ref{RTL q br}) for the VL and the RTL, respectively.
It would be important to find out, when do the analogs of the linear bracket 
(for the RTL) and of the cubic bracket (for both the VL and the RTL) hold,
and to find the corresponding expressions.

\subsection{Lax representation}
The natural Lax representation of the cKPL$(m)$ (\ref{cKP}) lives in 
$\g^{\otimes\,m}$ and is given in terms of the following matrices from $\g$:
\begin{eqnarray}
U_1(v^{(1)},\lambda) & = & \lambda\sum_{k=1}^N E_{k+1,k}+
\sum_{k=1}^N v_k^{(1)}E_{k,k}     \label{cKP U1}\\
V_j(v^{(j)},\lambda) & = & I+\sigma_j\lambda^{-1}\sum_{k=1}^N v_k^{(j)}E_{k,k+1},
\qquad j=1,2,\ldots,m\label{cKP Vj}
\end{eqnarray}
where 
\begin{equation}\label{cKP eps}
\sigma_j=\left\{\begin{array}{rcl}1 & \quad & \epsilon_j=0\\
-1 & \quad & \epsilon_j=1\end{array}\right\}=1-2\epsilon_j
\end{equation}

\begin{theorem}\label{cKP Lax in g+g}
The equations of motion {\rm(\ref{cKP})} are equivalent to
the following matrix differential equations:
\begin{eqnarray}
\dot{U}_1 & = & U_1B_{m}-B_1U_1\label{cKP Lax triads U}\\ \nonumber\\
\dot{V}_j & = & \left\{\begin{array}{ccl}
V_jB_{j-1}-B_jV_j & \quad & \epsilon_j=0\\ \\
V_jB_j-B_{j-1}V_j & \quad & \epsilon_j=1\end{array}\right\}
\qquad 2\le j\le m \label{cKP Lax triads V}
\end{eqnarray} 
where
\begin{equation}\label{cKP B}
B_j(v,\lambda)=\sum_{k=1}^N\Big(\sum_{i=1}^jv_k^{(i)}+\sum_{i=j+1}^{m}
v_{k-1}^{(i)}\Big)E_{k,k}+\lambda\sum_{k=1}^N E_{k+1,k}
\end{equation}
The evolution of the monodromy matrices
\begin{equation}\label{cKP Tj}
T_j(v^{(1)},\ldots,v^{(m)},\lambda)=V_j^{\sigma_j}(\lambda)\cdot\ldots\cdot
V_2^{\sigma_2}(\lambda)\cdot U_1(\lambda)\cdot V_{m}^{\sigma_m}(\lambda)
\cdot\ldots\cdot V_{j+1}^{\sigma_{j+1}}(\lambda)
\end{equation}
is governed by usual Lax equations:
\begin{equation}\label{cKP Lax for Tj}
\dot{T}_j=[T_j,B_j]
\end{equation}
The matrices $B_j$ allow the representation
\begin{equation}\label{cKP Bj thru Tj}
B_j=\pi_+(T_j)
\end{equation}
\end{theorem}
{\bf Proof} -- an easy check. \qed
\vspace{1.5mm}

It is easy to see that this Lax representation is exactly of the form
(\ref{Lax triads in recipe}), if one considers the $m$--tuple of matrices 
\[
(U_1,V_2^{\sigma_2},\ldots,V_m^{\sigma_m})\in\g^{\otimes\,m}
\]
as the Lax matrix. This Lax representation allows an $r$--matrix 
interpretation for the quadratic bracket of the previous theorem.
As a matter of fact, the corresponding quadratic bracket in $\g^{\otimes\,m}$
literally coincides with the bracket introduced for the BL1 in \cite{S12}.
The Lax representation (\ref{cKP Lax triads U}), (\ref{cKP Lax triads V})
serves as a starting point for applying the recipe of Sect. \ref{Sect recipe}.

\subsection{Discretization}

Taking in this recipe $F(T)=I+hT$, we come to the discrete time Lax equations
\begin{eqnarray}
\wU_1 & = & \mbB_1^{-1}U_1\mbB_{m}\nonumber\\ \nonumber\\
\wV_j & = & \left\{\begin{array}{ccl}
\mbB_j^{-1}V_j\mbB_{j-1} & \quad & \epsilon_j=0\\ \\
\mbB_{j-1}^{-1}V_j\mbB_j & \quad & \epsilon_j=1\end{array}\right\}
\qquad 2\le j\le m
\label{dcKP Lax triads}
\end{eqnarray} 
with $\mbB_j=\Pi_+(I+hT_j)$. 
\begin{theorem} The discrete time Lax equations {\rm(\ref{dcKP Lax triads})}
are equivalent to the map $v\mapsto\wv$ described by the equations
\begin{equation}\label{dcKP}
\wv_k^{(1)}=v_k^{(1)}\,\displaystyle\frac{\gb_k^{(m)}}{\gb_k^{(1)}}\,,
\qquad
\wv_k^{(j)}=\left\{\begin{array}{ccl}
v_k^{(j)}\,\displaystyle\frac{\gb_{k+1}^{(j-1)}}{\gb_k^{(j)}} & \quad &
\epsilon_j=0\\ \\
v_k^{(j)}\,\displaystyle\frac{\gb_{k+1}^{(j)}}{\gb_k^{(j-1)}} & \quad &
\epsilon_j=1\end{array}\right\}
\qquad(2\le j\le m)
\end{equation}
where the functions $\gb_k^{(j)}=\gb_k^{(j)}(v)=1+O(h)$ satisfy the following
equations:
\begin{eqnarray}
\displaystyle\frac{\gb_k^{(m)}}{\gb_k^{(1)}} & = &
\displaystyle\frac{\gb_{k+1}^{(1)}-hv_{k+1}^{(1)}}{\gb_k^{(1)}-hv_k^{(1)}}
\label{dcKP aux1}\\
\displaystyle\frac{\gb_{k+1}^{(j-1)}}{\gb_k^{(j)}} & = & 
\displaystyle\frac{\gb_{k+1}^{(j)}-hv_{k+1}^{(j)}}{\gb_k^{(j)}-hv_k^{(j)}},
\qquad \epsilon_j=0,\quad 2\le j\le m
\label{dcKP aux2}\\
\displaystyle\frac{\gb_{k+1}^{(j)}}{\gb_k^{(j-1)}} & = &
\displaystyle\frac{\gb_{k+1}^{(j-1)}+hv_{k+1}^{(j)}}{\gb_k^{(j-1)}+hv_k^{(j)}}
\qquad \epsilon_j=1,\quad 2\le j\le m
\label{dcKP aux3}
\end{eqnarray}
\end{theorem}
{\bf Proof.} First of all notice that the matrices $\mbB_j$ must have
the following structure:
\[
\mbB_j(\lambda)=\sum_{k=1}^N \gb_k^{(j)} E_{k,k}+h\lambda\sum_{k=1}^N E_{k+1,k}
\]
Now the equations of motion are derived straightforwardly. For example,
the $\epsilon_j=1$ variant of the last $m-1$ equations in (\ref{dcKP Lax 
triads}), i.e. the matrix equation $\mbB_{j-1}\wV_j=V_j\mbB_j$, 
is equivalent to the following system of scalar equations:
\[
\left\{\begin{array}{l}
\gb_k^{(j-1)}\wv_k^{(j)}=v_k^{(j)}\gb_{k+1}^{(j)}\\ \\
\gb_k^{(j-1)}-h\wv_{k-1}^{(j)}=\gb_k^{(j)}-hv_k^{(j)}
\end{array}\right. 
\]
This is equivalent to the corresponding variant of equations of motion 
(\ref{dcKP}) together with the relation (\ref{dcKP aux3}). \qed
\vspace{2mm}

{\bf Remark.} It is important to notice that the statement of the last theorem
deviates from the usual scheme in that it does not contain a system of
equations which determine $\gb_k^{(j)}$ {\it uniquely}. In fact, in order to
find such a system one has to deduce from the equations 
(\ref{dcKP aux1})--(\ref{dcKP aux3}) $m$ formulas of the type 
\[
\displaystyle\frac{\gb^{(j)}_{k+q}}{\gb^{(j)}_k}=
\displaystyle\frac{\Psi_k^{(j)}}{\Psi_{k+1}^{(j)}}
\]
Here the number $q$ does not depend on $j$;  $\Psi_k^{(j)}$ are certain 
expressions of the type $\prod_{i=1}^m \psi_{k+n_i}^{(i)}$,
and all $\psi_k^{(i)}=\gb_k^{(i)}-hv_k^{(i)}$ or 
$\psi_k^{(i)}=\Big(\gb_k^{(i-1)}+hv_k^{(i)}\Big)^{-1}$. This, in turn, implies
that there hold certain equations of the type 
\begin{equation}\label{dcKP aux hypo}
\gb_k^{(j)}\cdot\ldots\cdot\gb_{k+q-1}^{(j)}\Psi_k^{(j)}={\rm const}
\end{equation}
(here we assumed for definiteness that $q>0$). The value of the constant 
on the right--hand side is uniquely defined by the conditions
$\mbB_j=\Pi_+(I+hT_j)$. (As a matter of fact, it is easy to see that this 
constant does not depend on $j$). The value of the constant being determined,
the equations (\ref{dcKP aux hypo}) give the desired system which defines
$\gb_k^{(j)}$ uniquely. However, the outfit of this system depends heavily on 
the signature $\epsilon$ of the cKPL, and the general 
formulas would contain too many indices to be instructive enough. It is 
simpler to derive such formulas for each concrete signature separately. 
Nevertheless, the formulas (\ref{dcKP aux1})--(\ref{dcKP aux3}) completely
characterize the coefficients of the matrices which serve as the factors
$\Pi_+(\alpha I+hT_j)$ with {\it some} $\alpha$, so that every solution of 
{\it this} system leads to a discretization based on the factorization of 
$I+h'T$ with $h'=h/\alpha$, which enjoys all the positive properties 
of our general construction. We call the maps introduced in the previous 
theorem dcKPL$(m)$.

\subsection{Local equations of motion for dcKPL}

The dcKPL$(m)$ can be brought into the local form for an arbitrary signature 
$\epsilon$.
\begin{theorem} The change of variables $\cK_m(\bv)\mapsto\cK_m(v)$,
\begin{equation}\label{dcKP loc map}
v_k^{(j)}=\bv_k^{(j)}\prod_{i=1}^{j-1}\Big(1+h\bv_k^{(i)}\Big)\cdot
\Big(1+\epsilon_j h\bv_{k-1}^{(j)}\Big)\cdot
\prod_{i=j+1}^{m}\Big(1+h\bv_{k-1}^{(i)}\Big)
\end{equation}
conjugates {\rm dcKPL}$(m)$ with the following map:
\[
\widetilde{\bv}_k^{(j)}\prod_{i=1}^{j-1}\Big(1+h\widetilde{\bv}_k^{(i)}\Big)
\cdot\Big(1+\epsilon_j h\widetilde{\bv}_{k-1}^{(j)}\Big)\cdot
\prod_{i=j+1}^{m}\Big(1+h\widetilde{\bv}_{k-1}^{(i)}\Big)
\]
\begin{equation}\label{dcKP loc}
=\bv_k^{(j)}\prod_{i=1}^{j-1}\Big(1+h\bv_{k+1}^{(i)}\Big)\cdot
\Big(1+\epsilon_j h\bv_{k+1}^{(j)}\Big)\cdot
\prod_{i=j+1}^{m}\Big(1+h\bv_k^{(i)}\Big)
\end{equation}
\end{theorem}
{\bf Proof.} It is easy to calculate that if (\ref{dcKP loc map}) holds, and
if the quantities $\gb_k^{(j)}$ are {\it defined} by the formula 
\begin{equation}\label{dcKP beta in loc map}
\gb_k^{(j)}=\prod_{i=1}^{j}\Big(1+h\bv_k^{(i)}\Big)
\prod_{i=j+1}^{m}\Big(1+h\bv_{k-1}^{(i)}\Big)
\end{equation}
then 
\[
\gb_k^{(j)}-hv_k^{(j)}=\prod_{i=1}^{j-1}\Big(1+h\bv_k^{(i)}\Big)
\prod_{i=j+1}^{m}\Big(1+h\bv_{k-1}^{(i)}\Big),\qquad \epsilon_j=0
\]
\[
\gb_k^{(j-1)}+hv_k^{(j)}=\prod_{i=1}^{j}\Big(1+h\bv_k^{(i)}\Big)
\prod_{i=j}^{m}\Big(1+h\bv_{k-1}^{(i)}\Big),\qquad \epsilon_j=1
\]
and it is easy to check now that the equations (\ref{dcKP aux1})--(\ref{dcKP
aux3}) are satisfied. Indeed, for $\epsilon_j=0$ we find:
\[
\displaystyle\frac{\gb_k^{(j)}}{\gb_k^{(j)}-hv_k^{(j)}}=
\displaystyle\frac{\gb_{k+1}^{(j-1)}}{\gb_{k+1}^{(j)}-hv_{k+1}^{(j)}}=
1+h\bv_k^{(j)}
\]
which proves (\ref{dcKP aux2}), while for $\epsilon_j=1$ we find
\[
\displaystyle\frac{\gb_k^{(j-1)}+hv_k^{(j)}}{\gb_k^{(j-1)}}=
\displaystyle\frac{\gb_{k+1}^{(j-1)}+hv_{k+1}^{(j)}}{\gb_{k+1}^{(j)}}=
1+h\bv_k^{(j)}
\]
which proves (\ref{dcKP aux3}). The verification of (\ref{dcKP aux1}) is
completely analogous. The pull--back of the equations of motion is calculated
now straightforwardly. \qed
\vspace{2mm}

Unfortunately, we do not now a general formula for the second invariant 
Poisson structure for cKPL's. This prevents us from applying our general
scheme for finding the local invariant Poisson brackets for the localized maps.
However, by a direct analysis of equations of motion the following statement
can be proved.
\begin{theorem} The pull--back of the system {\rm(\ref{cKP})} under the 
change of variables {\rm(\ref{dcKP loc map})} is given by the formula
\begin{eqnarray}\label{cKP in dcKP loc map}
\dot{\bv}_k^{(j)} & = & \bv_k^{(j)}\Big(1+h\bv_k^{(j)}\Big)\Bigg(
\prod_{i=1}^{j-1}\Big(1+h\bv_{k+1}^{(i)}\Big)
\cdot\Big(1+\epsilon_j h\bv_{k+1}^{(j)}\Big)\cdot
\prod_{i=j+1}^{m}\Big(1+h\bv_k^{(i)}\Big)\nonumber\\
&&-\prod_{i=1}^{j-1}\Big(1+h\bv_k^{(i)}\Big)
\cdot\Big(1+\epsilon_j h\bv_{k-1}^{(j)}\Big)\cdot
\prod_{i=j+1}^{m}\Big(1+h\bv_{k-1}^{(i)}\Big)\Bigg)/h
\end{eqnarray}
\end{theorem}

\subsection{Example 1: $\epsilon=(0,0,0)$}
As illustrations we consider the systems with $m=3$ -- the simplest possible 
ones after VL and RTL+. We denote for simplicity
\[
v_k^{(1)}=u_k,\qquad v_k^{(2)}=v_k, \qquad v_k^{(3)}=w_k
\]
The Hamilton function is in all cases equal to
\[
\rH_1(u,v,w)=\sum_{k=1}^N (u_k+v_k+w_k)
\]
The nonvanishing Poisson brackets consist of the signature independent part,
\begin{eqnarray}
\{u_k,v_k\}_2=-u_kv_k,\qquad \{v_k,u_{k+1}\}_2=-u_{k+1}v_k \nonumber\\
\{u_k,w_k\}_2=-u_kw_k,\qquad \{w_k,u_{k+1}\}_2=-u_{k+1}w_k\label{cKP ex PB}\\
\{v_k,w_k\}_2=-v_kw_k,\qquad \{w_k,v_{k+1}\}_2=-v_{k+1}w_k\nonumber
\end{eqnarray}
supplemented by
\[
\{v_k^{(j)},v_{k+1}^{(j)}\}_2=-v_k^{(j)}v_{k+1}^{(j)}
\]
for those $j$ where $\epsilon_j=1$.

In particular, if all $\epsilon_j=0$, then all nonvanishing Poisson brackets
of the coordinate functions are exhausted by (\ref{cKP ex PB}), and we arrive
at the system
\begin{eqnarray}
\dot{u}_k & = & u_k(v_k+w_k-v_{k-1}-w_{k-1}) \nonumber\\
\dot{v}_k & = & v_k(u_{k+1}+w_k-u_k-w_{k-1}) \label{cKP BL1}\\
\dot{w}_k & = & w_k(u_{k+1}+v_{k+1}-u_k-v_k)\nonumber
\end{eqnarray}
which becomes the usual Bogoyavlensky lattice BL1(2) after the re--naming
\[
u_k\mapsto a_{3k-2},\qquad v_k\mapsto a_{3k-1},\qquad w_k\mapsto a_{3k}
\]
The localizing change of variables for its discretization:
\begin{eqnarray}
u_k & = & \bu_k(1+h\bv_{k-1})(1+h\bw_{k-1})  \nonumber\\
v_k & = & \bv_k(1+h\bu_k)(1+h\bw_{k-1})  \label{dcKP BL1 loc map}\\
w_k & = & \bw_k(1+h\bu_k)(1+h\bv_k) \nonumber 
\end{eqnarray}
The local discretization of the system (\ref{cKP BL1}):
\begin{eqnarray}
&&\widetilde{\bu}_k(1+h\widetilde{\bv}_{k-1})(1+h\widetilde{\bw}_{k-1})  = 
\bu_k(1+h\bv_k)(1+h\bw_k)  \nonumber\\
&&\widetilde{\bv}_k(1+h\widetilde{\bu}_k)(1+h\widetilde{\bw}_{k-1}) = 
\bv_k(1+h\bu_{k+1})(1+h\bw_k) \label{dcKP BL1 loc}\\
&&\widetilde{\bw}_k(1+h\widetilde{\bu}_k)(1+h\widetilde{\bv}_k)  = 
\bw_k(1+h\bu_{k+1})(1+h\bv_{k+1})  \nonumber
\end{eqnarray}

\subsection{Example 2: $\epsilon=(0,1,0)$}
In this example the signature dependent part of the Poisson brackets reads
\[
\{v_k,v_{k+1}\}_2=-v_kv_{k+1}
\]
and the equations of motion take the following form:
\begin{eqnarray}
\dot{u}_k & = & u_k(v_k+w_k-v_{k-1}-w_{k-1}) \nonumber\\
\dot{v}_k & = & v_k(u_{k+1}+v_{k+1}+w_k-u_k-v_{k-1}-w_{k-1}) \label{cKP ex1}\\
\dot{w}_k & = & w_k(u_{k+1}+v_{k+1}-u_k-v_k)\nonumber
\end{eqnarray}
The localizing change of variables for the discretization of this system:
\begin{eqnarray}\label{dcKP ex1 loc map}
u_k & = & \bu_k(1+h\bv_{k-1})(1+h\bw_{k-1}) \nonumber\\
v_k & = & \bv_k(1+h\bu_k)(1+h\bv_{k-1})(1+h\bw_{k-1}) \\
w_k & = & \bw_k(1+h\bu_k)(1+h\bv_k)
\end{eqnarray}
The local form of equations of motion for the discretization of (\ref{cKP ex1}):
\begin{eqnarray}
&&\widetilde{\bu}_k(1+h\widetilde{\bv}_{k-1})(1+h\widetilde{\bw}_{k-1})  = 
\bu_k(1+h\bv_k)(1+h\bw_k)  \nonumber\\
&&\widetilde{\bv}_k(1+h\widetilde{\bu}_k)(1+h\widetilde{\bv}_{k-1})
(1+h\widetilde{\bw}_{k-1}) = 
\bv_k(1+h\bu_{k+1})(1+h\bv_{k+1})(1+h\bw_k)  \nonumber\\
&&\widetilde{\bw}_k(1+h\widetilde{\bu}_k)(1+h\widetilde{\bv}_k)  = 
\bw_k(1+h\bu_{k+1})(1+h\bv_{k+1}) \label{dcKP ex1 loc}
\end{eqnarray}

Let us mention that the system (\ref{cKP ex1}) allows an interesting reduction
\begin{equation}\label{cKP ex1 reduction}
v_k=u_kw_k
\end{equation}
which is, moreover, compatible with the quadratic Poisson brackets. In this
reduction we arrive at the system PVL.
It is easy to check that in the variables $\bu_k,\bv_k,\bw_k$ the reduction
(\ref{cKP ex1 reduction}) takes the form $\bv_k=\bu_k\bw_k(1+h\bv_k)$,
so that
\[
\bv_k=\frac{\bu_k\bw_k}{1-h\bu_k\bw_k},\qquad 1+h\bv_k=
\frac{1}{1-h\bu_k\bw_k}
\]
This makes a link with the results of Sect. \ref{Sect perturbed Volterra}.

\subsection{Example 3: $\epsilon=(0,1,1)$}
In this case the Poisson brackets (\ref{cKP ex PB}) have to be supplemented by 
\[
\{v_k,v_{k+1}\}_2=-v_kv_{k+1},\qquad \{w_k,w_{k+1}\}_2=-w_kw_{k+1}
\]
and the equations of motion take the form:
\begin{eqnarray}
\dot{u}_k & = & u_k(v_k+w_k-v_{k-1}-w_{k-1}) \nonumber\\
\dot{v}_k & = & v_k(u_{k+1}+v_{k+1}+w_k-u_k-v_{k-1}-w_{k-1}) \label{cKP ex2}\\
\dot{w}_k & = & w_k(u_{k+1}+v_{k+1}+w_{k+1}-u_k-v_k-w_{k-1})\nonumber
\end{eqnarray}
The localizing change of variables for the discretization of this system:
\begin{eqnarray}\label{dcKP ex2 loc map}
u_k & = & \bu_k(1+h\bv_{k-1})(1+h\bw_{k-1}) \nonumber\\
v_k & = & \bv_k(1+h\bu_k)(1+h\bv_{k-1})(1+h\bw_{k-1}) \\
w_k & = & \bw_k(1+h\bu_k)(1+h\bv_k)(1+h\bw_{k-1}) \nonumber
\end{eqnarray} 
The local form of equations of motion for the discretization of (\ref{cKP ex2}):
\begin{eqnarray}
&&\widetilde{\bu}_k(1+h\widetilde{\bv}_{k-1})(1+h\widetilde{\bw}_{k-1})  = 
\bu_k(1+h\bv_k)(1+h\bw_k)   \nonumber\\
&&\widetilde{\bv}_k(1+h\widetilde{\bu}_k)(1+h\widetilde{\bv}_{k-1})
(1+h\widetilde{\bw}_{k-1}) = 
\bv_k(1+h\bu_{k+1})(1+h\bv_{k+1})(1+h\bw_k)  \nonumber\\
&&\widetilde{\bw}_k(1+h\widetilde{\bu}_k)(1+h\widetilde{\bv}_k)
(1+h\widetilde{\bw}_{k-1})  = 
\bw_k(1+h\bu_{k+1})(1+h\bv_{k+1})(1+h\bw_{k+1})\nonumber\\
\label{dcKP ex2 loc}
\end{eqnarray}

Let us discuss the following reduction of the system (\ref{cKP ex2}):
\begin{equation}\label{cKP ex2 reduction}
u_k=0
\end{equation}
It is compatible with the quadratic Poisson brackets, so that we arrive at the
following reduced system:
\begin{equation}\label{cKP ex2 red}
\begin{array}{l}
\dot{v}_k = v_k(v_{k+1}+w_k-v_{k-1}-w_{k-1})\\ \\
\dot{w}_k = w_k(v_{k+1}+w_{k+1}-v_k-w_{k-1})
\end{array}
\end{equation}
Interestingly enough, this system is again nothing but the usual Bogoyavlensky
lattice BL1(2), which becomes obvious after the re--naming
\[
v_k\mapsto a_{2k-1},\qquad w_k\mapsto a_{2k}
\]
So, we have found the {\it third} Lax representation for BL1(2). 

It is easy to see that the maps $\cM_{\pm}:\cK_3(0,v,w)\mapsto{\cal BC}(b,c)$ 
defined as
\begin{equation}\label{BC M+}
\cM_+:\qquad b_k=v_k+w_k, \quad c_k=v_kw_{k+1}
\end{equation}
and
\begin{equation}\label{BC M-}
\cM_-:\qquad b_k=v_{k+1}+w_k, \quad c_k=v_{k+2}w_k
\end{equation}
conjugate the flow (\ref{cKP ex2 red}) with the Belov--Chaltikian flow BCL, 
and are Poisson, if the both spaces are equipped with the brackets 
$\{\cdot,\cdot\}_2$. So, the system BCL is Miura related to BL1(2)
(this fact is similar to the Miura relation between the Toda and the 
Volterra hierarchy).

The localizing change of variables for the discretization of the reduced 
system (\ref{cKP ex2 red}) is given by
\begin{equation}\label{dcKP ex2 red loc map}
v_k=\bv_k(1+h\bv_{k-1})(1+h\bw_{k-1}),\qquad
w_k=\bw_k(1+h\bv_k)(1+h\bw_{k-1})
\end{equation}
and the corresponding local equations of motion read:
\begin{equation}\label{dcKP ex2 red loc}
\begin{array}{l}
\widetilde{\bv}_k(1+h\widetilde{\bv}_{k-1})(1+h\widetilde{\bw}_{k-1}) = 
\bv_k(1+h\bv_{k+1})(1+h\bw_k) \\ \\
\widetilde{\bw}_k(1+h\widetilde{\bv}_k)(1+h\widetilde{\bw}_{k-1})  = 
\bw_k(1+h\bv_{k+1})(1+h\bw_{k+1})
\end{array}
\end{equation}
So, the discretizations of BL1(2) based on different Lax representations, 
agree with one another.  

It turns out that the Miura maps $\cM_{\pm}$ are still given by
nice local formulas, when translated to the localizing variables. Namely,
the following diagram is commutative:

\begin{center}
\unitlength1cm
\begin{picture}(9,6.5)
\put(3.5,1.1){\vector(1,0){2}}
\put(3.5,5.1){\vector(1,0){2}}
\put(2,4.1){\vector(0,-1){2}}
\put(7,4.1){\vector(0,-1){2}}
\put(1,0.6){\makebox(2,1){$\cK_3(0,v,w)$}} 
\put(1,4.6){\makebox(2,1){$\cK_3(0,\bv,\bw)$}}
\put(6,4.6){\makebox(2,1){${\cal BC}(\bb,\bc)$}}
\put(6,0.6){\makebox(2,1){${\cal BC}(b,c)$}}
\put(0,2.6){\makebox(2,1){{\rm(\ref{dcKP ex2 red loc map})}}}
\put(7,2.6){\makebox(2,1){{\rm(\ref{dBC loc map})}}}
\put(3.8,-0.2){\makebox(1.4,1.4){$\cM_{\pm}$}}
\put(3.8,5.0){\makebox(1.4,1.4){$\bM_{\pm}$}}
\end{picture}
\end{center}
if the maps $\bM_{\pm}$ are defined by the formulas
\begin{equation}\label{dBC M+}
\bM_+: \qquad 1+h\bb_k=(1+h\bv_k)(1+h\bw_k),\quad 
\bc_k=\bv_k\bw_{k+1}(1+h\bw_k)(1+h\bv_{k+1}) 
\end{equation}
and
\begin{equation}\label{dBC M-}
\bM_-:  \qquad 1+h\bb_k=(1+h\bv_{k+1})(1+h\bw_k),\quad 
\bc_k=\bv_{k+2}\bw_k(1+h\bv_{k+1})(1+h\bw_{k+1}) 
\end{equation}
This statement may be verified by a simple calculation.



\setcounter{equation}{0}
\section{Bruschi--Ragnisco lattice}\label{Chapter Bruschi--Ragnisco}

The Bruschi--Ragnisco lattice (hereafter BRL)
was introduced in \cite{BR2}:
\begin{equation}\label{br}
\dot{b}_k=b_{k+1}c_k-b_kc_{k-1}, \qquad \dot{c}_k=c_k(c_k-c_{k-1})
\end{equation}
It may be considered either under open--end boundary conditions
($b_{N+1}=c_0=c_N=0$), or under periodic ones (all the subscripts are
taken (mod $N$), so that $c_0\equiv c_N$, $b_{N+1}\equiv b_1$). 
The phase space of the Bruschi--Ragnisco lattice: 
\[
{\cal BR}={\Bbb R}^{2N}(b_1,c_1,\ldots,b_N,c_N)
\]
Two compatible brackets may be defined on $\cal BR$ such that the system 
(\ref{br}) is Hamiltonian with respect to each one of them. The linear Poisson
bracket is given by
\begin{equation}\label{br l br}
\{b_k,c_k\}_0=-\{b_{k+1},c_k\}_0=-c_k
\end{equation}
while the quadratic one - by
\begin{equation}\label{br q br}
\{b_k,b_{k+1}\}_1=-b_{k+1}c_k,\qquad \{b_k,c_k\}_1=c_k^2,\qquad
\{b_k,c_{k+1}\}_1=-c_kc_{k+1}
\end{equation}
The corresponding Hamilton functions are: 
\begin{equation}\label{br H1}
\rH_1(b,c)=\sum_{k=1}^Nb_{k+1}c_k\qquad{\rm and}
\qquad
\rH_0(b,c)=\sum_{k=1}^N b_k
\end{equation}
for the brackets $\{\cdot,\cdot\}_0$ and $\{\cdot,\cdot\}_1$, respectively.

It has been pointed out in \cite{BR2} that this system allows a  very complete
study by different methods of the soliton theory. As was demonstrated
in \cite{S2}, this is due to its extreme simplicity. Namely, in a certain 
gauge the Lax representation of this system is a linear matrix equation.
Namely, if the entries of the Lax matrix $T=T(b,c)\in\g=gl(N)$ are defined as
\begin{equation}\label{br T}
T_{kj}=\left\{\begin{array}{lcl}b_j\prod_{i=k}^{j-1}c_i &\quad  &k\le j\\ \\
b_j\left(\prod_{i=j}^{k-1}c_i\right)^{-1} & \quad  & k>j\end{array}\right.
\end{equation}
then the system (\ref{br}) is equivalent to the equation
\begin{equation}\label{br Lax}
\dot{T}=[T,M]
\end{equation}
with the constant matrix $M$
\begin{equation}\label{br M}
M=\sum_{k=1}^{N-1} E_{k,k+1}\qquad {\rm or}
\qquad\sum_{k=1}^{N-1} E_{k,k+1}+CE_{N,1}
\end{equation}
for the open--end and periodic case, respectively (in the latter case it
is supposed that the dynamics of the BRL is restricted to the set
$c_1\ldots c_N=C$). The whole hierarchy of the BRL consists of equations
\[
\dot{T}=[T,M^m]
\]
which are linear and may be immediately integrated:
\begin{equation}\label{br solution}
T(t)=\exp(-tM^m)\cdot T(0)\cdot\exp(tM^m)
\end{equation}
The brackets (\ref{br l br}), (\ref{br q br}) were shown in \cite{S2} to give a
coordinate representation of certain Lie--Poisson brackets on $\g$, restricted
to the subset of the Lax matrices $T(b,c)$.
\vspace{1.5mm}

Obviously, the recipe of Sect. \ref{Sect recipe} cannot be literally
applied to the BRL. However, the philosophy behind
this recipe is, of course, applicable, and requires to seek for the discrete 
time Bruschi--Ragnisco lattice in the same hierarchy. It should share 
the Lax matrix with the continuous time system, and its explicit solution
should be given by
\begin{equation}\label{dbr solution}
T(nh)=(I+hM)^{-n}T(0)(I+hM)^n
\end{equation}
(cf. (\ref{br solution})).
Hence the corresponding discrete Lax equation should have the form
\begin{equation}\label{dbr Lax}
\widetilde{T}=(I+hM)^{-1}T(I+hM)
\end{equation}
\begin{theorem}
The discrete time Lax equation {\rm(\ref{dbr Lax})} is equivalent to the
following map on the space ${\cal BR}$:
\begin{equation}\label{dbr}
\wb_k(1+h\wc_{k-1})=b_k+hb_{k+1}c_k, \qquad
\wc_k=c_k\;\frac{1+h\wc_k}{1+h\wc_{k-1}}
\end{equation}
\end{theorem}
{\bf Proof} -- an easy calculation. \qed
\vspace{2mm}

By construction, this map is Poisson with respect to the both brackets 
(\ref{br l br}), (\ref{br q br}).
We see that the extreme simplicity of the BRL allows to find its local 
discretization in the original variables. The localizing change of
variables is not necessary for this system.


\setcounter{equation}{0}
\section{Conclusion}

This paper contains a rich collection of examples illustrating the 
procedure of constructing local integrable discretizations for 
integrable lattice systems. The construction is based on the notion of
the $r$-matrix hierarchy and consists of three steps of a rather different
nature. 

The first step is to find a Lax representation for a given lattice system,
living in an associative algebra $\g$. This Lax representation has to 
be a member of a hierarchy governed by an $R$--operator on $\g$ 
satisfying the modified Yang--Baxter equation. In all examples
treated here this operator is simply a difference of projections to two
complementary subalgebras. 

The second step is an application of a general recipe for integrable 
discretization. This step is almost algorithmic, the only non--formalized
(and, probably, non--formalizable) point being the choice of the function
$F(T)$ approximating $\exp(hT)$ for $T\in\g$ (cf. Sect. \ref{Sect recipe}).
In all examples treated here the simplest possible choice $F(T)=I+hT$ works
perfectly. The difference equations obtained on this step share the invariant
Poisson structures, the integrals of motion, the Lax matrices, etc. with the 
underlying continuous time systems. However, as a rule, they are non--local. 
This feature is unpleasant from both the esthetical and the
practical point of view, because it makes the equations ugly and not well 
suited for practical realization.

The third step is finding the localizing change of variables. This step is again
absolutely non--algorithmic (at least, at our present level of knowledge).
These changes of variables have remarkable properties: they often produce
one--parameter local deformations of Poisson brackets algebras, and always 
produce one--parameter integrable deformations of the lattice systems themselves.
At the moment we cannot provide a rational explanation neither for these 
properties nor for the mere existence of localizing changes of variables. 
However, our collection seems to be 
representative enough to convince that these phenomena are very general.
We feel that they are connected with the Poisson geometry of certain 
$r$--matrix brackets on associative algebras and of monodromy maps, but
we prefer to stop at this point.


\end{document}